\documentclass[a4paper,aps,prx,reprint,notitlepage,superscriptaddress,nofootinbib,twocolumn,notitlepage,floatfix]{revtex4-1}
\usepackage[utf8]{inputenc}
\usepackage[T1]{fontenc}

\usepackage{graphicx}
\usepackage{amsmath}  
\usepackage{braket}
\usepackage{amsfonts} 
\usepackage{amssymb} 
\usepackage{csquotes}
\usepackage[x11names]{xcolor}
\usepackage{siunitx}

\usepackage{bm}
\usepackage{stmaryrd}
\usepackage{booktabs}
\usepackage{enumitem}
\usepackage{mathrsfs}

\usepackage{microtype}

\usepackage{xpatch}
\makeatletter
\xpatchcmd{\@ssect@ltx}{\@xsect}{\protected@edef\@currentlabelname{#8}\@xsect}{}{}\xpatchcmd{\@sect@ltx}{\@xsect}{\protected@edef\@currentlabelname{#8}\@xsect}{}{}\makeatother

\usepackage{hyperref}
\usepackage{bookmark}

\definecolor{tab_blue}{HTML}{1F77B4}
\definecolor{tab_orange}{HTML}{FF7F0E}
\definecolor{tab_green}{HTML}{2CA02C}
\definecolor{tab_red}{HTML}{D62728}
\definecolor{tab_purple}{HTML}{9467BD}
\definecolor{tab_brown}{HTML}{8C564B}
\definecolor{tab_pink}{HTML}{E377C2}
\definecolor{tab_gray}{HTML}{7F7F7F}
\definecolor{tab_olive}{HTML}{BCBD22}
\definecolor{tab_cyan}{HTML}{17BECF}

\hypersetup{breaklinks=true,
  colorlinks=true,
  linkcolor=tab_green,
  filecolor=tab_green,
  urlcolor=tab_green,
  citecolor=tab_green,
  pdfborder=0 0 0,
}

\def\sign{\text{sign}}

\def\ee{\text{e}}
\def\ii{\text{i}}
\def\dd{\text{d}}
\def\absvec#1{\big\lVert#1\big\rVert}

\def\RR{\ensuremath{\mathbb{R}}}

\def\ZZ{\ensuremath{\mathbb{Z}}}

\def\Id{\ensuremath{\text{Id}}}

\def\strong#1{\textbf{#1}}

\def\div{\operatorname{div}}
\def\grad{\operatorname{grad}}

\def\figurename{Fig.}
\makeatletter
\def\fnum@figure{\textbf{\figurename~\thefigure}}
\makeatother
\makeatletter
\def\fnum@table{\textbf{\tablename~\thetable}}
\makeatother

\def\siref#1#2{\ref{#1}}
\def\mainref#1#2{\ref{#1}}

\def\methodsection#1{\section{#1}}
\def\methodsref#1#2{\ref{#1}~\emph{\nameref{#1}}}

\makeatletter\begin{document}

\title{Non-reciprocal phase transitions}

\author{Michel Fruchart}
\thanks{These authors contributed equally to this work.}
\affiliation{James Franck Institute and Department of Physics, University of Chicago, Chicago IL 60637, USA}

\author{Ryo Hanai}
\thanks{These authors contributed equally to this work.}
\affiliation{James Franck Institute and Department of Physics, University of Chicago, Chicago IL 60637, USA}
\affiliation{Department  of  Physics,  Osaka  University,  Toyonaka  560-0043,  Japan}
\affiliation{Pritzker School of Molecular Engineering, University of Chicago, Chicago IL 60637, USA}

\author{Peter B. Littlewood}
\affiliation{James Franck Institute and Department of Physics, University of Chicago, Chicago IL 60637, USA}

\author{Vincenzo Vitelli}
\affiliation{James Franck Institute and Department of Physics, University of Chicago, Chicago IL 60637, USA}
\affiliation{Kadanoff Center for Theoretical Physics, University of Chicago, Chicago IL 60637, USA} \date{\today}

\begin{abstract}
Out of equilibrium, the lack of reciprocity is the rule rather than the exception. 
Non-reciprocal interactions occur, for instance, in networks of neurons~\cite{Dayan2001,Amir2016,Krakauer2017,Montbri2018,Parisi1986,Derrida1987,Sompolinsky1986}, directional growth of interfaces~\cite{Cross1993}, and synthetic active materials~\cite{Marchetti2013,Kotwal2019,Bricard2013,Palacci2013,Geyer2018,Bain2019,vanZuiden2016,Dayan2001,Caussin2014}.
While wave propagation in non-reciprocal media has recently been under intense study~\cite{Fleury2014,Estep2014,Coulais2017,Miri2019,Scheibner2020,Ghatak2019,Lee2019,Helbig2020,Lin2011,Schindler2011,Weidemann2020,Hofmann2019,Brandenbourger2019,Bender2007,Ramos2020}, less is known about the consequences of non-reciprocity on the collective behavior of many-body systems.
Here, we show that non-reciprocity leads to time-dependent phases where spontaneously broken symmetries are dynamically restored. 
The resulting phase transitions are controlled by spectral singularities called exceptional points~\cite{Kato1984}.
We describe the emergence of these phases using insights from bifurcation theory~\cite{Golubitsky2002,Crawford1991} and non-Hermitian quantum mechanics~\cite{Hatano1996,Bender1998}.
Our approach captures non-reciprocal generalizations of three archetypal classes of self-organization out of equilibrium: synchronization, flocking and pattern formation. 
Collective phenomena in these non-reciprocal systems range from active time-(quasi)crystals to exceptional-point enforced pattern-formation and hysteresis. 
Our work paves the way towards a general theory of critical phenomena in non-reciprocal matter. \end{abstract}

\maketitle

\begin{figure*}[p]
  \centering
  \hspace*{0cm}
  \includegraphics{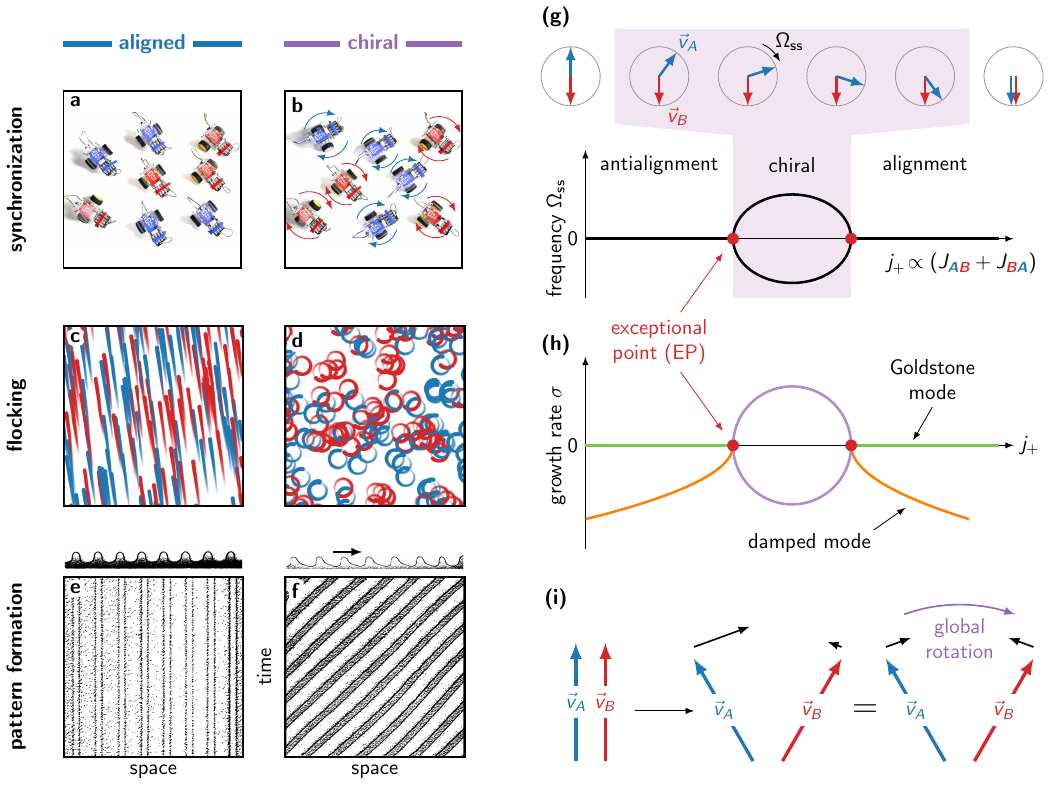}
  \caption{\label{figure_bots_phases}\strong{Exceptional transitions from non-reciprocity and continuous symmetries.}
  Non-reciprocal interactions between two species $A$ and $B$ ($J_{AB} \ne J_{BA}$) induce a phase transition from static alignment to a chiral motion that spontaneously breaks parity.
  (a-b) Non-reciprocal synchronization. The oscillators represented by robots (programmed as non-reciprocal spins) spontaneously rotate either clockwise or counterclockwise, despite having no average natural frequency, i.e. $\omega_m = 0$ in Eq.~\eqref{kuramoto}.
  (See SI Movie~\ref{movie_robots} for a demonstration.)
  (c-d) Non-reciprocal flocking. The oscillators are replaced by self-propelled particles. They run in circles, either clockwise or counterclockwise, despite the absence of any external torque.
  See Fig.~\ref{figure_phase_diagram} for details and SI Movie \ref{movie_vicsek} for movies of the simulations.
  (e-f) Non-reciprocal pattern formation. A one-dimensional pattern starts traveling, either to the left or to the right, see Methods for details.
  The figure represents an experimental observation of viscous fingering at an oil-air interface adapted with permission from Ref.~\cite{Pan1994} (Copyright 1994 by the American Physical Society).
  In this case, the species correspond to different harmonics of the same field.
  (g) Schematic bifurcation diagram of the exceptional transition showing the frequency of the steady-state $\Omega_{\text{ss}}$.
  Between the static phases with alignment and antialignment (in which $\Omega_{\text{ss}} = 0$), we find an intermediate chiral phase that spontaneously breaks parity. 
  Two equivalent steady-states (clockwise and counterclockwise, corresponding to opposite values of $\Omega_{\text{ss}}$) are present in this time-dependent phase, which can be seen as a manifestation of spontaneous PT-symmetry breaking.
  The chiral phase continuously interpolates between the antialigned and aligned phases, both through $|\Omega_{\text{ss}}|$ and through the angle between the order parameters $\vec{v}_A$ and $\vec{v}_B$.
  (h) The transition between (anti)aligned and chiral phases occurs through the coalescence of a damped mode (orange) and a Goldstone mode (green) at an exceptional point (EP, marked by a red circle).  
  (i) Non-reciprocity leads to an exceptional point.
  In a reciprocal system, rotating the order parameter $\vec{v}_{A}$ has the same effect on $\vec{v}_{B}$ as rotating $\vec{v}_{B}$ would have on $\vec{v}_{A}$. 
  By definition, this is not the case in non-reciprocal systems. 
Hence, a purely antisymmetric perturbation (for which the velocities $\vec{v}_A$ and $\vec{v}_B$ are rotated by equal amounts in opposite directions) does not generate a purely antisymmetric response: $\vec{v}_A$ is rotated back by a greater amount than $\vec{v}_B$ (black arrows in middle panel). 
This response is equivalent to pushing back $\vec{v}_A$ and $\vec{v}_B$ towards each other by equal amounts (black arrows in right panel), plus applying a global rotation (purple arrow). 
The purely symmetric perturbation (solid-body rotation) is a Goldstone mode, so it does not generate any response.
As the non-reciprocal coupling increases, it reaches a critical value where all perturbations result in a global rotation of $\vec{v}_A$ and $\vec{v}_B$.
This corresponds to an exceptional point and marks the onset of the chiral phase.
  }
\end{figure*}

Non-reciprocity naturally arises in nonequilibrium many-body systems~\cite{Nagy2010,Morin2015,Ginelli2015,Yllanes2017,Durve2018,Barberis2016,Lavergne2019,Bonilla2019,Costanzo2019,Dadhichi2019,AgudoCanalejo2019,Lahiri1997,Lahiri2000,Chajwa2020,Gupta2020,Soto2014,Ivlev2015,Kryuchkov2018,Saha2019,Maitra2020,Uchida2010,Hanai2019,Hanai2020,Metelmann2015} ranging from inhibitory and excitatory neurons~\cite{Dayan2001,Amir2016,Krakauer2017,Montbri2018,Parisi1986,Derrida1987,Sompolinsky1986} to conformist and contrarian members of social groups~\cite{Hong2011,Hong2011b,Pluchino2005}.
Our goal is to explore how non-reciprocity affects phase transitions.
To do so, we consider multiple species or fields (interacting non-reciprocally with one another) modeled by a vector order parameter $\vec{v}_a(t, \vec{x})$ for each species $a$.
These can encode, for instance, the average velocities of self-propelled particles, the average phases of coupled oscillators such as biological clocks or firing neurons, or the amplitude and position of a periodic pattern (Fig.~\ref{figure_bots_phases}a-d). 
The order parameters $\vec{v}_a(t, \vec{x})$ could either be distinct fields, or different harmonics of the same physical field such as in directional interface growth experiments~\cite{Malomed1984,Coullet1989,Flesselles1991,Pan1994}, see Fig.~\ref{figure_bots_phases}e-f.

All of these systems are described by the evolution equation
\begin{equation}
	\label{general_SO2_invariant_main}
	\partial_t \vec{v}_a = \mathbb{A}_{a b} \, \vec{v}_b + \mathbb{B}_{a b c d} \, ({\vec{v}_b \cdot \vec{v}_c}) \, \vec{v}_d + \mathcal{O}(\nabla)
\end{equation}
where summation over repeated indices is implied. 
Equation~\eqref{general_SO2_invariant_main} is (up to third order in $\vec{v}_a$) the most general dynamical system invariant under rotations.
In flocking, rotational symmetry naturally arises from the isotropy of space, while in synchronization and pattern formation it emerges from the underlying time or space translation invariance. 
Different symmetries or other representations of rotations can be similarly enforced~\cite{Golubitsky2002} leading to variants of Eq.~\eqref{general_SO2_invariant_main}, see Methods and SI Sec.~\siref{sec_O3}{VIII}. 
The quantities $\mathbb{A}_{a b}$ and $\mathbb{B}_{a b c d}$ are arrays of parameters that couple the different species of fields. 
As we shall see, they are also matrices operating on the vectors $\vec{v}_a$ that reduce to the identity for full rotation symmetry, but have more complicated forms when parity (mirror reflection) is broken.

While our ultimate goal is to model spatially extended systems, we have temporarily omitted in Eq.~\eqref{general_SO2_invariant_main} terms with spatial derivatives and only retained nonlinearities essential to discuss transitions between distinct non-equilibrium steady-states. 
Here, we allow the macroscopic coefficients to be asymmetric, e.g. $\mathbb{A}_{a b} \ne \mathbb{A}_{b a}$. 
In the Landau theory of equilibrium phase transitions, these coefficients would be fully symmetric because the dynamics of the order parameter $\partial_t \vec{v}_a = - \partial_{\vec{v}_a} F$ is derived from a free energy $F$.
Removing this symmetry constraint allows to extend the theory of critical phenomena~\cite{Hohenberg1977} to fields with non-reciprocal interactions.

Dynamical systems described by Eq.~\eqref{general_SO2_invariant_main} arise in various forms of non-reciprocal matter.
Consider, for instance, a collection of coupled oscillators described by the Kuramoto model~\cite{Kuramoto1984,Acebron2005}
\begin{equation}
  \label{kuramoto}
  \partial_t \theta_m = \omega_m + \sum_{n} J_{m n} \sin(\theta_n - \theta_m) + \eta(t).
\end{equation}
The metronome or neuron labeled by $m$ in Eq.~\eqref{kuramoto} ticks or fires when its phase~$\theta_m(t)$ crosses a fixed value (say zero). 
The oscillator $m$ has a natural frequency $\omega_m$ and is coupled to the other oscillators by $J_{m n}$.
In the Kuramoto model, the oscillators are typically all coupled and the random noise $\eta(t)$ is usually ignored.
Equation~\eqref{kuramoto} with $\omega_m=0$ and with random noise also captures the Vicsek model of flocking~\cite{Vicsek1995,Toner1995}, provided that the oscillators are replaced by self-propelled particles moving at constant speed $v_0$ in the plane in the direction $\theta_m$. Their positions $\vec{r}_m$ then follow the equation
\begin{equation}
    \label{vicsek_r}
    \partial_t \vec{r}_m = v_0 \, \begin{pmatrix}
    \cos\theta_m \\
    \sin\theta_m
\end{pmatrix}.
\end{equation}
The couplings $J_{m n}$ are short-ranged and depend on $\vec{r}_m$.
In both models, the agent $m$ tries to align (be in phase) with the agent $n$ when $J_{m n}$ is positive, and to antialign with it when $J_{m n}$ is negative. 

Above a critical coupling, both models exhibit a phase transition from incoherent motion (incoherent oscillations) to flocking (synchronization) heralded by a non-vanishing order parameter $\vec{v}_a$ (Fig.~\ref{figure_bots_phases}a,c).
We show in the Methods and SI Sec.~\siref{sec_hydro}{I} that coarse-graining these microscopic models leads to an evolution equation of the form of Eq.~\eqref{general_SO2_invariant_main}, with the addition of rotationally invariant  terms with spatial derivatives [see Eqs.~\eqref{eom_with_gradients_two_populations} and \eqref{kuramoto_ott_antonsen}]. 
As expected, the coefficients of the order parameter equations become asymmetric when the interaction is non-reciprocal (i.e., when $J_{mn} \ne J_{nm}$), see also Refs.~\cite{Daido1992,Sakaguchi1986,Pal2017,Hong2011,Hong2011b,Martens2009,Bonilla1998,Uchida2010,Hong2012,Ott2008,Abrams2008,Pikovsky2008,Choe2016,Gallego2017,Bonilla2019,OKeeffe2017,Levis2019,Das2002,Das2004,Matheny2019} for situations in which this can occur. Equation~\eqref{general_SO2_invariant_main}, viewed as an amplitude equation, also describes non-reciprocal pattern formation far from equilibrium~\cite{Cross1993} (Fig.~\ref{figure_bots_phases}e).
For example, the non-reciprocal Swift-Hohenberg model~\cite{Swift1977}
\begin{equation}
	\label{coupled_swift_hohenberg_main}
	\partial_t u_{a} = r_{a b} u_{b} - (1 + \nabla^2)^2 u_{a} - g u_{a}^3
\end{equation}
with $r_{a b} \ne r_{b a}$ reduces to Eq.~\eqref{general_SO2_invariant_main} by letting $u_a(x) = A_a(x) \ee^{\ii k x} + \text{c.c.}$, where $k$ is a wavevector and the complex amplitude is decomposed as $A_{a} \equiv v_{a}^x + \ii v_{a}^y$ (see Methods).

Let us start with only two species A and B and consider models where parity is not explicitly broken.
When the inter-species interactions are reciprocal, we find (besides a disordered phase) two {\it static} phases where the order parameters $\vec{v}_A$ and $\vec{v}_B$ (red and blue arrows in Fig.~\ref{figure_bots_phases}g-i) are either aligned or antialigned, in analogy with (anti)ferromagnetism.
When the inter-species interactions are non-reciprocal, the macroscopic coefficients in Eq.~\eqref{general_SO2_invariant_main} can become asymmetric and we find a time-dependent chiral phase {with no equilibrium analogue} that emerges between the static phases (Fig.~\ref{figure_bots_phases}g-h).
In this intermediate chiral phase, parity is spontaneously broken:  $\vec{v}_A$ and $\vec{v}_B$ rotate either clockwise or counterclockwise at a constant speed~$\Omega_{\text{ss}}$, while keeping a fixed relative angle. 
{We stress that this rotation of the order parameters is different from the precession of spins in a magnetic field, that can be simply captured with a Hamiltonian.}

Figure~\ref{figure_bots_phases}a-f illustrates the aligned and chiral phases in synchronization, flocking and pattern formation.
{
Panels e-f show a manifestation of the aligned to chiral transition in viscous fingering experiments~\cite{Pan1994,Rabaud1990,Cummins1993,Bellon1998}, see Methods for a detailed treatment.
We have identified additional examples in naturally occurring phenomena ranging from directional solidification of liquid-crystals~\cite{Simon1988,Flesselles1991,Melo1990,Oswald1987} and growth of lamellar eutectics~\cite{Faivre1989,Faivre1992,Kassner1990,Ginibre1997} to overflowing fountains~\cite{Counillon1997,Brunet2001}. 
Unlike the examples in panels a-d where two species are present, here the non-reciprocal (asymmetric) couplings in the amplitude equations occur between two different harmonics $A$ and $B$ of the {\it same} physical field, see Methods.
}

Intuitively, the chiral phase is caused by the frustration experienced by species with opposite goals.
For concreteness, consider two populations of agents $A$ and $B$ where
$A$ want to align with $B$ but $B$ want to antialign with $A$.
As the agents are never satisfied, they start running in circles.
This dynamical frustration results macroscopically in a constant \enquote{chase and runaway} motion of the order parameters ${\vec v}_a$: this is the chiral phase. 

\begin{figure*}[p]
  \centering
  \includegraphics{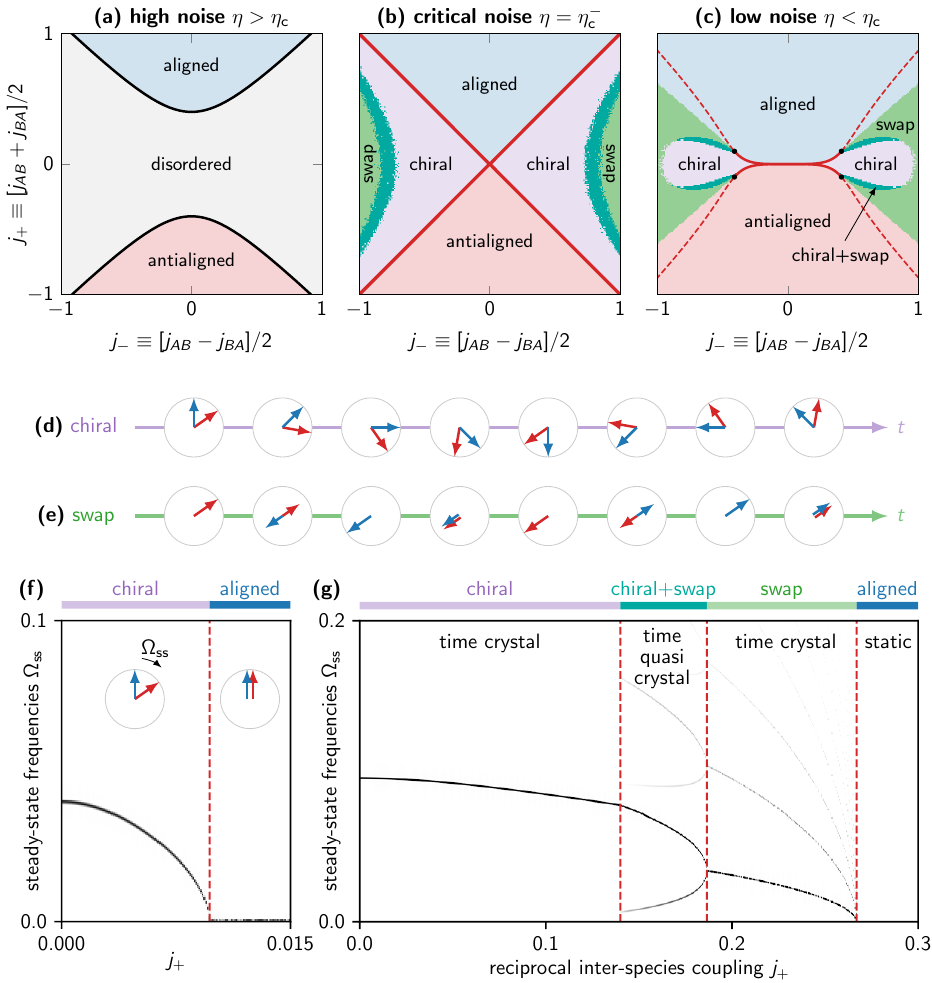}
  \caption{\label{figure_phase_diagram}\strong{Many-body exceptional points and active time (quasi)crystals.}
  (a-c) Slices of the phase diagram of the non-reciprocal flocking model Eq.~\eqref{eom_with_gradients_two_populations} for different values of the noise strength $\eta$.
  The parameters $j_{ab}$ (entering $\alpha_{a b}$ and $\beta_{a b c d}$ in Eq.~\eqref{general_SO2_invariant_main}) are coarse-grained versions of the microscopic couplings $J_{a b}$.
  The red (resp. black) lines correspond to the analytically-determined  phase transition lines from the (anti)aligned phase to the chiral (resp. disordered) phase.
  The red lines correspond to lines of exceptional points.
  In (b) and (c), the analytical prediction for the (anti)aligned/chiral transition is in excellent agreement with the numerical phase diagram up to tetracritical points marked by black dots where new phases emerge (see SI Sec.~\siref{sec_mean_field}{II}).
  (d) Schematic representation of one period of the chiral phase: $\vec{v}_A$ and $\vec{v}_B$ rotate in block at a constant angular velocity $\Omega_{\text{ss}}$, see SI Movie~\ref{movie_order_parameter_phases}.
  (e) Schematic representation of one period of the swap phase: $\vec{v}_A$ and $\vec{v}_B$ oscillate along a fixed direction, 
  see SI Movie~\ref{movie_order_parameter_phases}.
  (f,g) Plot of the frequencies present in the steady-state solution as a function of $j_{+}$, (f) for $j_{-} = \num{-0.6}$ and (g) for $j_{-} = \num{-0.25}$.
  In the chiral phase, a single frequency is present in the spectrum (at each point), which corresponds to the solid-body rotation.
  In the swap phase, a single frequency accompanied by harmonics are present.
  In contrast, in the mixed chiral/swap phase, two independent frequencies are present (with their harmonics).
  These frequencies are not harmonics of each other, leading to a quasiperiodic phase.
  The aligned phase is static ($\Omega_{\text{ss}} = 0$). Similar plots would be observed with the antialigned phase.
  In (g), a direct transition between aligned and chiral phases is observed. 
  The phase diagrams are determined by solving Methods Eq.~\eqref{eom_with_gradients_two_populations} numerically from random initial conditions, with $\rho_A = \rho_B = \num{1}$, $j_{A A} = j_{B B} = \num{1}$, and (a) $\eta/\eta_{\text{c}}=\num{1.5}$, (b) $\eta/\eta_{\text{c}}=\num{0.99}$, (c) $\eta/\eta_{\text{c}}=\num{0.5}$. The parameters in (f,g) are the same as in (c).
  }
\end{figure*}

Less intuitively, but crucially, the chiral phase hinges on a subtle interplay between noise and many-body effects.
Let us first consider an exactly solvable example: the Kuramoto model in Eq.~\eqref{kuramoto} with noise set to zero and identical frequencies within each species. In this case, the dynamics of the two populations can be mapped to a system with only two agents (SI Sec.~\siref{sec_non_reciprocal_kuramoto}{VI}). Unless the interaction is precisely non-reciprocal (i.e., $J_{AB}\equiv - J_{BA}$), this system always converges to a static phase as the two effective agents will eventually align or antialign when one catches up with the other (SI Sec.~\siref{two_agents}{V}).
However, the presence of microscopic noise or frequency disorder in Eq.~\eqref{kuramoto} constantly resets and restarts the chase and runaway motion of A/B pairs. 
In a many-body system, the noise-activated motions of individual agents become macroscopically correlated through their interactions and the chiral motion is stabilized.
We verified this numerically by computing the standard deviations of the order parameters in the chiral phase that decrease as $1/\sqrt{N}$ with the number of agents $N$, see SI Fig.~\siref{standard_deviations_sqrt_N}{S6}.
While not exactly solvable, the flocking model exhibits the same phenomenon; noise enlarges the size of the region in which time-dependent phases exist (compare Fig.~\ref{figure_phase_diagram}b and Fig.~\ref{figure_phase_diagram}c).
This is reminiscent of so-called order-by-disorder transitions in frustrated many-body systems~\cite{Villain1980,Daido1992}.

To elucidate the many-body character of the aligned to chiral transitions, it is instructive to contrast them to analogous phenomena occurring for only two coupled nonlinear ring-resonators in PT-symmetric lasers~\cite{Hassan2015,Miri2019,Liertzer2012,Lumer2013} (see SI Sec.~\siref{lasers}{IX}). 
In this case, the state of the system randomly switches between clockwise and counterclockwise under the effect of noise~\cite{Hassan2015}, destroying the chiral phase.
This noise-activated process would also occur in any of the systems we consider {if too few constituents (e.g. agents or resonators) were present}. We can model it by adding a hydrodynamic noise to the right-hand side of Eq.~\eqref{general_SO2_invariant_main}, see SI Sec.~\siref{arrhenius}{VII}.
The average time $\tau$ between successive chirality flips follows an Arrhenius law $\tau = \tau_0 \exp(\Delta/\sigma^2)$ where $\Delta$ is the height of the effective barrier between the clockwise and counterclockwise states, $\sigma$ is the standard deviation of the hydrodynamic noise, and $\tau_0$ is a constant determined from microscopics. 
When a large number $N$ of constituents is considered, the central limit theorem suggests that $\sigma \sim \sigma_0/\sqrt{N}$ 
($\sigma_0$ is the strength of the noise acting on a single element), consistent with our numerical observations (see SI Secs.~\siref{two_agents}{V}-\siref{arrhenius}{VII}).
As a result, the time~$\tau$ between chirality flips grows exponentially with~$N$: the chiral phase is salvaged by many-body effects.
In optics, this scenario could be realized by implementing non-reciprocal couplings~\cite{Miri2019} in photonic networks of many coupled lasers~\cite{Nixon2013,Pal2017,Mahler2020,Parto2020,Ramos2020,HonariLatifpour2020,HonariLatifpour2020b,Acebron2005}.

Unlike the familiar paramagnet to ferromagnet transition, the aligned to chiral transition (and the chiral phase itself) cannot occur in thermodynamic equilibrium. Here, we develop a general approach to describe this class of time-dependent phases and transitions unique to non-reciprocal matter.
Our starting point is a general principle that applies both in and out of equilibrium: phases of a many-body system can be identified from the steady-states of the corresponding dynamical system (such as Eq.~\eqref{general_SO2_invariant_main}).
Phase transitions then occur when one steady-state becomes unstable, i.e. when perturbations around it are no longer damped.
We therefore linearize Eq.~\eqref{general_SO2_invariant_main}, by separating the order parameters $\vec{V}\equiv(\vec{v}_A, \vec{v}_B, \dots)$ into steady-state components $\vec{V}_{\text{ss}}$ and fluctuations $\delta\vec{V}$ around them. We obtain the equation
\begin{equation}
    \label{linearized}
    \partial_t \delta\vec{V} = L \; \delta\vec{V}.
\end{equation}
Crucial to our approach is the following fact: non-reciprocity implies that the linear operator $L$ can be non-Hermitian. Consider first a conservative system deriving from a free energy $F$. In this case, we would have 
$L_{a b} = - \partial_{\vec v_b} \partial_{\vec v_a} F 
= L_{b a}$. This implies that $L$ would be Hermitian, i.e. $L_{a b} = L_{b a}$, as it is real valued. 
In contrast, a non-reciprocal system cannot be described by a free energy. As a result, $L$ can be and generally is non-Hermitian, i.e. $L_{a b} \neq L_{b a}$.

The linear operator $L$ determines the nature of phase transitions in non-reciprocal matter through the dynamics of fluctuations.
The $2M$ eigenvectors of $L \equiv L(\vec{V}_{\text{ss}})$ (for $M$ species) describe the time evolution of fluctuations around the steady-state. 
They do not have to be orthogonal because $L$ is not Hermitian.
The corresponding eigenvalues $s_i = \sigma_i + \ii \omega_i$ can be decomposed in growth rates $\sigma_i$ and frequencies $\omega_i$ where $i=1, \dots, 2M$. 
Here, one of the eigenmodes of $L$ always has a vanishing eigenvalue $s = 0$, because it corresponds to a global rotation of the $\vec{v}_a$ (i.e., the Goldstone mode of broken rotation invariance), green line in Fig.~\ref{figure_bots_phases}h.
In the static (anti)aligned phase, the other modes are always damped ($\sigma_i < 0$).
The aligned-to-chiral phase transitions occur when the damped mode with the growth rate $\sigma_i$ closest to zero
\footnote{Modes that are more damped do not play any role in the transition mechanism. This allows to immediately generalize our conclusions to more than two populations provided that two transitions do not occur at the same time. See SI Sec.~\siref{sec_multiple_populations}{X} for further discussions.} 
(orange line in Fig.~\ref{figure_bots_phases}h) coalesces with the Goldstone mode (green line) at special points labeled in red. 

This coalescence of two eigenmodes is known as an exceptional point (EP)~\cite{Kato1984}. 
In addition to having the same eigenvalues (that crucially vanish in our case), the two eigenmodes become parallel at the exceptional point~\cite{Hanai2020}.
In a many-body system, such a mode coalescence defines a class of phase transitions that we dub exceptional transitions.
{
As an illustration, imagine a ball constantly kicked by noise at the bottom of a potential shaped as a Mexican hat. In this simplified picture, the ball represents the order parameter of a many-body system.
In a reciprocal system, the ball rolls back to the bottom when perturbed in the uphill direction.
In a non-reciprocal system, there are non-conservative forces in addition to the potential energy landscape, that lead to transverse responses.
When you kick the ball in the uphill direction, it also moves perpendicular to it along the bottom of the potential, but not vice-versa. 
This non-reciprocal response is described mathematically by the non-orthogonality of the eigenmodes of $L$.
At the exceptional point, the ball moves along the bottom of the potential irrespective of whether it is kicked along or perpendicular to it. This corresponds to the coalescence between the Goldstone mode and the damped mode with growth rate $\sigma$ closest to zero.
After the exceptional transition, the ball starts running along the bottom of the potential by itself, at a speed set by the now positive growth rate $\sigma$ (Fig. \ref{figure_bots_phases}h). 
}

More generally, exceptional transitions can be viewed as the dynamical restoration of a spontaneously broken continuous symmetry: the Goldstone mode is actuated by the noise, and after the transition the system runs along the manifold of degenerate ground states. 
Figure~\ref{figure_bots_phases}i provides an almost mechanical picture of how non-reciprocal interactions lead to an exceptional point and to the onset of the chiral phase.
From the point of view of non-Hermitian quantum mechanics~\cite{Hatano1996,Bender1998}, these transitions are manifestations of the spontaneous PT symmetry breaking (see Methods). 
From the point of view of dynamical systems, they are instances of Bogdanov-Takens bifurcations~\cite{Kuznetsov2004}, with the peculiarity that one of the modes involved is a Goldstone mode (see Methods).

As a concrete example, let us consider the continuum theory of non-reciprocal flocking [Methods Eq.~\eqref{eom_with_gradients_two_populations}]. Analogous treatments of synchronization and pattern formation are detailed in Methods.
Figure~\ref{figure_phase_diagram}a-c shows the key result of our analysis based on Eq.~\eqref{eom_with_gradients_two_populations}: the phase diagrams as a function of the reciprocal and non-reciprocal parts of the rescaled inter-species interactions $j_{\pm} = [j_{AB} \pm j_{BA}]/2$ respectively (see SI Sec.~\siref{sec_mean_field}{II}).
These phase diagrams exhibit a disordered phase (gray), an aligned phase (blue), an antialigned phase (red), and a chiral phase (purple, see also Fig.~\ref{figure_phase_diagram}d). 
In the SI Sec.~\siref{sec_excitation_supp}{III}, we prove that these phases are linearly stable against velocity fluctuations in the hydrodynamic theory over large ranges of parameters.
This theory further predicts that the phase boundary between the chiral phase and the (anti)aligned phase is marked by exceptional points (red lines in Fig.~\ref{figure_phase_diagram}c).

The chiral phase is not the only time-dependent phase induced by non-reciprocity.
We have also identified a \emph{swap phase} (green region in Fig.~\ref{figure_phase_diagram}b-c), where $\vec{v}_A$ and $\vec{v}_B$ oscillate between two values along a fixed direction.
In contrast with the chiral phase where a continuous symmetry is restored on average, the swap phase dynamically restores a \emph{discrete} symmetry through the loss of time-translation invariance. 
We also predict and observe a mixed chiral+swap phase, where both swap and chiral motions occur at the same time (dark green region).
The dynamical nature of these phases is illustrated in Fig.~\ref{figure_phase_diagram}d-e and SI Movie~\ref{movie_order_parameter_phases}.
All these phases break time translation invariance, in a way reminiscent of time crystals~\cite{Shapere2012,Wilczek2012,Khemani2019,Yao2018,Prigogine1968,Winfree2001} and quasicrystals~\cite{Giergiel2018,Autti2018}, see Fig.~\ref{figure_phase_diagram}f-g.
The existence of all the phases found in non-reciprocal flocking follows from general symmetry principles~\cite{Golubitsky2002}.
Hence, they transcend specific models. 
We also observe them in our non-reciprocal synchronization (Extended Data Fig.~\ref{si_kuramoto_oa_phase_diagram}) and pattern formation models (Extended Data Fig.~\ref{figure_non_reciprocal_swift_hohenberg}), see Methods.

\begin{figure*}[p]
  \centering
  \includegraphics{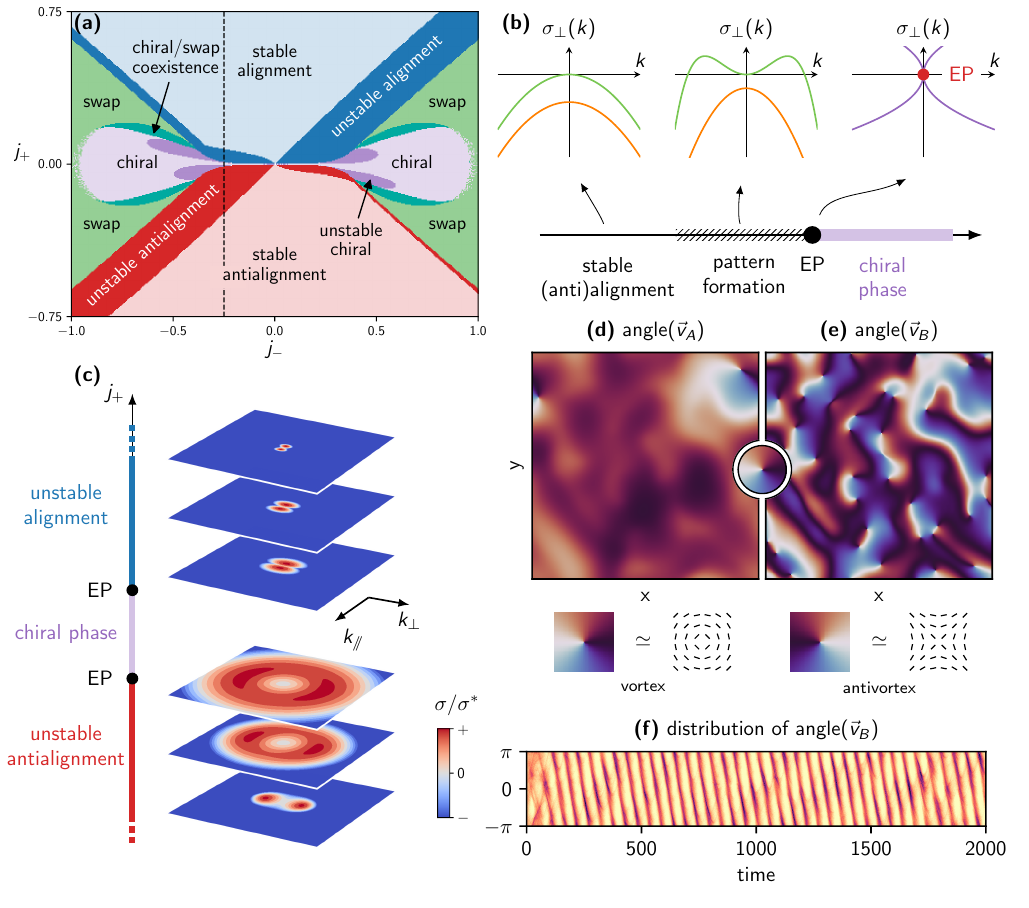}
  \caption{\label{figure_stability}\strong{Stability of spatially extended systems: exceptional-point enforced pattern formation.}
  The combination of convective terms with exceptional point at the (anti)aligned to chiral transitions gives rise to pattern formation near the transition lines.
(a) Numerical phase diagram including the linear stability analysis of the (anti)aligned and chiral phases in the non-reciprocal Toner-Tu model (the stability of the swap and chiral+swap phases is not analyzed).
  (b) Mechanism of exceptional transition and of exceptional-point enforced pattern formation.
  The coalescence of the Goldstone mode with a damped mode lead to instabilities at finite momentum. The growth rate of transverse perturbations $\sigma_{\bot}(k)$ becomes positive, and has a maximum $\sigma^*$ at a finite wavevector $k^*$.
  (c) Plot of the normalized growth rate $\sigma(k)/\sigma^*$ as a function of wavevector, for different values of $j_{+}$ at fixed $j_{-}$ [along the dashed line in (a)].
  (c-f) Fully non-linear simulation of the pattern formation in the unstable aligned regime. We show snapshots of the angle with a fixed direction (d-e) of the order parameters $\vec{v}_A$ and $\vec{v}_B$ (see SI Movie~\ref{movie_patterns} of the time evolution). Note the existence of numerous topological defects (vortices and antivortices) in the order parameters.
  In (f), we show the evolution in time of the histogram of the angle between $\vec{v}_A$ and a fixed direction (the corresponding plot for $\vec{v}_B$ has identical features). An approximate periodicity emerges amidst the spatio-temporal chaos, ushering in the chiral phase.
  We have used the same parameters as in Fig.~\ref{figure_phase_diagram} with $\eta/\eta_{\text{c}} = \num{0.5}$.
  In (a-b), $v^0_A = \num{0.06}$ and $v^0_B = \num{0.01}$. 
  The simulations in (c-g) are performed on a $2L \times 2L$ box with periodic boundary conditions, and $L=\num{0.32}$, $j_{+}=\num{0.1}$, $j_{-}=\num{0.2}$, $v^0_A = \num{0.06}$ and $v^0_B = \num{0.01}$ (see Methods for details on the simulation). 
}
\end{figure*}

We now show that exceptional transitions must necessarily be accompanied by pattern-forming instabilities when the non-reciprocally interacting constituents are moving, i.e. when there is a steady-state flow $\vec{v}_{\text{ss}}$.
Figure~\ref{figure_stability}a shows that pattern formation occurs around the critical lines marking the mean-field phase transition (bright red and blue regions).
This phenomenon can be captured by the following model
\begin{equation}
    \partial_t \, \delta\vec{V}
    = \big[ L_{\text{EP}} + M (\vec{v}_{\text{ss}} \cdot \nabla) + D \, \nabla^2 \big] \delta\vec{V} .
    \label{ep+flow}
\end{equation}
Equation (\ref{ep+flow}) is a minimal extension of Eq.~\eqref{linearized} valid near the exceptional transition where $L_{\text{EP}}$ is the singular matrix accounting for the presence of the exceptional point.
It contains an additional convective term in the linearized matrix form $M (\vec{v}_{\text{ss}} \cdot \nabla)$ that mixes the two species $A$ and $B$ as well as a diffusive term $D \, \nabla^2$. 
Since the eigenvalues of a perturbed exceptional point typically diverge as a square root of the perturbation~\cite{Kato1984}, the momentum space \emph{complex} growth rate behaves as $s_{\pm}(k) \simeq \pm \ii\sqrt{\ii v_{\text{ss}} \, k}$ at small wavevector $k$, and as $- D \, k^2$ at large~$k$. 
This leads to a positive maximum in the growth rate, corresponding to a linear instability at finite momentum (Fig.~\ref{figure_stability}b-c).

Figure~\ref{figure_stability}d-f and SI Movie~\ref{movie_patterns} provide glimpses into the non-linear regime of pattern formation in which vortices and antivortices continuously unbind and annihilate. 
The density of topological defects is different in $\vec{v}_A$ and $\vec{v}_B$ because each density decreases with the self-propulsion speed of the respective species (Fig.~\ref{figure_stability}d-e). 
While the system does not coarsen towards an ordered state, there is a clear precursor of the chiral phase within its chaotic dynamics: the angular distribution of the order parameters plotted in Fig.~\ref{figure_stability}f is approximately periodic.

We have seen that parity is {\it spontaneously} broken in the chiral phase. 
It is also natural to consider systems in which parity is {\it explicitly} broken. Concrete examples include: (i) the Kuramoto model Eq.~\eqref{kuramoto} with nonvanishing natural frequencies (ii) the Vicsek model Eqs.~(\ref{kuramoto}--\ref{vicsek_r}) with external torques or (iii) the Swift-Hohenberg model Eq.~\eqref{coupled_swift_hohenberg_main} with broken up-down ($u \to - u$) symmetry. 
All these systems still fit in the framework of Eq.~\eqref{general_SO2_invariant_main}, provided that
the matrices $\mathbb{A}^{\mu \nu}_{a b}$ and $\mathbb{B}^{\mu \nu}_{a b c d}$ acting on~$\vec{v}_a^{\,\nu}(t,x)$ are appropriately chosen ($\mu$ and $\nu$ denote the vector components).
By imposing the rotational symmetry constraint, we can fully determine the form of these matrices: $\mathbb{A}_{a b}^{\mu \nu} = \alpha_{a b} \, \delta^{\mu \nu} + {\alpha}^\star_{a b} \, \epsilon^{\mu \nu}$ and $\mathbb{B}_{a b c d}^{\mu \nu} = \beta_{a b c d} \delta^{\mu \nu} + {\beta}^\star_{a b c d} \, \epsilon^{\mu \nu}$.
Here, $\delta^{\mu \nu}$ is the identity and $\epsilon^{\mu \nu}$ is an antisymmetric matrix that rotates the vectors by~\ang{90}.
Using this decomposition, we can distinguish (I) parity symmetric systems in which $\alpha^\star$ and $\beta^\star$ vanish and (II) systems in which parity is explicitly broken, in which $\alpha^\star$ and $\beta^\star$ can be nonzero. 
The classes~I and~II correspond to the symmetry groups of general rotations~$O(2)$ and proper rotations~$SO(2)$, respectively.

\begin{figure*}[p]
  \centering
  \includegraphics[width=15cm]{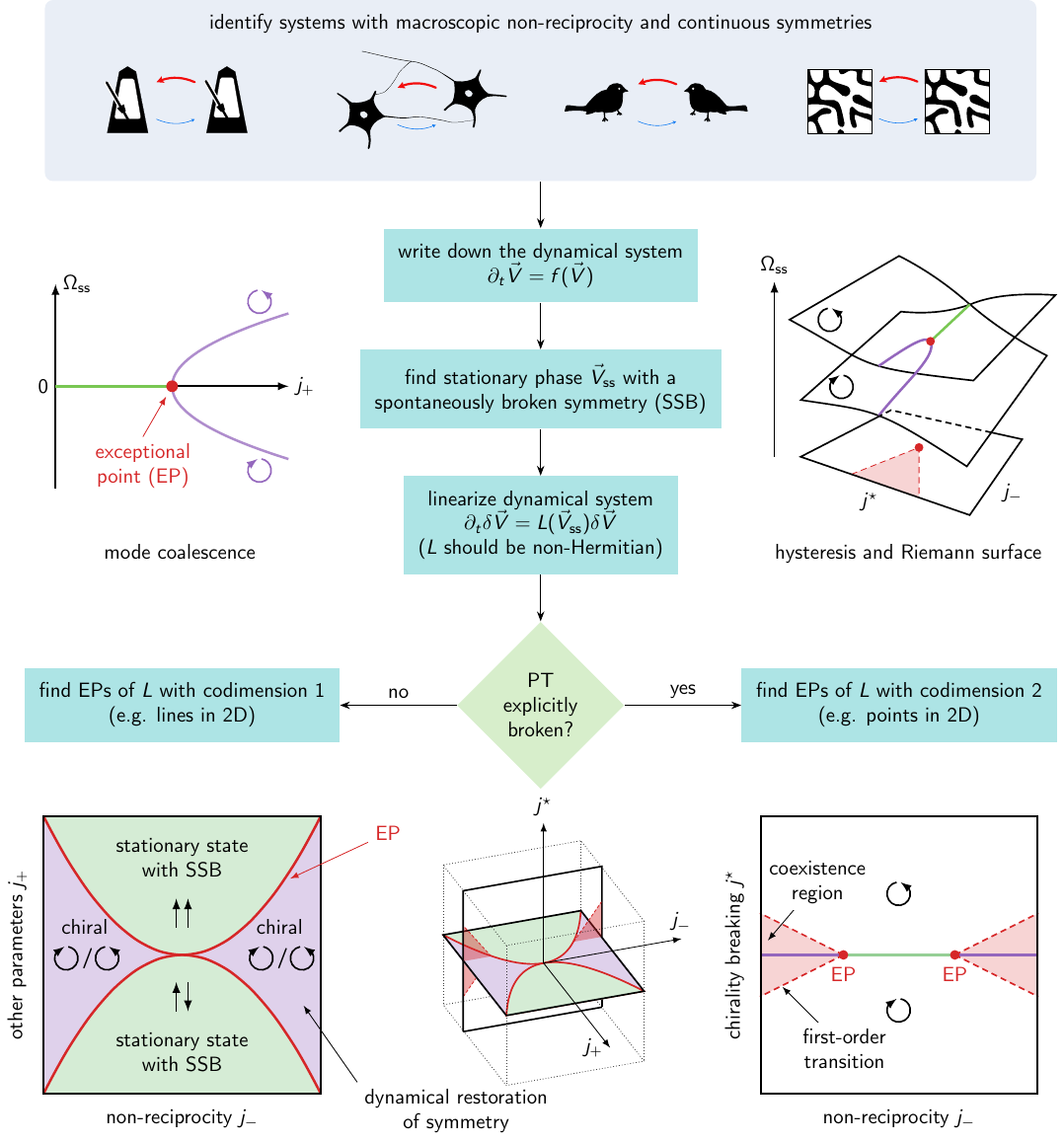}
  \caption{\label{figure_recipe_main}\strong{A visual procedure to identify and analyze exceptional transitions.}
  We start by identifying a system with (i) non-reciprocal interactions between two or more components and (ii) a continuous symmetry that can be spontaneously broken.
  A dynamical system describes the evolution of the relevant order parameter $V$ (an effective hydrodynamic theory), obtained either purely by symmetry considerations, or from more microscopic considerations. 
  We then look for a time-independent (stationary) steady-state $V_{\text{ss}}$ in which a continuous symmetry is spontaneously broken (SSB, green regions).  
  The dynamical system is then linearized above this time-independent  steady-state. 
  Because of the spontaneously broken symmetry, the corresponding linear operator (Jacobian matrix) $L(V)$ always has a vanishing eigenvalue.
  The exceptional points (EP, red points and continuous red lines) of $L(V)$ with zero eigenvalue correspond to exceptional transitions.
  The parameters of the system are separated in two classes: (I) those who do not explicitly break parity, collectively labeled by $j$, and (II) those who explicitly break parity, labeled by $j^\star$. 
  We further split the parameter set $j$ into two subclasses, (a) $j_{-}$ for parameters encoding the non-reciprocity and (b) $j_{+}$ for the others. When $j^\star = 0$ (left panel), the EP have codimension 1 (they form lines in a 2D parameter space).
  At the exceptional transition, the Goldstone mode collides with a damped mode. This leads to a spontaneous dynamical restoration of symmetry (at the price of losing time-translation invariance) in which the Goldstone mode is actuated by the fluctuations.
  In the corresponding chiral phase (in purple), the space of equivalent steady-states (corresponding to the broken symmetry) is traveled in a direction (clockwise or counterclockwise) chosen at random. 
  When $j^\star \neq 0$, this mechanism competes with an explicit rotation of the order parameter set by $j^\star$.
  This leads to extended regions (in light red) starting at the EP (red point, now of codimension 2) in which the (counter)clockwise states coexist, marked by first-order transitions and accompanied by hysteresis: the complex frequencies of the (stable) equilibrium states are organized on a Riemann surface (see Extended Data Fig.~\ref{si_figure_hysteresis} for a numerically computed version in the non-reciprocal Kuramoto model).
  }
\end{figure*}

In the language of non-Hermitian quantum mechanics~\cite{Bender1998,Mostafazadeh2002,Miri2019},
class I systems exhibit PT-symmetry, which may be \emph{spontaneously} broken in the chiral phase.
In this case, reaching an exceptional transition requires tuning only one parameter (the transitions have codimension one, see e.g. Fig.~\ref{figure_phase_diagram} where they are lines in a 2D phase diagram).
In class II, on the other hand, PT symmetry is explicitly broken: as a result, the equivalent of the aligned and antialigned phases become time-dependent, and reaching an exceptional point requires tuning two parameters (the exceptional points have codimension two, see e.g. Extended Data Fig.~\ref{si_figure_hysteresis} and Fig.~\ref{figure_recipe_main} where they are isolated points in 2D).

In the Methods, we present a detailed analysis of the non-reciprocal Kuramoto model~\eqref{kuramoto} with non-vanishing $\omega_m$ that explicitly breaks parity. 
The combination of non-reciprocity and explicit PT-symmetry breaking leads to hysteresis and discontinuous transitions between (i) regions where the clockwise and counterclockwise states coexist and (ii) regions where only one state exists (Extended Data Fig.~\ref{si_figure_hysteresis}). 
These results reveal a remarkable similarity between non-reciprocal synchronization and driven Bose-Einstein condensates~\cite{Hanai2019,Hanai2020}.

We conclude with a visual procedure to extend our approach to other systems (Figure~\ref{figure_recipe_main}). 
The key ingredients are (i) macroscopic non-reciprocity (manifested in the asymmetry of macroscopic coefficients) and (ii) a spontaneously broken continuous symmetry. 
Although we have focused mostly on systems with circular symmetry $O(2)$, exceptional transitions could occur for any continuous group, see SI Sec.~\siref{sec_O3}{VIII} for an extension to spherical symmetry $O(3)$ relevant to three-dimensional vector order parameters such as in 3D flocks. 
Our analysis was illustrated with vector order parameters whose evolution is not the expression of a conservation law. 
This paradigmatic case is known as \enquote{model A} in the Hohenberg-Halperin classification of dynamical critical phenomena~\cite{Hohenberg1977} but the same approach applies also to other order parameters and classes, see Ref.~\cite{Scheibner2020} for a non-reciprocal active elasticity that conserves linear momentum and Refs.~\cite{You2020,Saha2020} for non-reciprocal models of phase separation that conserve mass (both illustrate \enquote{model B} in Ref.~\cite{Hohenberg1977}). 

Our work lays the foundation for a general theory of critical phenomena in non-reciprocal matter from driven quantum condensates to biological and artificial neural networks.
These systems are marked by the interplay between the non-reciprocal enhancement of the fluctuations and the rigidity bestowed by many-body effects.
Our field-theoretical approach inspired by non-Hermitian quantum mechanics captures these effects and builds new bridges between many-body physics and bifurcation theory.

\medskip

\noindent\strong{Acknowledgments.} We thank Andrea Alù, Denis Bartolo, Demetrios Christodoulides, Aashish Clerk, Alexander Edelman, Alexey Galda, Ming Han, Kabir Husain, Tsampikos Kottos, Zhiyue Lu, M. Cristina Marchetti, Mohammad-Ali Miri, Benjamin Roussel, Colin Scheibner, David Schuster, Jonathan Simon, and Benny van Zuiden.
MF acknowledges support from a MRSEC-funded Kadanoff-Rice fellowship (DMR-2011854) and the Simons foundation. 
RH was supported by a Grand-in-Aid for JSPS fellows (Grant No. 17J01238). 
VV was supported by the Complex Dynamics and Systems Program of the Army Research Office under grant no. W911NF-19-1-0268 and the Simons foundation. 
This work was partially supported by the University of Chicago Materials Research Science and Engineering Center, which is funded by National Science Foundation under award number DMR-2011854. 
This work was completed in part with resources provided by the University of Chicago’s Research Computing Center. 
Some of us benefited from participation in the KITP program on Symmetry, Thermodynamics and Topology in Active Matter supported by grant no. NSF PHY-1748958.

\clearpage

\begin{center}{\bf METHODS}\end{center} 
\def\figurename{Methods Fig.}
\setcounter{figure}{0}

\def\tablename{Methods Table}
\setcounter{table}{0}
\methodsection{Non-reciprocity out of equilibrium}
\label{non_reciprocity_out_of_equilibrium}

In this section, we discuss the relations between non-reciprocity, the non-Hermitian (non-normal) character of the Jacobian of a dynamical system, which allows it to exhibit an exceptional point, and the non-equilibrium character of the system.

A dynamical system 
\begin{equation}
	\partial_t X = f(X)
\end{equation}
is said to be conservative when it derives from some potential $F$ (e.g., a free-energy), such that $f_a(X) = \partial_{X_a} F$.
(Here, we assume that $X$ is real.)
The Jacobian of the dynamical system is defined as the real matrix $L_{a b} = \partial_{X_b} f_a$.
When the dynamical system is conservative, $L_{a b} = - \partial_{X_b} \partial_{X_a} F = L_{b a}$ is symmetric, and hence it is a normal operator.
When the dynamical system is not conservative, it is possible to have $L_{a b} \neq L_{b a}$, i.e. $L$ is not symmetric. 
This is our operative definition of non-reciprocity.
In the language of quantum mechanics, we would say that the operator is non-Hermitian (because it could be complex-valued).
The key point is that $L$ is not a normal operator (a matrix $N$ is normal when it is unitarily diagonalizable; equivalently, $N^\dagger N = N N^\dagger$, see Ref.~\cite{Trefethen2005}).
Deviations from normality allow the eigenvectors of $L$ not to be orthogonal, leading to a variety of physical consequences related to an enhanced sensitivity to fluctuations, in hydrodynamics~\cite{Trefethen1993,Grossmann2000,Chomaz2005,Schmid2007,Chajwa2020}, (general and neural) networks~\cite{Murphy2009,Hennequin2012,Amir2016,Asllani2018a,Asllani2018b,Baggio2020,Nicolaou2020}, ecological systems~\cite{Neubert1997,Nelson1998,Neubert2004,TOWNLEY2007,Ridolfi2011,Biancalani2017}, photonics~\cite{Miri2019,Feng2017}, and quantum systems~\cite{Ashida2020,Hatano1996,Bender1998,Hanai2020,Ashida2017,Tripathi2016,Bernier2014,Aharonyan2019}.
In particular, the presence of exceptional points requires a non-normal operator.

Note that the notion of normality depends on the choice of the scalar product and the associated norm (this is also true for symmetry and Hermiticity).
Equivalently, these notions are not invariant under a generic invertible change of basis.
Depending on the context, a certain scalar product might be selected by physical considerations (such as the presence of noise, see below).
For example, consider the one-dimensional harmonic oscillator described by the linear system
\begin{equation}
	\label{harmonic_oscillator}
	\partial_t \begin{pmatrix} x \\ p \end{pmatrix}
	=
	\begin{pmatrix} 0 & 1/m \\ -k & 0 \end{pmatrix}
	\begin{pmatrix} x \\ p \end{pmatrix}
\end{equation}
in which $x$ is the position and $p$ the linear momentum of a particle of mass $m$ in a harmonic well with stiffness $k$. 
The matrix in Eq.~\eqref{harmonic_oscillator} is not normal with respect to the standard scalar product on $\RR^2$, but it is normal with respect to the scalar product associated with the energy (defined such that $\lVert (x, p) \rVert^2 = k x^2/2 + p^2/2m$).
In contrast, the notions of spontaneously/explicitly/not-broken generalized PT-symmetry (see section \methodsref{generalized_PT_symmetry}{Generalized PT symmetry and dynamical systems}) and the presence of an exceptional point are independent of the choice of the basis.

Non-reciprocity is also related to the breaking of detailed balance (i.e., microscopic reversibility).
The notion of detailed balance deals with stochastic processes, hence we have to consider a stochastic dynamical system $\partial_t X = f(X) + \eta(t)$, where $\eta(t)$ is a noise.
This can either represent a microscopic system or a fluctuating hydrodynamic equation.
Assuming that the noise is scalar, detailed balance implies (i.e., requires) that $\partial_{X_a} f_b = \partial_{X_b} f_a$ (i.e. $L_{a b} = L_{b a}$).
(See SI Sec. IV and references therein. When the noise is not scalar, this equality is weighted by the corresponding diffusion constants.)

\medskip

The concepts of \enquote{non-conservative dynamical systems} and \enquote{non-conserved order parameters} discussed in the main text are completely unrelated. 
(Their names are borrowed from conservative forces and conserved quantities.)
Following the classification of Ref.~\cite{Hohenberg1977}, we talk about a conserved order parameter (\enquote{model B} in Ref.~\cite{Hohenberg1977}) when its dynamics is the expression of a conservation law (e.g., conservation of mass).
When this is not the case, we say that the order parameter is not conserved (\enquote{model A} in Ref.~\cite{Hohenberg1977}).
In parallel, a dynamical system is conservative when it derives from a potential, as discussed in the previous paragraph.

\medskip

Let us now discuss connections between the notions discussed above and more general notions of non-reciprocity.
Broadly, non-reciprocity occurs when $A$ does not have the same effect on $B$ than $B$ has on $A$. 
This can often lead to a non-conservative dynamical system as defined above, but the connection is not systematic.

Newton's third law~\cite{Newton1687} states that the force $f_{i j}$ that an object $i$ exerts on an object $j$ is exactly opposite to the force $f_{j i}$ than $j$ exerts on $i$ (i.e., $f_{i j} = - f_{j i}$).
This symmetry between action and reaction can be violated when the interaction between the objects is mediated by an non-equilibrium environment.
Such non-reciprocal interactions can arise in various contexts: particles in fluids~\cite{Ermak1978,DiLeonardo2008,Lahiri1997,Lahiri2000,Chajwa2020}, non-equilibrium plasma~\cite{Ivlev2015,Kryuchkov2018}, chemically and biologically active matter~\cite{Soto2014,AgudoCanalejo2019,Saha2019,Uchida2010}, optical matter~\cite{Dholakia2010,Yifat2018,Peterson2019}, etc. 
The symmetry between action and reaction has no particular reason to occur in complex systems in which the interactions summarize the decisions of agents/algorithms: it is explicitly violated in active matter~\cite{Morin2015,Ginelli2015,Bonilla2019,Dadhichi2019,Yllanes2017}, e.g. for biological reasons such as a limited vision cone~\cite{Durve2018,Costanzo2019} or hierarchical relationships~\cite{Nagy2010}, as well as in systems with synthetic physical interactions~\cite{Soto2014,Ivlev2015,Kryuchkov2018,Saha2019,Scheibner2020} or programmable robotic interactions~\cite{Coulais2017,Brandenbourger2019,Ghatak2019,Rosa2020,Chen2020}. The non-equilibrium character of such non-conservative forces leads to diverse but crucial consequences on the behavior of the corresponding systems~\cite{Ivlev2015,Dadhichi2019,Tasaki2020,Fodor2016,Loos2020,Loos2019}. 

In condensed matter, in particular in the context of topological insulators, non-symmetric tight-binding Hamiltonians $H \neq H^T$ with non-symmetric hopping terms (leading to non-Hermitian Hamiltonians in momentum space) are called non-reciprocal, see Refs.~\cite{Malzard2015,Lee2019,Helbig2020,Lee2020,Okuma2020,Lee2019b,Zhang2020,Okuma2020,Hatano1996,Nelson1998}.
In an elastic network, the Hamiltonian is replaced by a dynamical matrix $D$ such that the force $F_i^{\mu}$ on the particle $i$ is $F_i^{\mu} = - D_{i j}^{\mu \nu} u_j^{\nu}$ (with $u_j^{\nu}$ the displacement of the particle $j$ with respect to its equilibrium position). 
The overdamped dynamics of such a system is ruled by an equation of the form $\gamma \partial_t u_i^\mu = D_{i j}^{\mu \nu} u_{j}^{\nu}$. In this case, the symmetry of the dynamical matrix indeed corresponds to reciprocity as we have defined in the first paragraph, and is associated with a non-conservative dynamical system.
Non-reciprocity ($D \neq D^T$) can occur through a violation of Newton's third law by which $D_{i j} \neq D_{j i}$ (see e.g. Ref.~\cite{Brandenbourger2019}), or through \enquote{odd springs} with transverse responses by which $D^{\mu \nu} \neq D^{\nu \mu}$ (see Ref.~\cite{Scheibner2020}). 
In both cases, some degree of activity is required (energy is not conserved), and the noisy overdamped dynamics again exhibits broken detailed balance.

At the level of responses, reciprocity is captured by various notions that share many similarities, such as Maxwell–Betti reciprocity in elasticity and acoustics~\cite{Achenbach2004,Nassar2020}, Lorentz reciprocity in optics~\cite{Potton2004,Caloz2018}). Similar relations appear also in fluid dynamics~\cite{Masoud2019}, etc. 
For instance, the non-reciprocal elastic networks (with $D \neq D^T$) in Refs.~\cite{Scheibner2020,Scheibner2020b,Zhou2020,Chen2020,Brandenbourger2019} violate Maxwell–Betti reciprocity.
A similar notion exists in non-equilibrium thermodynamics: Onsager reciprocal relations are the statement that the matrix of response coefficients $L$ relating thermodynamic fluxes $J_i$ and forces $\mathcal{F}_k$ through $J_i = L_{i k} \mathcal{F}_k$ is symmetric (more precisely, the Onsager-Casimir relations state that $L_{i k}(B) = \epsilon_i \epsilon_k L_{k i}(-B)$ where $\epsilon_i = \pm 1$ depending on whether the quantity $i$ is even/odd with respect to time-reversal, and where $B$ represents all external time-reversal breaking fields such as magnetic fields and rotations)~\cite{DeGrootMazur}. This relation is also a consequence of microscopic reversibility. Depending on the system, $L$ contains the diffusion coefficients, electric conductivities, viscosities, etc. For example, Hall conductivity and odd viscosity~\cite{DeGrootMazur,Avron1998,Banerjee2017,Souslov2019,Soni2018,Han2020} are instances of antisymmetric components, that require broken detailed balance.

\methodsection{Demonstration with programmable robots}
\label{experimental_demonstration}

We demonstrate the effect of non-reciprocal interactions using programmable robots evolving according to a modified version of Eq.~\eqref{kuramoto}. The main differences are that (i) the evolution is discrete in time, (ii) the term $\sin(\theta_n - \theta_m)$ is replaced by $\sign \sin(\theta_n - \theta_m)$ and (iii) we do not add artificial noise. 
Hence, Eq.~\eqref{kuramoto} is replaced by
\begin{equation}
  \label{vicsek_robots}
  \theta_m(t+T_{\text{mv}}) = \sum_{n} [J_{m n} \, T_{\text{mv}}] \, \sign \sin(\theta_n(t) - \theta_m(t))
\end{equation}
In practice, additional differences such as delays and noises are also present due to imperfections in the implementation. This motivates the lack of artificial noise.
We use two programmable robots (GoPiGo3, Dexter Industries). 
Each robot is connected to a magnetic sensor (Bosh BNO055 packaged in Dexter Industries IMU Sensor) as a compass to measure its absolute orientation.
The magnetic sensor is attached to the body of the robot at a distance from the motors to reduce electromagnetic interferences.
Each robot communicates its respective orientation to the other via Wi-Fi every $T_{\text{ms}} = \SI{0.1}{\second}$.
The communication is implemented through a central server, which allows to easily record the angles of each robot as a function of time (see Fig.~\ref{figure_bots_phases}c).
Every $T_{\text{mv}} = \SI{0.5}{\second}$, each robot computes the left hand side of the modified Eq.~\eqref{kuramoto} described above, and actuates its two motors with opposite angular velocities for a given time in order to perform a rotation of $\pm \theta_0$ where $\theta_0 = \ang{15}$, depending on the result of the computation. 
The change in angle is not instantaneous, because the rotation speed of the motors cannot be arbitrarily large. (Orientation measurements and communication are not instantaneous either, but they are much faster.) Hence, we have chosen to make new decisions only at discrete times. 
Performing rotations with very small angles (lower than $\simeq \ang{4}$) is not effective; this is compensated by the replacement of $\sin(\theta_n - \theta_m)$ by $\sign \sin(\theta_n - \theta_m)$ to avoid the presence of arbitrarily small angle increments.
The robots and the server are implemented in Python, using the GoPiGo3 Python package (version 1.2.0) to control the robots, the DI\_Sensors package (version 1.0.0) to access the magnetometer data and ZeroMQ (version 4.3.2) as a messaging library.

We show an example of behavior in SI Movie~\ref{movie_robots}, in which we observe a rotation of the two populations.
This behavior is reproducible, but not in a systematic way: depending on the initial conditions, the robots can also quickly align (or antialign). 
We observed chiral motions over relatively long time; however, the robots have a tendency to eventually align as a result of imperfections (not shown in the movie; this can be qualitatively understood from the analysis of SI Sec.~V and VI).

The movie was modified in postproduction to color half of the robots in blue, partially hide imperfections in the substrate, and enhance the color balance.

\begin{figure}
  \centering
  \includegraphics{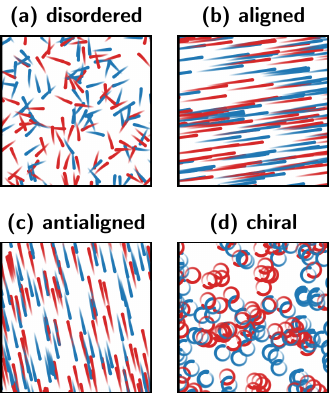}
  \caption{\label{si_figure_vicsek_snapshots}\strong{Snapshots of simulations of the Vicsek model.}
  See text for a description and for the parameters used.
  }
\end{figure}

\methodsection{Molecular dynamics simulations of the Vicsek model}
\label{molecular_dynamics_vicsek}

We perform simple molecular dynamics simulations of a moderately large number of active agents following Eqs.~(\ref{kuramoto}--\ref{vicsek_r}) in order to visually demonstrate the disordered, flocking, antiflocking, and chiral behaviours, as shown in Extended Data Fig.~\ref{si_figure_vicsek_snapshots} and SI Movie~\ref{movie_vicsek}.

We simulate $N$ agents following Eqs.~(\ref{kuramoto}--\ref{vicsek_r}) discretized using the Euler–Maruyama scheme with a timestep $\delta t$, with a ratio $N_{A}/N_{B}$ between populations A and B, in a $L \times L$ box with periodic boundary conditions, for a duration $T_{\text{sim}}$.
We choose the couplings in Eq.~\eqref{kuramoto} to be $J_{m n} = J_{s(m) s(n)} H[\lVert r_i - r_j \rVert - R_0]$ where $s(m)$ represents the species of particle $m$ ($A$ or $B$) and $H$ is the Heaviside step function.
We set $N = \num{512}$, $N_{A}/N_{B} = \num{1}$, $v_0 = \num{0.5}$ $R_0 = \num{2}$, $L = 8$, $T_{\text{sim}}/\delta t = \num{8000}$ with $\delta t = \num{0.01}$.
Figure~\ref{figure_phase_diagram}a-d and SI Movie~\ref{movie_vicsek} show simulations exhibiting (a) disordered, (b) flocking, (c) antiflocking and (d) chiral behaviors.
For these simulations, the noise is (a) $\eta = \num{200e-2}$ and (b,c,d) $\eta = \num{2e-2}$ and the coupling matrices $(J_{A A}, J_{A B} ; J_{B A}, J_{B B})$ are
(a) $\num{1e-4} \times (1, 1 ; 1, 1)$;
(b) $(1, 1 ; 1, 1)$;
(c) $(1, -1 ; -1, 1)$;
(d) $\num{0.39} \times (1, -0.25 ; 0.25, 1)$.

\methodsection{Hydrodynamic theory for non-reciprocal flocking}
\label{hydrodynamic_theory_flocking}

We have derived hydrodynamic equations for the densities $\rho_a(t,r)$ and polarizations $\vec{P}_a(t,r)$ (or equivalently the velocities $\vec{v}_a(t,r)$; in the main text, we denote the polarization fields by $\vec{v}_a(t,r)$ for simplicity) of an arbitrary number of populations from Eqs.~(\ref{kuramoto}--\ref{vicsek_r}).
Our derivation, presented in the SI Sec. I, follows the methods described in Refs.~\cite{Dean1996,Bertin2006,Bertin2009,Farrell2012,Yllanes2017,Marchetti2013,Chate2019}. The set of hydrodynamic equations obtained generalize the Toner-Tu equations~\cite{Toner1995,Toner1998} to several populations with non-reciprocal interactions, and are the basis of the analysis in the main text.

Our results for two populations also generalize the situation considered in Ref.~\cite{Yllanes2017}, which considers aligners A (standard Vicsek-like self-propelling particles) and dissenters B that do not align at all with anyone (neither A or B), but with which the population A aligns. 
With our notations, this corresponds to $j_{A A}, j_{A B} > 0$ but $j_{B B} = j_{B A} = 0$.

Several methods of deriving continuum hydrodynamic equations from microscopics were applied to active matter, going from (i) approaches based on the Fokker–Planck (Smoluchowski) equation for the hydrodynamic variables~\cite{Dean1996,Farrell2012,Marchetti2013}, to (ii) kinetic theory approaches based on the Boltzmann equation~\cite{Bertin2006,Bertin2009,Peshkov2014}, or (iii) directly from the Chapman-Kolmogorov equation~\cite{Ihle2011} (in increasing order of complexity).
Although coarse-graining microscopic models provides invaluable qualitative insights on the behaviour of the system, even current state-of-the-art coarse-graining procedures only provide a qualitative agreement, at best semi-quantitative, with the microscopic starting point~\cite{Chate2019,Mahault2019}. With this in mind, we use the easiest coarse-graining method (i) along with several simplifying approximations (see SI Sec. I). This procedure has the benefit of simplicity and allows to highlight the key features of a non-reciprocal multi-component fluid.
However, the correspondence between the resulting hydrodynamic equations and the microscopic model is only qualitative, in the sense that the values of the coefficients might be inaccurate.

Here, we write the hydrodynamic equations for two populations $a=A,B$.
We refer to SI Sec. I for the general case of an arbitrary number of populations and its derivation from the microscopic equations.

The continuity equation reads
\begin{equation}
	\label{density_eom}
	\partial_t \rho_a + v_0^a \div(\vec{P}_a) = 0.
\end{equation}
while the equation of motion for the polarizations reads
\begin{widetext}
\begin{equation}
\begin{split}
	\label{eom_with_gradients_two_populations}
	\partial_t \vec{P}_A
	= 
	\left[
	j_{A A} \rho_A
	- \eta
	- \frac{1}{2 \eta}  \lVert j_{A A} \vec{P}_A + j_{A B} \vec{P}_B \rVert^2
	\right]
	 \vec{P}_A
	+ j_{A B} \rho_A \vec{P}_B \\
- \frac{v_0^A}{2} \nabla \rho_A + D_A \nabla^2 \vec{P}_A \\
	+ \lambda_{A A} \left[
	 5/2 \nabla(\vec{P}_A \cdot \vec{P}_A) 
	 - 3 (\vec{P}_A \cdot \nabla) \vec{P}_A
	 - 5 \vec{P}_A \div(\vec{P}_A)
	 \right] \\
	 + \lambda_{A B} \left[
		     (\vec{P}_B \cdot \nabla) \vec{P}_A 
		 - 2 (\vec{P}_A \cdot \nabla) \vec{P}_B
		 - 2 \vec{P}_B \div(\vec{P}_A)
		 +   (\vec{P}_{B}^{*} \cdot \nabla) \vec{P}_{A}^{*}
		 + 2 (\vec{P}_{A}^{*} \cdot \nabla) \vec{P}_{B}^{*}
		 + 2 \vec{P}_{B}^{*} \div(\vec{P}_{A}^{*})
	 \right]
\end{split}
\end{equation}
\end{widetext}
Here, $j_{a b} = \frac{R_0^2}{2} \, J_{a b}$ where $R_0$ is a characteristic length scale.
The equation for $B$ is obtained by exchanging the indices.
In this equation, the notation $\vec{U}^*$ denotes the 2D vector $\vec{U}=(U_x, U_y)$ rotated by \and{90} in the clockwise direction, namely $\vec{U}^*=(U_y, -U_x)$.
The polarizations denoted by $\vec{P}_A$ and $\vec{P}_B$ here and in the SI are called $\vec{v}_A$ and $\vec{v}_B$ in the main text.
The hydrodynamic parameters in Eq.~\eqref{eom_with_gradients_two_populations} are related to the microscopic parameters, see SI Sec. I.
When considering uniform fields, Eq.~\eqref{eom_with_gradients_two_populations} reduces to
\begin{equation}
  \label{eom_mean_field_two_populations}
  \partial_t \begin{pmatrix} \vec{P}_A \\ \vec{P}_B \end{pmatrix}
  = \begin{pmatrix}
\alpha_{A}[\vec{P}_A,\vec{P}_B]  & 
  \ j_{AB} \rho_A\\
  j_{BA} \rho_B & 
\alpha_{B}[\vec{P}_A,\vec{P}_B] \end{pmatrix} \begin{pmatrix} \vec{P}_A \\ \vec{P}_B \end{pmatrix}
\end{equation}
where
\begin{equation}
	\alpha_{A} = j_{A A} \rho_A	- \eta - \frac{1}{2 \eta}  \lVert j_{A A} \vec{P}_A + j_{A B} \vec{P}_B \rVert^2.
\end{equation}
Again, the equation for $B$ is obtained by exchanging the indices.
This equation is used to construct the phase diagrams of Fig.~\ref{figure_phase_diagram}, see SI Sec. II.
{
We find (a) a disordered regime where the order parameter vanishes, (b) a flocking regime where the order parameters are parallel, (c) an antiflocking regime where the order parameters are antiparellel (sharing some similarities with Refs.~\cite{Oza2016,Suzuki2015,Nishiguchi2017}), (d) a periodic chiral regime where the order parameter have circular trajectories (sharing similarities with chiral active matter~\cite{Tsai2005,Liebchen2017,Banerjee2017,Souslov2019,Soni2018,Han2020}), (e) a periodic swap regime where the order parameter oscillate along a fixed direction, (f) a quasiperiodic chiral+swap regime in which the order parameter oscillates along a rotating direction.
}
Equation~\eqref{eom_with_gradients_two_populations} is then linearized above the uniform solution of Eq.~\eqref{eom_mean_field_two_populations} to obtain the stability diagram of Fig.~\ref{figure_stability} (see SI Sec. III for details on the computations).

\methodsection{Simulations of the continuum equations in exceptional point induced pattern formation}
\label{simulations_continuum_equations}

To explore the pattern formation beyond linear stability, we directly solve the hydrodynamic equation \eqref{eom_with_gradients_two_populations} under periodic boundary conditions using the open-source pseudospectral solver Dedalus~\cite{Burns2020}.
For simplicity, we focus on Eq.~\eqref{eom_with_gradients_two_populations} only and assume that the fluctuations of the densities $\rho_a$ are high-frequency modes that can be ignored and integrated out and set $\rho_a \approx \text{const.}$, but we do not enforce the incompressibility constraint $\div(\vec{v}_{a}) = 0$ that would arise from Eq.~\eqref{density_eom} in a system where mass is conserved.
In the SI Sec. XI, we show that pattern formation also occurs when the incompressibility constraint $\div(\vec{v}_{a}) = 0$ is enforced, both through a linear stability analysis and simulations of the non-linear hydrodynamic equations.
The simulations in Fig.~\ref{figure_stability}(c-g) are performed on a box of size $2L \times 2L$ with $L = \num{0.32}$. Each dimension is discretized with $N={2^6}$ modes. A random initial condition (in space, with the value at each point drawn independently from a uniform distribution on $[-1,1]$) is evolved in time with the time-stepper SBDF2 (second-order semi-implicit backwards differentiation formula) implemented in Dedalus~\cite{Burns2020,Wang2008} with a constant time step $\delta t = \num{0.01}$ for a total time $T_{\text{sim}} \approx \num{2000}$.

\methodsection{Phase transitions and bifurcations}
\label{phase_transitions_bifurcations}

In this section, we first review standard relations between phase transitions and bifurcations of dynamical systems.
We then discuss our results from the point of view of bifurcation theory~\cite{Arnold1988,Kuznetsov2004,Golubitsky1985b,Golubitsky1988,Golubitsky2002,Chossat2000}. The exceptional transitions analysed in the main text are closely related to Bogdanov-Takens bifurcations~\cite{Bogdanov1981a,Bogdanov1981b,Takens1974,Kuznetsov2004}.
At first sight, a striking difference is present: the Bogdanov-Takens (BT) bifurcation has codimension two (i.e., two parameters have to be adjusted to get to the bifurcation), while the exceptional transitions in our work have codimension one (i.e., a single parameter has to be adjusted; hence, we observe transition \emph{lines} in a 2D phase diagram).
We argue that this apparent tension is solved because the Goldstone theorem effectively reduces the codimension of BT bifurcations from phases with a spontaneously broken continuous symmetry.

\subsection{Phase transitions and dynamical systems}

Equilibrium phase transitions are usually described in terms of a free energy. The minimum of the free energy corresponds to the current phase, and a phase transition occurs when it ceases to be a global minimum, or a minimum at all.
Although this landscape picture is static, it relies on an underlying dynamics that shepherds the system into the global minimum in a way or another~\cite{Landau1954,Hohenberg1977,Cross1993}.
For instance, it arises naturally when one considers a Ginzburg-Landau-Wilson Hamiltonian obtained from renormalizing a microscopic Hamiltonian~\cite{Wilson1974,Hohenberg2015} instead of a phenomenological Ginzburg-Landau free energy.

The most simple dynamics is relaxational. In this case, the dynamical system that describes the time evolution of the order parameter~$\phi$ near its equilibrium value reads~\cite{Landau1954,VanHove1954,Hohenberg1977,Cross1993,Hohenberg2015}
\begin{equation}
  \label{TDGL}
  \frac{\partial \phi}{\partial t} = - \frac{\delta F}{\delta \phi}
\end{equation}
for a system described by the Ginzburg-Landau free energy $F[\phi]$. 
Phase transitions can be seen as bifurcations of this dynamical system~\cite{Lagues2003,Muoz2018,Sornette2000}.

Let us immediately note that this point of view naturally encompasses out-of-equilibrium systems, for they have an equivalent of equation \eqref{TDGL} even when they are not described by a free energy (and more generally, when they do not possess a Lyapunov function).
For instance, this occurs in non-equilibrium pattern formation~\cite{Cross1993,Saarloos1994,Aranson2002,Hohenberg2015} (we refer to \cite[III.A.5]{Cross1993} for a discussion on the difference between bifurcations and thermodynamic phase transitions; here, we will use liberally the term \enquote{phase transition} to describe both situations).

As an example, consider the paramagnet/ferromagnet transition of the Ising model, for which the Ginzburg-Landau free energy density reads $f(\phi) = (a/2) \phi^2 + (b/4) \phi^4 + (c/2) (\nabla \phi)^2$.
This free energy describes any real-valued scalar order parameter $\phi$ with inversion symmetry $\phi \to - \phi$, which also include other systems such as incompressible symmetric binary mixtures~\cite{Barrat2003}. Applying equation \eqref{TDGL} gives
\begin{equation}
	\partial_t \phi = - a \phi - b \phi^3 + c \Delta \phi.
\end{equation}
For uniform order parameters, the Laplacian $\Delta \phi$ vanishes and we recognize the normal form of a supercritical pitchfork bifurcation, up to rescaling of the parameters. 
It is instructive to analyze the stability of the equilibrium solutions $\phi_{0} = 0$ and $\phi_{\pm} \equiv \sqrt{-a/b}$ of this dynamical system by linearizing \eqref{TDGL} around equilibria $\phi_\text{ss}$ and computing a Fourier transform to momentum space.
Writing $\phi = \phi_\text{ss} + \delta \phi$, we find
\begin{equation}
	\frac{\partial \delta \phi}{\partial t} = \left[ - a - 3 b \phi_\text{ss}^2 + c \nabla^2 \right] \delta \phi + \mathcal{O}(\delta \phi^2).
\end{equation}
In momentum space (where $k$ is the momentum),
\begin{equation}
	\frac{\partial \delta \phi}{\partial t} = - \left[ a + 3 b \phi_\text{ss}^2 + c k^2 \right] \delta \phi.
\end{equation}
Hence, the growth rates of the Fourier modes $\delta \phi(k)$ of the perturbation are $\sigma_0 = - a - c k^2$ for the trivial equilibrium $\phi = 0$, and $\sigma_{\pm} =  2 a - c k^2$ for the two non-trivial equilibria (when they exist).

At the transition, the growth rate of the long-wavelength perturbations (i.e., at $k=0$) vanishes (in the ordered phase, it is negative, meaning that the perturbations are damped).
In the language of high-energy physics, a massive (i.e, gapped) mode becomes massless (i.e., gapless).
This is a standard mechanism for phase transitions: global fluctuations (i.e., with arbitrarily large wavelength) becomes less and less damped in time, until the transition where they are not damped anymore.

\subsection{Relations with bifurcation theory}

In the main text, we show that the phase transition between the aligned phase (or antialigned) and the chiral phase are marked by the presence of an exceptional point (EP) in the linearized dynamical system (hence, we refer to these as \emph{exceptional transitions}).
From the point of view of bifurcation theory, this is a Bogdanov-Takens (BT) bifurcation~\cite{Bogdanov1981a,Bogdanov1981b,Takens1974}, precisely characterized by the occurrence of an exceptional point (equivalently, a Jordan block of size two), see Ref.~\cite{Kuznetsov2004}.
A direct computation of the linearized operator at a typical point in the exceptional transition line (in red in Fig.~\ref{figure_phase_diagram}b-c) confirms that its (real) eigenvalues satisfy $\lambda_1 = \lambda_2 = 0 < \lambda_3 < \lambda_4$ (in this section, we order eigenvalues by decreasing real part, so $\lambda_1$ is the most unstable), see also SI Sec. II for an analytical proof on the occurrence of EPs.
However, the BT bifurcation has codimension two (i.e., two parameters have to be adjusted to get to the bifurcation), so BT bifurcations are typically points in a two-dimensional parameter space. In contrast, the exceptional transitions in our work have codimension one (i.e., a single parameter has to be adjusted), and we observe transition \emph{lines} in the two-dimensional phase diagram Fig.~\ref{figure_phase_diagram}b-c. We note that the phase diagram in Fig.~\ref{figure_phase_diagram}b-c is not fine-tuned (besides the $O(2)$ symmetry of the dynamical system): the existence of transition lines persists even if we perturb the dynamical system, see below. 

\begin{figure}
  \centering
  \includegraphics[width=\columnwidth]{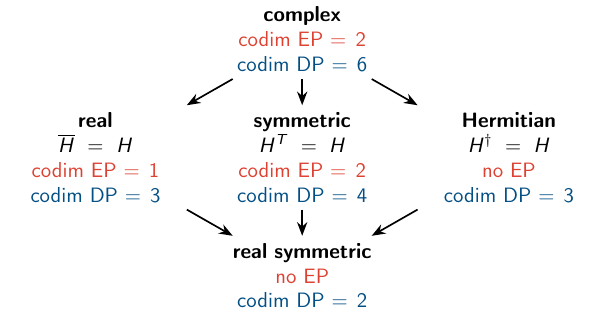}
  \caption{\label{figure_codimensions}\strong{Codimensions of eigenvalue degeneracies.}
  This graph gives the codimension (codim) of two-fold degeneracies of eigenvalues in different matrix spaces, see Ref.~\cite{Seyranian2005}.
  These degeneracies can be exceptional points (EP) or diabolic points (DP, also known as Dirac points).
  An identical graph can be drawn by replacing \enquote{real symmetric} with \enquote{purely imaginary symmetric}, \enquote{Hermitian} with \enquote{anti-Hermitian} and \enquote{real} with \enquote{imaginary}.
  }
\end{figure}

To solve this puzzle, let us analyze the bifurcation conditions that characterize the BT bifurcation, and how they usually lead to a codimension~two.
The codimension of subspaces with equal eigenvalues is a nontrivial problem, see Refs.~\cite{vonNeumann1929,Arnold1972,Arnold1995,Seyranian2005}.
The BT transition occurs at an equilibrium point where the Jacobian has a vanishing eigenvalue of algebraic multiplicity two $\lambda_1 = \lambda_2 = 0$.
Such a degeneracy can occur in two ways: at an exceptional point (EP) where the eigenvectors become collinear, or at a diabolic point (DP, also known as Dirac point) where the eigenvectors stay linearly independent.
DPs have a considerably higher codimension that EPs (so they can essentially be ignored), see Extended Data Fig.~\ref{figure_codimensions}; they do not correspond to the BT bifurcation (see e.g. Refs.~\cite{Julien1994,Renardy1999} for the corresponding codimension 4 bifurcation). 
For real matrices, the codimension of EPs is one; combined with the condition that the degenerate eigenvalues must vanish, this gives a codimension two to the BT bifurcation.

The reason why the codimension is different here lies in symmetry. A direct inspection shows that the dynamical system Eq.~\eqref{eom_with_gradients_two_populations} is invariant under the group $O(2)$ of orthogonal transformations (acting diagonally, i.e. on all populations at the same time).
We first note that the phase transitions in Fig.~\ref{figure_phase_diagram}b-c are not Takens-Bogdanov bifurcations with $O(2)$ symmetry in the sense of Ref.~\cite{Guckenheimer1986,Dangelmayr1987}, because the stable eigenvalues $0 < \lambda_3 \neq \lambda_4$ are generally different.
This is because we are considering the departure from a (time-independent) ordered phase that spontaneously breaks the $O(2)$ symmetry (such as the aligned phase), not from a fully symmetric steady-state (such as the disordered phase). (This could be analyzed as a secondary bifurcation through mode interactions~\cite{Golubitsky1985b,Golubitsky1988,Golubitsky2002} or with the formalism of Refs.~\cite{Krupa1990,Field1980}.)
A crucial consequence of the spontaneous breaking of a continuous symmetry is the appearance of modes with vanishing frequency and growth rate at large wavelength called Nambu-Goldstone modes. This property is known as the Goldstone theorem, see Refs.~\cite{Nambu1960,Goldstone1961,Goldstone1962,Hidaka2013,Watanabe2020,Watanabe2012,Nielsen1976}, and we note that it applies to dynamical systems (not only Hamiltonian systems), see Refs.~\cite{Leroy1992,Minami2018,Hongo2019,Watanabe2020}.
Because of the Goldstone theorem, one eigenvalue $\lambda_1 = 0$ always vanishes in the static phases with a spontaneously broken continuous symmetry (such as the aligned and antialigned phases). In this situation, the codimension of the BT transition is simply the codimension of EPs, which is one. (More precisely, it is the codimension of EPs in the space of matrices with at least one zero eigenvalue.)
To summarize, the existence of a Goldstone mode associated with a spontaneously broken continuous symmetry effectively reduces the codimension of BT bifurcations by one, leading to lines of BT bifurcations from the phase with broken symmetry in a two-dimensional phase diagram,

\begin{figure}
  \centering
  \includegraphics[width=\columnwidth]{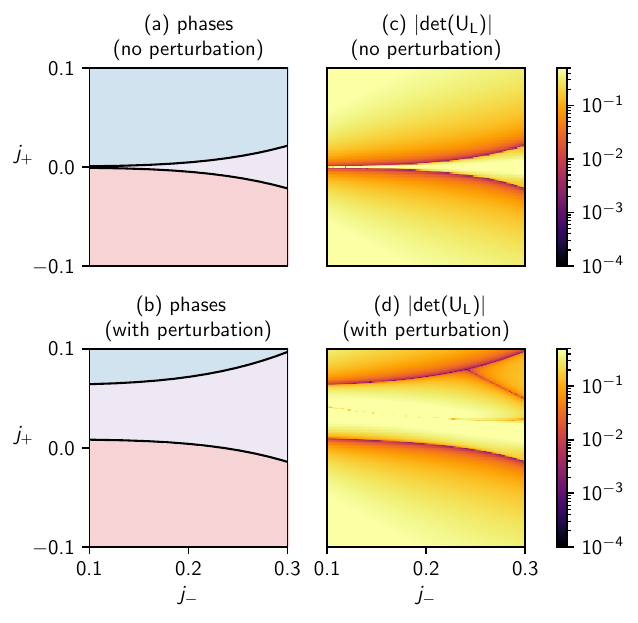}
  \caption{\label{si_figure_effect_perturbations_O2}\strong{Effect of generic symmetry-preserving perturbations.}
  We compare (a,c) a portion of the phase diagram from Fig.~\ref{figure_phase_diagram} to (b,d) the same portion of the phase diagram, after a generic $O(2)$-symmetric perturbation.
  We plot (a,b) the phase diagrams and (c,d) the absolute value of the determinant of the matrix $U_L$ of eigenvectors of $L(k=0)$.
  The vanishing of the determinant in (c,d) marks exceptional points (EPs).
  The phase boundaries are modified, but the topology of the diagram is not modified by small perturbations, and lines of EPs still mark the transition between (anti)aligned and chiral phases. Direct inspection of the spectrum of $L$ shows that the EPs indeed correspond to the coalescence of the two most unstable eigenvalues $\lambda_1$ and $\lambda_2$.
  We also note the existence of additional lines of EPs inside the chiral phase in the perturbed case. These are not the subject of the present analysis.
  We have used the same parameters as in figure~\ref{figure_phase_diagram} with $\eta/\eta_{\text{c}} = \num{0.5}$.
  For the perturbed case (b,d), the deviations $\Delta p$ of the parameters $p$ from the unperturbed case are $\Delta\beta_{AAAA}$ = \num{-0.015}, $\Delta\beta_{AABA}$ = \num{-0.035}, $\Delta\beta_{ABBA}$ = \num{-0.045}, $\Delta\beta_{BAAB}$ = \num{-0.025}, $\Delta\beta_{BABB}$ = \num{-0.045}, $\Delta\beta_{BBBB}$ = \num{-0.065}, $\Delta\beta_{AAAB}$ = \num{-0.01}, $\Delta\beta_{AABB}$ = \num{-0.03}, $\Delta\beta_{ABAB}$ = \num{-0.05}, $\Delta\beta_{ABBB}$ = \num{-0.07}, $\Delta\beta_{BAAA}$ = \num{-0.02}, $\Delta\beta_{BABA}$ = \num{-0.04}, $\Delta\beta_{BBAA}$ = \num{-0.06}, $\Delta\beta_{BBBA}$ = \num{-0.08}, $\Delta{j}_{AA}$ = \num{0.05}, $\Delta{j}_{BB}$ = \num{0.01}, $\Delta\rho_{A}$ = \num{-0.05}, $\Delta\rho_{B}$ = \num{0.04}.
  }
\end{figure}

If the existence of Bogdanov-Takens lines is indeed due to the $O(2)$ symmetry through the Goldstone theorem, BT lines should persist under any (small) perturbation that preserves the $O(2)$ symmetry of the dynamical system. A full analysis using the methods of equivariant bifurcation theory is outside of the scope of this work; instead, we now provide numerical evidences that support our hypothesis. To do so, we consider the dynamical system
\begin{equation}
	\label{general_O2_invariant}
	\partial_t \vec{P}_a = \alpha_{a b} \vec{P}_b + \beta_{a b c d} \braket{\vec{P}_b, \vec{P}_c} \vec{P}_d
\end{equation}
which includes all $O(2)$-symmetric terms up to order three in $\vec{P}_a$ (see e.g. Ref~\cite{Dangelmayr1987}; we can choose $\beta_{a b c d} = \beta_{a c b d}$ by symmetry of the Euclidean scalar product, so there are a total of $n^2 + n \times n(n+1)/2 \times n$ parameters in this dynamical system for $n$ populations [\num{16} for $n=\num{2}$], two of which might be removed by rescalings).
We start from values of the parameters corresponding to the dynamical system Eq.~\eqref{eom_with_gradients_two_populations}, and add (small) perturbations, see
Extended Data Fig.~\ref{si_figure_effect_perturbations_O2}. The figure shows that lines of exceptional points marking (anti)aligned/chiral transitions persist under a (small enough) generic $O(2)$-preserving perturbation. This strongly suggests that this phenomenon is not the result of a fine-tuning of some parameters not accounted for in our particular model.

Our argument assumes that time-independent steady-states form a submanifold of codimension zero. 
This is not always the case. For instance, this is not true when the symmetry group is reduced to $SO(2)$ (instead of $O(2)$). 
Correspondingly, the codimension of the BT bifurcation becomes higher (we find codimension 2 points in the systems analyzed in the main text, but not all possible terms are considered).
Physically, this is because there are parameters (e.g., external torques) that drive the Goldstone mode in a given direction (we interpret this as an explicit PT-symmetry breaking).

\medskip

In the main text, we have described the appearance of time-dependent phases as a dynamical restoration of (part of) the spontaneously broken symmetries.
The main idea is that a continuous group of symmetries $G$ that is spontaneously broken to a subgroup $H \subset G$. 
A time-independent steady-state can be represented by a point in $G/H$ (the \enquote{manifold of degenerate ground states}), while a limit cycle corresponds to a loop $\mathcal{C}$ in $G/H$ (we assume that the motion can be made harmonic, e.g. by a nonlinear change of variables and a reparameterization of time; in the instances of the chiral phase analyzed in the main text, the motion is already harmonic and hence the loop is traveled at constant velocity).
With this picture in mind, the class I defined in the main text corresponds to situations in which some operation in $H$ exchanges clockwise and counterclockwise motions on $\mathcal{C}$ (like parity does in the case of $O(2)$), while class II corresponds to situations in which there is no such operation. 
In class II, there is a predefined sense of rotation on the loop $\mathcal{C}$ (corresponding to an explicitly broken PT symmetry), while there is not in class I (in which the onset of a rotation either clockwise or counterclockwise along $\mathcal{C}$ would be a manifestation of a spontaneous PT symmetry breaking).

The partial restoration of spontaneously broken symmetries through the appearance of time-dependent phases also occurs for discrete groups (though not through an exceptional point).
For instance, the swap phase corresponds to a $\ZZ_2$ symmetry. Near the transition, the order parameters approach square functions corresponding to discrete jumps between the two states related by the broken $\ZZ_2$ symmetry (orthogonal to the common direction of the order parameters in the (anti)aligned phase), leading to the rich harmonic content in Fig.~\ref{figure_phase_diagram}g. 
In contrast, the continuous $SO(2)$ symmetry restored by the chiral phase corresponds to a harmonic motion (i.e., there is a fundamental frequency without higher harmonics), by which the continuous manifold of degenerate ground state is continuously traveled, with all points equivalent.

\methodsection{Non-reciprocal Kuramoto model}
\label{non_reciprocal_kuramoto_model}

In this section, we provide details on the analysis of the non-reciprocal Kuramoto model~\cite{Kuramoto1984,Acebron2005,Sakaguchi1986,Daido1992,Daido1987,Omata1988,Hong2011,Hong2011b,Martens2009,Bonilla1998,Uchida2010,Hong2012,Ott2008,Abrams2008,Pikovsky2008,Martens2016,Choe2016,Gallego2017}.
Depending on whether the system is in class I or in class II (PT-symmetric or not), we find codimension 2 or codimension 1 exceptional points around which the phase diagram is organized.
In class I, the exceptional line (in a 2D phase diagram) separate static (aligned or antialigned) phases from a chiral phase where parity (equivalent here to PT-symmetry) is spontaneously broken.
In class II, an exceptional point structures the phase diagram: the stable steady-states are organized on a truncated version of the Riemann surface of the square root. 
This leads to discontinuous transitions marked by hysteresis between regions where two stable states coexist and regions where only one state exists, in a similar manner to driven-dissipative quantum fluids~\cite{Hanai2019,Hanai2020}.
We first present analytic self-consistency arguments where the existence of static or harmonic steady-state is assumed.
We then resort to numerical simulations of a reduced dynamical system to confirm our analytic predictions and explore the full phase diagram, including non-harmonic time-dependent phases.

\subsection{General considerations}

\def\kuramotoj{j}

We start from Eq.~\eqref{kuramoto} of the main text. 
Following the standard Kuramoto model, we consider globally coupled (all-to-all) oscillators, and neglected the noise $\eta(t) \equiv 0$. 
The oscillators are separated into two communities $A$ and $B$, and the dynamics of the oscillators separated into two populations reads
\begin{equation}
\label{kuramoto_for_species}
\partial_t \theta_m^a 
= \omega_m^a 
+ \sum_{b} \sum_{n=1}^{N_b}
J_{a b} \sin(\theta_n^b-\theta_m^a)
\end{equation}
where $a,b=A,B$ represent the two species (or communities), and where we neglected the noise for simplicity. 
Hence, the coupling constants $J_{m n}$ can be $J_{A A}$, $J_{A B}$, $J_{B A}$, $J_{B B}$ depending on which populations $m$ and $n$ belong to. 
The distribution of the natural frequencies $\omega^a_m$ in different groups may be different in general.
The conventional Kuramoto model~\cite{Kuramoto1984,Acebron2005} is recovered by setting the coupling strength to be identical, i.e. $J_{AA}=J_{AB}=J_{BA}=J_{BB}$.

Note that the similarity between Fig.~\ref{si_kuramoto_oa_phase_diagram} and the Fig.~\ref{figure_phase_diagram} of the main text can already be anticipated from the equations of motion.
The Kuramoto model with the Vicsek model are very similar besides the obvious differences summarized in Table~\ref{table_translation}, when all the oscillators are identical (i.e., they have the same natural frequencies $\omega_m^A=\omega_m^B\equiv \omega_0$),  as the common frequency $\omega_0$ can be absorbed by a transformation of the degrees of freedom (where the oscillators are observed in a rotating frame).
As a result, we can expect similar phases as those found in Fig.~\ref{figure_phase_diagram} (such as the flocking and chiral phases) to arise.  
Indeed, from numerical simulation the dynamics using the methods introduced in the next section, we indeed find a surprisingly similar phase diagram to the flocking model (Fig.~\ref{si_kuramoto_oa_phase_diagram}).
In terms of synchronization, the disordered phase corresponds to a desynchronized phase, while the flocking, antiflocking, and chiral phases all correspond to synchronized oscillators, respectively in phase, completely out of phase, and with an arbitrary phase delay.
They are respectively named incoherent state, coherent state, $\pi$-state, and traveling wave state in Refs.~\cite{Hong2011,Choe2016}, see also Table~\ref{table_translation}.

\subsection{Self-consistency equation for the steady-states}

Following Kuramoto~\cite{Kuramoto1984}, we introduce an order parameter that characterizes synchronization for each species
\begin{equation}
z_a(t) \equiv r_a(t) \ee^{\ii\phi_a(t)} 
=\frac{1}{N_a}
\sum_{m=1}^{N_a} \ee^{\ii\theta_m^a(t)}
\end{equation}
which becomes finite when the oscillators synchronize. 
Here, $r_a\ge 0$ and $\phi_a$ respectively characterize the phase coherence and the average phase of the component $a$.

Introducing $\kuramotoj_{a b} = J_{a b} N_b$ and 
\begin{equation}
\label{Ra}
R_a \ee^{\ii\alpha_a} = \sum_b \kuramotoj_{a b} z_b
\end{equation}
allows to express Eq.~\eqref{kuramoto_for_species} as
\begin{equation}
\label{Kuramoto_2}    
\partial_t \theta_m^a = \omega_m^a +R_a \sin(\alpha_a-\theta_m^a).
\end{equation}

In order to obtain a continuum equation for a large number of oscillators $N_a \to \infty$, we introduce the distribution of natural frequencies
$g_a(\omega)=\frac{1}{N_a}\sum_{m=1}^{N_a}\delta(\omega-\omega_m^a)$
and the density of angles
$\rho_a(\theta;\omega,t)=\delta(\theta-\theta_a(\omega,t))$
where $\theta_a(\omega,t)$ is a solution of Eq.~\eqref{Kuramoto_2} in which $\omega_m^a$ is replaced by $\omega$ and $\theta^a_m$ by $\theta_a(\omega)$, with an initial condition specified by $\rho_a(\theta;\omega,t=0)$.
The order parameter then becomes
\begin{equation}
z_a(t)=\int_{-\infty}^\infty d\omega \int_{-\pi}^\pi \dd\theta \ee^{\ii\theta}
g_a(\omega) \rho_a(\theta;\omega,t)
\end{equation}
This equation provides a self-consistency condition for the order parameter. 

We focus on steady-states of the form
\begin{equation}
\label{Kuramoto steady state}
    z_a(t) = z_a \ee^{\ii \Omega t}
    \quad
    \text{with}
    \quad
    z_a = r_a \ee^{\phi_a^0}.
\end{equation}
Those must satisfy the self-consistency equation~\cite{Acebron2005}
\begin{equation}
z_a = R_a e^{i\alpha_a} F_a[z_A,z_B],
\end{equation}
where we have introduced the functions
\begin{equation}
\begin{split}
\label{Fa}
&F_a[z_A,z_B]=\int_{-\pi/2}^{\pi/2} \dd\theta \cos\theta \, \ee^{\ii\theta} 
g_a(R_a\sin\theta +\Omega)
\\
&+\int_{-\pi}^\pi \dd\theta\int_{|x|>1} \dd x
\, \ee^{\ii\theta} g_a(R_a x + \Omega)
\frac{\sqrt{x^2-1}}{|x-\sin \theta|}.
\end{split}
\end{equation}
The first (second) term in Eq.~\eqref{Fa} is the contribution from the synchronized (unsynchronized) oscillators. 
Using Eq.~\eqref{Ra}, this condition can be written in the form,
\begin{equation}
\label{Kuramoto non-Hermitian}
\hat{M}[z_A,z_B,\Omega] \begin{pmatrix} z_A \\ z_B \end{pmatrix} = 0,
\end{equation}
with
\begin{equation}
\hat{M}[z_A,z_B,\Omega] =
\begin{pmatrix} 
F_A^{-1} - \kuramotoj_{AA} & - \kuramotoj_{AB} \\ 
-\kuramotoj_{BA} & F_B^{-1} - \kuramotoj_{BB}
\end{pmatrix}.
\end{equation}

The structure of Eq.~\eqref{Kuramoto non-Hermitian} is similar to Eq.~\eqref{eom_mean_field_two_populations} (obtained for non-reciprocal flocking), in the sense that the steady state condition is controlled by a nonlinear non-Hermitian $2\times 2$ matrix. 
This is more easily illustrated in the particular case in which $g_A(\omega)=g_B(\omega)$ is a symmetric distribution centered at $\omega_0$.
When $|z_a|$ is small, we can expand $F_a^{-1}=\eta + \beta R_a^2 + \ii\zeta\Omega + \mathcal{O}(\Omega^2, \Omega R_a, R_a^2)$ where $\eta$, $\beta$, $\zeta$ are real numbers and set $\omega_0=0$ without loss of generality. 
We then find
\begin{equation}
 \label{Kuramoto zero detuning}
  \ii \Omega \zeta
 \begin{pmatrix} z_A \\ z_B \end{pmatrix}
 =\begin{pmatrix}
 \tilde{\alpha}_{A}[z_A,z_B] & \!\!  \kuramotoj_{AB}\\
 \kuramotoj_{BA} \!\! & \tilde{\alpha}_{B}[z_A,z_B]  
 \end{pmatrix}
 \begin{pmatrix} z_A \\ z_B \end{pmatrix}
\end{equation}
where $\tilde{\alpha}_a = -\eta + \kuramotoj_{a a} - \beta R_a^2$.
This equation has a remarkable resemblance to Eq.~\eqref{eom_mean_field_two_populations} (with $\partial_t \to \ii \Omega$).
This suggests that in this particular case, the transition from the phase corresponding to the flocking/antiflocking phase ($\Omega=0$) to one corresponding to the chiral phase ($\Omega\ne 0$) takes place at an exceptional point of $\hat M$. 

However, an important difference arises in the general case where $g_A(\omega)\ne g_B(\omega)$:
as the (generalized) PT symmetry is explicitly broken by the presence of natural frequencies, the matrix $\hat M$ in Eq.~\eqref{Kuramoto non-Hermitian} is \emph{complex}-valued, in contrast with the matrix in Eq.~\eqref{eom_mean_field_two_populations} that is real-valued.
As a consequence, exceptional points should occur at \emph{points} in a two-dimensional parameter space, rather than along lines such as in Figs.~\ref{figure_phase_diagram}b-c (i.e., their codimension is two; see also the discussion in the section \methodsref{phase_transitions_bifurcations}{Phase transitions and bifurcations}). 
This is consistent with the occurrence of Bogdanov-Takens points in generalized Kuramoto models~\cite{Choe2016,Gallego2017,Pazo2009,Pietras2018,Martens2016,Bonilla1998}, in which hysteresis can be present~\cite{Pazo2009,Pietras2018}.
This situation is similar to the case of quantum fluids analyzed in Refs.~\cite{Hanai2019,Hanai2020}, where a first-order-like phase transition associated with a jump in physical quantities may arise with an exceptional point marking the endpoint of the phase boundary. 
To further investigate it, we now analyze the mean-field dynamics of the non-reciprocal Kuramoto model.

\subsection{Mean-field dynamics in the Ott-Antonsen manifold}

The dynamics of the generalized Kuramoto model in Eq.~\eqref{kuramoto} can exactly be captured by a small number of coupled differential equations in the limit of a large number of oscillators, see Refs.~\cite{Ott2008,Ott2009,Watanabe1993,Watanabe1994,Marvel2009,Pikovsky2011,Pikovsky2008,Hong2011,Hong2011b,Hong2012,Tyulkina2018,Montbrio2015} and the review Ref.~\cite{Bick2019}.
In SI Sec. VI, we have demonstrated that this mean-field dynamics is quantitatively consistent with direct simulation of microscopic model Eq.~\eqref{kuramoto}.

Through this mean-field reduction, the evolution of the complex order parameter $z_a(t)$ for each community $a$ is described by~\cite{Ott2008}
\begin{equation}
	\label{kuramoto_ott_antonsen}
	\partial_t z_{a} = (\ii \omega_a - \Delta_a) z_a + \frac{1}{2} \sum_{b} \kuramotoj_{a b} \left[ z_{b} - z_a^2 \, \overline{z_b} \right].
\end{equation}
where $\overline{z_b}$ is the complex conjugate of $z_b$.
We assumed that the natural frequencies of the oscillators in the community $a$ follow a Lorentzian distribution 
$g_{a}(\omega) = \pi^{-1} \, [(\omega - \omega_a)^2 + \Delta_a^2]^{-1}$.
The term $\ii \omega_a z_a$ in Eq.~\eqref{kuramoto_ott_antonsen} explicitly breaks the mirror symmetry $z_a \to \overline{z_a}$ (and hence, breaks parity), but is invariant under rotations $z_a \to \ee^{\ii \theta} z_a$. 

\begin{figure}
  \centering
  \includegraphics[width=\columnwidth]{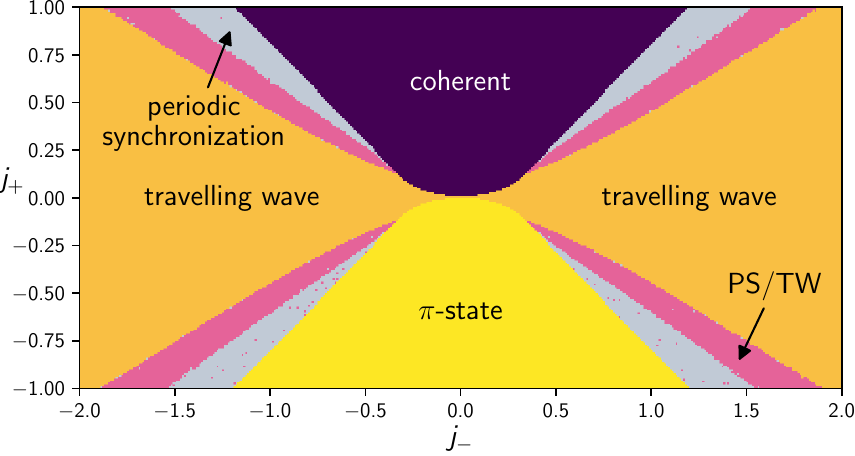}
  \caption{\label{si_kuramoto_oa_phase_diagram}\strong{Phase diagram of the PT-symmetric non-reciprocal Kuramoto model.}
  The states are defined in Table~\ref{table_definition_phases}.
  We have set $\kuramotoj_{AA} = \kuramotoj_{BB} = \num{1}$, $\Delta_{A} = \Delta_{B} = \num{0.25}$ and $\omega_{A} = \omega_{B} = \num{0}$.
  }
\end{figure}

\begin{figure}
  \centering
  \includegraphics[width=\columnwidth]{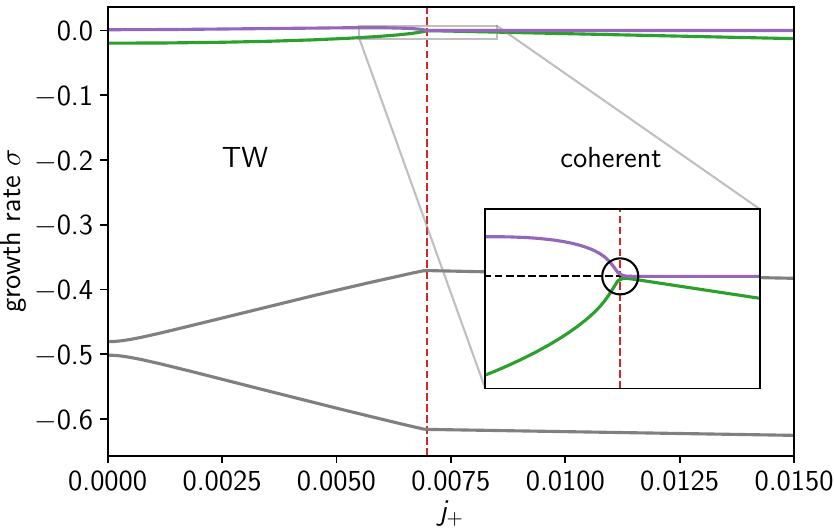}
  \caption{\label{si_kuramoto_oa_growth_rate_Kplus}\strong{Exceptional point in the spectrum of the Jacobian in the Kuramoto model.}
  The two most unstable eigenvalues $\lambda_i = \sigma_i + \ii \omega_i$ of $L$ coalesce at $\kuramotoj_{+} \approx \num{0.007}$. 
  This value coincides with the transition from traveling waves (TW) to coherent states, marked by a red dashed line.
  Note that this coalescence occurs at $\lambda = 0$ (not at finite frequency and/or growth rate).
  The corresponding eigenvectors become collinear (this can be verified, for instance, by computing the determinant of the matrix of eigenvectors, that vanishes at the EP).
  The imaginary parts $\omega_i$ (not shown) are all zero.
  We used the same parameters as in Extended Data  Fig.~\ref{si_kuramoto_oa_phase_diagram} with $\kuramotoj_{-} = \num{0.1}$. 
  A similar behavior is observed for neighboring values of $\kuramotoj_{-}$.
  }
\end{figure}

When $\omega_a = 0$ for all the communities, the system has a full $O(2)$ symmetry, and one observes phases with spontaneously broken parity.
In this paragraph, we focus on this situation.
To mirror the analysis in the main text, we define $\kuramotoj_{\pm} = [\kuramotoj_{AB} \pm \kuramotoj_{BA}]/2$ and determine a numerical phase diagram of the system in the $(\kuramotoj_{-},\kuramotoj_{+})$ plane, see Extended Data Fig.~\ref{si_kuramoto_oa_phase_diagram}. This phase diagram shares several qualitative features with the flocking phase diagram in Fig.~\ref{figure_phase_diagram} in the main text.
In particular, we find that the phase boundaries between the (anti)synchronized state (labeled coherent and $\pi$-state in Extended Data Fig.~\ref{si_kuramoto_oa_phase_diagram}) and the chiral state (labeled traveling wave in Extended Data Fig.~\ref{si_kuramoto_oa_phase_diagram}) are marked by exceptional points in the Jacobian $L$ of the dynamical system Eq.~\eqref{kuramoto_ott_antonsen}.
Writing the right-hand side of Eq.~\eqref{kuramoto_ott_antonsen} as $f_a(z_b)$, the $4 \times 4$ Jacobian matrix $L$ has blocks
\begin{equation}
	\label{jacobian_complex}
	L_{a b} = \begin{pmatrix}
		{\partial f_b}/{\partial z_a} & {\partial f_b}/{\partial \overline{z_a}} \\
		{\partial \overline{f_b}}/{\partial z_a} & {\partial \overline{f_b}}/{\partial \overline{z_a}} \\
	\end{pmatrix}
\end{equation}
for $a,b=A,B$, where the derivatives are evaluated at the steady-state.
A direct numerical evaluation of this matrix shows that the two most unstable eigenvalues indeed coalesce (i.e., form an exceptional point) at the transition, see Extended Data Fig.~\ref{si_kuramoto_oa_growth_rate_Kplus} for an example.

\begin{figure*}
  \centering
  \includegraphics[width=\textwidth]{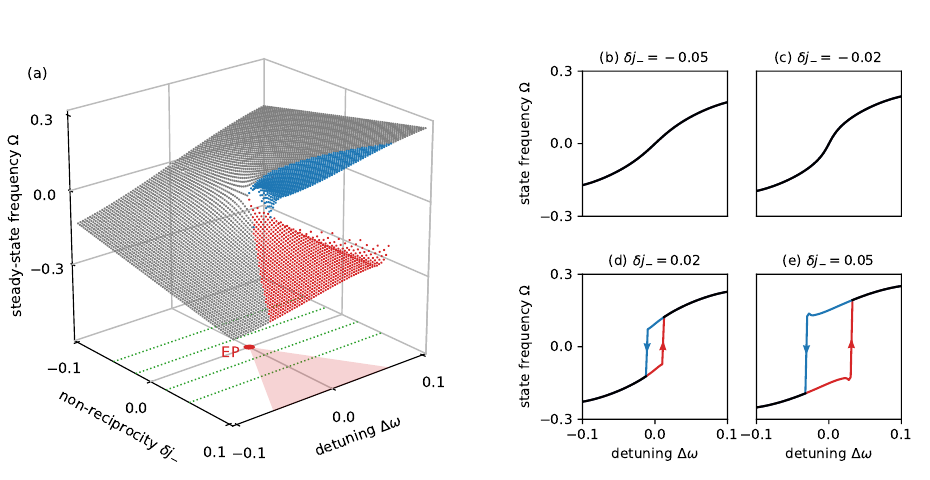}
  \caption{\label{si_figure_hysteresis}\strong{Hysteresis in the chiral Kuramoto model.}
  When chirality is explicitly broken, exceptional points have codimension two, i.e. they are typically points in a two-dimensional parameter space.
  We plot the frequency $\Omega$ of the steady-state of the Kuramoto model with explicitly broken PT symmetry as a function of the difference $\Delta \omega = \omega_{A} - \omega_{B}$ between the two communities (also called detuning) and the deviation $\delta \kuramotoj_{-} = \kuramotoj_{-} - \kuramotoj_{-}^{\text{EP}}$ of the non-reciprocal part $\kuramotoj_{-}$ of the coupling between the communities from its value $\kuramotoj_{-}^{\text{EP}}$ at the exceptional point.
  The system exhibits a region where two possible steady-states with different properties coexist (the two steady-states are the continuation of the clockwise and counterclockwise chiral phases present in the PT-symmetric case $\Delta \omega = 0$).
  This region (red triangle) starts at the EP (red point) and its size increases with the amount of non-reciprocity (here $\kuramotoj_{-}^{\text{EP}} \simeq \num{0.2915} > 0$).
  The system exhibits hysteresis in the coexistence region (red points). 
  In (b-e), we show slices at fixed $\delta \kuramotoj_{-}$ (marked by blue dotted lines in (a).
  After the EP, there is hysteresis/first-order (discontinuous) behavior.
  In (d), the hysteresis curve bends outwards near the transition. This is due to the oscillation of the norm of the order parameter (that we refer to swap or periodic synchronization elsewhere) for large enough $\delta \kuramotoj_{-}$. This additional complication does not occur for moderate values of $\delta \kuramotoj_{-}$, such as in (c).
  The solution of the dynamical system Eq.~\eqref{kuramoto_ott_antonsen} were computed along lines at fixed $\delta \kuramotoj_{-}$, starting at large $|\delta \omega|$ (in a region without phase coexistence) from a random initial condition.
  The solution (after convergence) was used as an initial value for the next point in the line with fixed $\delta \kuramotoj_{-}$. This procedure was carried out two times, starting from positive and negative large $|\delta \omega|$.
  We have set $\kuramotoj_{+} = \num{0.08}$, $\kuramotoj_{AA} = \kuramotoj_{BB} = \num{1}$, $\Delta_{A} = \Delta_{B} = \num{0.25}$, $\omega_{A} = - \omega_{B} = \Delta\omega/2$. 
  }
\end{figure*}

To analyze the situation with explicitly broken PT symmetry (class II), we introduce a finite detuning $\Delta \omega = \omega_A - \omega_B$ between the natural frequencies of the two communities (we keep $\omega_A + \omega_B = 0$ for simplicity). The numerical simulation of Eq.\eqref{kuramoto_ott_antonsen} show that there are regions of the phase diagram in which two states (clockwise and counterclockwise) coexist, as well as regions in which a single state is present. This can be understood as the result between the spontaneous PT-symmetry breaking at $\Delta \omega = 0$ (in which the two states are equivalent, and mapped to each other by PT symmetry) and the detuning that explicitly breaks PT symmetry. At the boundary between these regions, the properties of the steady-states (such as their frequency $\Omega_{\text{ss}} \equiv \Omega$) change in a discontinuous way (like in a first-order phase transition). This is illustrated in Extended Data Fig.~\ref{si_figure_hysteresis}. In Extended Data Fig.~\ref{si_figure_hysteresis}a, we show the manifold of stable steady-states obtained from numerical simulations, which is a truncated version of the Riemann surface of the square root characteristic of exceptional points. There is coexistence between two states (blue and red dots) in the red region in parameter space. 
In Extended Data Fig.~\ref{si_figure_hysteresis}b, we show hysteresis curved corresponding to slices of the manifold represented Extended Data Fig.~\ref{si_figure_hysteresis}a.

\medskip 

This behavior shares some features with the dynamical encircling of an exceptional point in a linear system~\cite{Doppler2016,Dembowski2004,Milburn2015,Mailybaev2005}.
However, a crucial difference is that here, we are dealing with the steady-state (i.e., many-body phase) of the system, which is possible only because of the non-linearity (similar situations occur in Refs.~\cite{Hanai2019,Galda2016,Galda2019,Kepesidis2016,Graefe2010,Cartarius2008,Guthrlein2013,Strack2013}).
In addition, the breakdown of the adiabatic theorem plays a crucial role in the situations analyzed in Refs.~\cite{Doppler2016,Dembowski2004,Milburn2015,Mailybaev2005}, but it is not the case in the first-order-like transitions and hysteretic behavior described here.
In particular, the hysteresis observed in Extended Data Fig.~\ref{si_figure_hysteresis} does not depend on the speed at which the parameters are changed ($\Delta \omega$ in Extended Data Fig.~\ref{si_figure_hysteresis}b), provided that the change is slow enough (so that the system is always in a steady-state). 
The hysteresis curve is then independent of the arbitrarily small rate of change. 
This is in sharp contrast with the situations analyzed in Refs.~\cite{Doppler2016,Dembowski2004,Milburn2015,Mailybaev2005}, ruled by a linear dynamical system, in which the most unstable state (i.e., the one with the largest positive growth rate) always eventually dominates given enough time: in this situation, there is no hysteresis in the limit of an arbitrarily small rate of change.

\begin{table}
\centering 
\begin{tabular}{ccc} 
  \toprule
  {\bf flocking} & {\bf synchronization}  \\
  \midrule
  active agents & oscillators \\
  typically short-range & typically all-to-all \\
  external torque & natural frequency \\
  self-propulsion & n/a \\
  \midrule
  disordered & incoherent \\
  flocking & coherent \\
  antiflocking & $\pi$-state  \\
  chiral & traveling wave state (TW) \\
  swap & periodic synchronization (PS) \\
  swap+chiral & PS+TW \\
  \bottomrule \end{tabular}
\caption{\label{table_translation}\strong{A flocking-synchronization Rosetta stone.}
We compare the ingredients in the generalized Vicsek (flocking) and Kuramoto (synchronization) models, as well as some of their states.
For synchronization, we mostly followed the nomenclature of Refs.~\cite{Acebron2005,Hong2011,Hong2011b,Hong2012,Hong2014,Kemeth2016,Choe2016}.
The literature is not entirely consistent in the choice of names and definitions of the states; our definitions are summarized in Table~\ref{table_definition_phases}.
}
\end{table}

\begin{table}
\centering 
\begin{tabular}{ccc} 
  \toprule
  {\bf state} & $r_a$ & $\phi_B - \phi_A$ \\
  \midrule
incoherent & $0$ & n/a \\
coherent & constant $\neq 0$ & $0$ \\
$\pi$-state & constant $\neq 0$ & $\pi$ \\
traveling wave state & constant $\neq 0$ & constant $\neq 0,\pi$ \\
periodic synchronization & time-dependent & constant \\
PS+TW & time-dependent & time-dependent \\
  \bottomrule \end{tabular}
\caption{\label{table_definition_phases}\strong{Definition of the states.}
The complex order parameters $z_a = r_a \ee^{\ii \phi}$ are decomposed in amplitude $r_a = |z_a|$ and phase $\ee^{\ii \phi_a} = z_a/|z_a|$.
The label \enquote{PS+TW} corresponds to \enquote{periodic synchronization + traveling wave}.
Here, we make the following choices: (i) the system is observed in the lab frame (not in the rotating frame); (ii) we do not distinguish fully coherent states ($r=1$) from partially coherent ones ($0 \neq r < 1$), both are called \enquote{coherent}. 
}
\end{table}

\methodsection{Non-reciprocal pattern-forming instabilities}
\label{non_reciprocal_pattern_formation}

In this section, we apply our general strategy to pattern-forming instabilities within the formalism of amplitude equations~\cite{Cross1993,Saarloos1994,Aranson2002,Hoyle2006,Cross2009,Meron2015}. 
These describe a variety of physical systems ranging from fluid convection and lasers to ecological and chemical reaction-diffusion systems.

To clear any misunderstanding, let us warn the reader: this section is \emph{not} about the exceptional-point enforced pattern formation in Fig.~\ref{figure_stability}!
Instead, we consider a toy model of pattern formation with two fields that are coupled in a non-reciprocal way.
Here, the pattern formation is the spontaneous symmetry breaking (the Euclidean group $E(d)$ of isometries of space is broken by the appearance of the pattern).

In addition to the formalism of amplitude equations which allows for a direct parallel with the discussion in the main text, we perform direct simulations of two coupled copies of the Swift-Hohenberg equation~\cite{Swift1977}, a simple model of pattern formation. 

We then review a slightly more complicated situation, in which a single field is present, but two Fourier modes with non-reciprocal couplings are relevant (the non-reciprocity occurs between the harmonics), in which patterns with spontaneously broken parity also occur~\cite{Malomed1984,Coullet1985b,Brachet1987,Douady1989,Coullet1989,Coullet1990,Fauve1991} (see also Refs.~\cite{Knobloch1995,Armbruster1988,Proctor1988,Dangelmayr1997}). 
This situation has several experimental realizations in directional solidification of liquid crystals \cite{Simon1988,Flesselles1991,Melo1990,Oswald1987}, directional solidification of lamellar eutectics~\cite{Faivre1989,Faivre1992,Kassner1990,Ginibre1997}, directional viscous fingering~\cite{Rabaud1990,Cummins1993,Pan1994,Bellon1998}, and in overflowing fountains \cite{Counillon1997,Brunet2001}.
We show that in this situation too, the transition is marked by an exceptional point where the Goldstone mode of the spontaneously broken translation symmetry coalesces with a damped mode.

Without any attempt at completeness, we also refer to Refs.~\cite{Knobloch1981,Cross1988,Cross1986,Coullet1983,Cross1988b,Brand1983,Brand1984,Guckenheimer1984,Moses1986,Walden1985,Coullet1985,Bensimon1989,Knobloch1990,Schopf1993} on binary convection and to Refs.~\cite{Bressloff2001,Bressloff2002,Cho2004,Schnabel2008,Butler2011,Curtu2004,Adini1997,Hensch2005} on the visual cortex, and to Refs.~\cite{Chossat1994,Riecke1992,Tennakoon1996,Mutabazi1995,Bot1998,Wiener1992,Andereck1986,Altmeyer2010,Pinter2006} on Taylor-Couette/Dean flows.

\subsection{Coupled amplitude equations}

Let us first consider the one-dimensional Ginzburg-Landau/amplitude equation
\begin{equation}
	\label{simple_GL}
	\partial_t A = \epsilon A - g |A|^2 A + D \partial_x^2 A
\end{equation}
where $A$ is a complex amplitude.
This equation describes, for instance, rolls in Rayleigh-Bénard convection. 
The physical field $u$ (such as velocity or temperature) reads $u(t, x) = A(t,x) \ee^{\ii (q_{\text{c}} x - \omega t)} + \text{c.c.}$, where $q_{\text{c}}$ is the wavenumber of the convection rolls, and $A(t,x)$ is a slowly varying envelope.
The apparition of a pattern is marked by $A \neq 0$, and corresponds to the spontaneous breaking of translation symmetry.
The amplitude equation \eqref{simple_GL} satisfies translation symmetry by which $A \to A \ee^{\ii \phi}$, corresponding to a translation of the pattern by a distance $\phi/q_{\text{c}}$ in the $x$ direction; as well as inversion symmetry $x \to -x$ by which $A \to \overline{A}$ (overbar is complex conjugation). The reflection does not commute with the translations, so overall we do not have the direct product of these groups, but instead the semidirect product $U(1) \rtimes Z_{2} \simeq O(2)$. 
This symmetry prohibits terms such as $A^2$ in the right-hand side of Eq.~\eqref{simple_GL}, and guarantees that the coefficients are real.

Let us now introduce non-reciprocity: to do so, we consider two coupled amplitudes $A_1$ and $A_2$ (describing two different coupled fields), and write the most general equation of motion compatible with the symmetry, up to third order (like in Eq.~\eqref{simple_GL}). The only terms allowed are first order terms, as well as third order terms of the form $(\overline{A_b} A_c + \overline{A_c} A_b) A_d$, in both cases with real coefficients.
Hence, our amplitude equation reads
\begin{equation}
	\label{coupled_GL_O2}
	\partial_t A_a = \epsilon_{a b} A_b - g_{a b c d} (\overline{A_b} A_c + \overline{A_c} A_b) A_d + D_{a b} \partial_x^2 A_b
\end{equation}
where all the coefficients are real. In the following, we will focus on spatially uniform fields and ignore the diffusive term in Eq.~\eqref{coupled_GL_O2}.
In hindsight, we recognize Eq.~\eqref{general_O2_invariant} upon representing the complex amplitude $A_a$ as a two-dimensional $P_a = (\text{Re} A_a, \text{Im} A_a)$, owing to the fact that the symmetry groups are isomorphic.
We note, however, that the \emph{physical interpretation} of the symmetries are quite different in both case.
Having identified Eq.~\eqref{coupled_GL_O2} with Eq.~\eqref{general_O2_invariant} (in the uniform case), we can immediately predict that all the phases described in the main text should appear in the current context.
Our last task is then to provide a physical interpretation for each of them:
\begin{enumerate}[nosep,label=(\alph*)]
	\item disordered: there is no pattern, the amplitude decays to zero
	\item aligned: a pattern is present (and spontaneously breaks translational invariance, leading to a Goldstone mode often known as phase diffusion), here, the patterns for both fields are superimposed (they are in-phase)
	\item antialigned: same as flocking, except that the maxima of a field now coincide with the minima of the other (they are completely out-of-phase)
	\item chiral: the patterns move along $x$ (with a spontaneously chosen direction and at constant velocity), and they are partially out-of-phase (neither in-phase nor completely out-of-phase)
	\item swap: the amplitude of the patterns oscillates (usually not sinusoidally)
	\item chiral/swap: the patterns move along $x$ while their amplitudes fluctuate.
\end{enumerate}

\subsection{Coupled Swift-Hohenberg equations}

\begin{figure}
  \centering
  \includegraphics[width=\columnwidth]{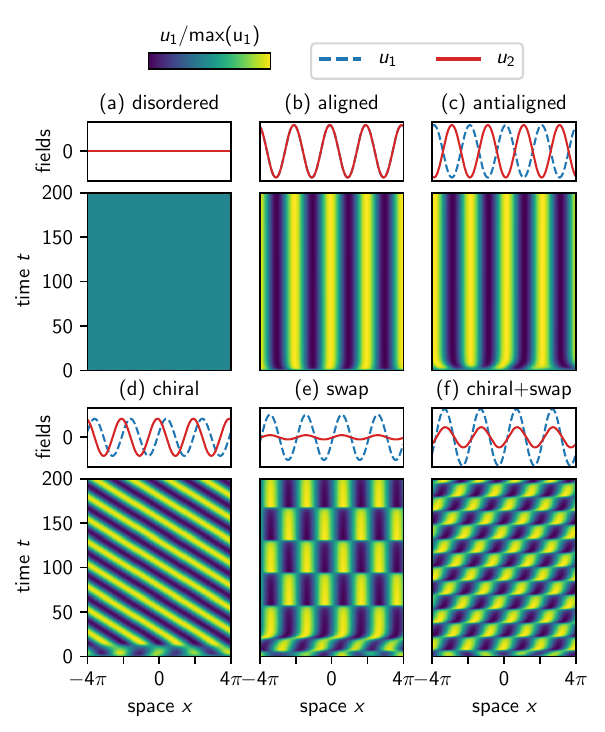}
  \caption{\label{figure_non_reciprocal_swift_hohenberg}\strong{Non-reciprocal pattern formation}
  We show a space-time density plot of the field $u_1(x,t)$ in different phases, as well as snapshots of the fields $u_1(x,t)$ and $u_2(x,t)$ at time $t=\num{200}$.
  We observe (a) a disordered phase where both field vanish, (b) an aligned phase where both patterns are static and in phase, (c) an antialigned phase where the patterns are static and completely out-of-phase, (d) a chiral phase where the patterns move at constant velocity, either to the left or to the right, and in which the fields have a finite phase difference, usually neither zero nor $\pi$, (e) a swap phase where the patterns essentially jump by a phase $\pi$ every period and (f) a mix of the chiral and swap behaviors (as in the chiral phase (d), there is a spontaneously broken symmetry between left and right movers).
  The fields are obtained by direct numerical simulation of the coupled Swift-Hohenberg equations on a one-dimensional domain of size $2 L$ with periodic boundary conditions, starting from random initial conditions. 
  The simulations are performed using the open-source pseudospectral solver Dedalus~\cite{Burns2020}.
  We have used $g=\num{0.25}$ in all cases.
  In (a) $r_{11}=r_{22}=\num{-0.5}$ and $r_{+}=r_{-}=\num{0.00}$.
  In the other cases (b-f), we have set $r_{11}=r_{22}=\num{0.5}$ and
  (b) $r_{+}=\num{0.50}$, $r_{-}=\num{0.00}$
  (c) $r_{+}=\num{-0.50}$, $r_{-}=\num{0.00}$
  (d) $r_{+}=\num{0.00}$, $r_{-}=\num{0.25}$
  (e) $r_{+}=\num{0.87}$, $r_{-}=\num{1.00}$
  (f) $r_{+}=\num{0.85}$, $r_{-}=\num{1.00}$.
  }
\end{figure}

To further support our claims and illustrate the phases described above, we consider two coupled Swift-Hohenberg equations~\eqref{coupled_swift_hohenberg_main}~\cite{Swift1977}
describing the dynamics of the real fields $u_a(t,x)$, with $a=1,2$ (we also define $r_{\pm} = [r_{1 2} \pm r_{2 1}]/2$).
An explicit version of the amplitude equations~\eqref{coupled_GL_O2} (obtained from symmetry considerations) could be derived from Eq.~\eqref{coupled_swift_hohenberg_main}, following e.g. Ref.~\cite{Saarloos1994}.
Instead, we solve Eq.~\eqref{coupled_swift_hohenberg_main} numerically on a one-dimensional domain of size $2 L$ with periodic boundary conditions using the open-source pseudospectral solver Dedalus~\cite{Burns2020}, starting from random initial conditions. 
The results confirm our predictions based on the coupled amplitude equations~\eqref{coupled_GL_O2}.
In Extended Data Fig.~\ref{figure_non_reciprocal_swift_hohenberg}, we show snapshots of the numerical results, in which all the phases described above appear.
In this case, Eq.~\eqref{coupled_swift_hohenberg_main} have the full Euclidean group $E(1)$ as a symmetry group, that is broken by pattern formation. 
(The $O(2)$ symmetry of Eq.~\eqref{coupled_GL_O2} pertains to the amplitude equation description, in which additional knowledge about the pattern is taken into account.)

\subsection{Discussion}

As we have emphasized in the introduction of this section, the pattern formation appears at different levels here compared to the main text.
Here, pattern formation (as a spontaneous breaking of the Euclidean symmetry) is our starting point; we couple two pattern-forming systems in a non-reciprocal way, and observe exceptional transitions as a consequence.
In particular, there is no convective term in the amplitude equation \eqref{simple_GL}. Only diffusive terms are present.
This is in contrast with the situation presented in Fig.~\ref{figure_stability} in the main text, 
where the interplay between convective terms and exceptional transitions is the origin of pattern-forming instabilities.
Besides, we emphasize that we did not assume non-reciprocal cross-diffusion, in contrast with Refs.~\cite{You2020,Saha2020} (another difference with these references is that we consider a non-conserved order parameter in the language of Ref.~\cite{Hohenberg1977}). Our analysis focuses on the mean-field transitions, and our conclusions remain valid as long as the growth rates are negative at finite $k$ (this is in particular the case where $D_{a b} = D \delta_{a b}$, so the growth rates are of the form $\sigma_i(k) = \sigma_i(0) - D k^2$; but this is especially not guaranteed when $D_{a b}$ is not symmetric).

We also mention that upon lifting the constraint put upon Eq.~\eqref{coupled_GL_O2} by reflection symmetry, one is left with an $U(1)$-equivariant system with explicitly broken PT symmetry, as Eq.~\eqref{coupled_GL_O2} becomes a complex Ginzburg-Landau equation, in which we expect the analysis of the section \methodsref{non_reciprocal_kuramoto_model}{Non-reciprocal Kuramoto model} to hold.

\subsection{Directional interface growth}

Following Refs.~\cite{Malomed1984,Douady1989,Coullet1989,Coullet1990,Fauve1991}, we now consider a single scalar field decomposed as
\begin{equation}
    u(t, x) = A_1(t,x) \ee^{\ii (q_{\text{c}} x - \omega t)} + A_2(t,x) \ee^{\ii (2 q_{\text{c}} x - \omega t)} + \text{c.c.}
\end{equation}
As in the previous case, the transition and reflection symmetry of the underlying system endows the amplitude equation with $O(2)$ symmetry. 
However, note that while $A_1$ transforms as $A_1 \to A_1 \ee^{\ii \theta}$ when the field $u$ is translated in space, $A_2$ transforms as $A_2 \to A_2 \ee^{2 \ii \theta}$.
This is a different representation of the $SO(2)$ group compared to the previous paragraph. (The $\ZZ_2$ part is unchanged, and still corresponds to $A_1 \to \overline{A_1}$ and $A_2 \to \overline{A_2}$.)
Because the representation is different, the general form of the amplitude is different, and reads~\cite{Malomed1984,Douady1989,Coullet1989,Coullet1990,Fauve1991}
\begin{equation}
\label{amplitude_equation_interface_growth}
\begin{split}
	\partial_t A_1 &= \mu_1 A_1 - \overline{A_1} A_2 - \alpha |A_1|^2 A_1 - \beta |A_2|^2 A_1 \\
	\partial_t A_2 &= \mu_2 A_2 + \epsilon A_1^2 - \gamma |A_1|^2 A_2 - \delta |A_2|^2 A_2
\end{split}
\end{equation}
The coefficients $\alpha$, $\beta$, $\gamma$ and $\delta$ are usually assumed to be positive to ensure stability, and the coefficient of $\overline{A_1} A_2$ is set to $-1$ by rescaling.
The non-reciprocity is then captured by the coefficient $\epsilon$ being positive, which is necessary for the apparition of traveling patterns~\cite{Douady1989}.
As the amplitudes $A_1 = r_1 \ee^{\ii \phi_1}$ and $A_2 = r_2 \ee^{\ii \phi_2}$ correspond to different Fourier components, the relevant phase difference between them is $\Delta \phi = 2 \phi_1 - \phi_2$.

\begin{figure}
  \centering
  \includegraphics[width=\columnwidth]{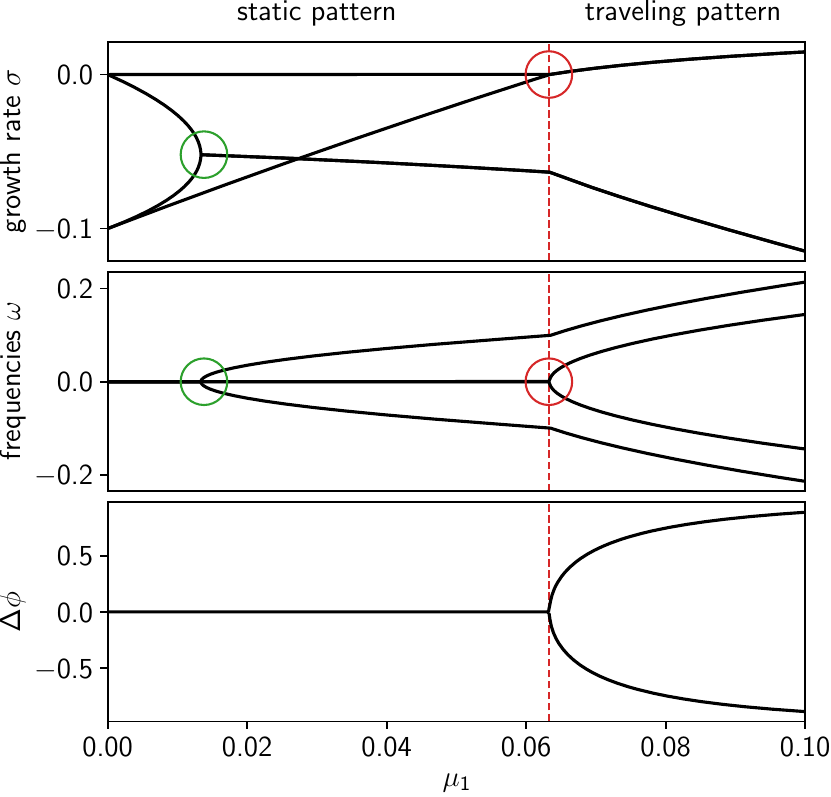}
  \caption{\label{linear_operator_spectrum_fingering}\strong{Exceptional point in directional interface growth.}
  The spectrum of the Jacobian $L$ corresponding to Eq.~\eqref{amplitude_equation_interface_growth} exhibits an exceptional point at the transition between static patterns and traveling patterns with spontaneous parity breaking (i.e., the patterns travel with equal probability to the left or to the right).
  The two most unstable eigenvalues $\lambda_i = \sigma_i + \ii \omega_i$ of $L$ coalesce at $\mu_1 \approx \num{0.064}$ (red circle).
  This value coincides with the transition from a constant solution to traveling waves (TW), marked by a red dashed line.
  The coalescence occurs at $\lambda = 0$ (not at finite frequency and/or growth rate), and the corresponding eigenvectors become collinear.
  Note that another exceptional point occurs near $\mu_1 \approx \num{0.014}$ (green circle), but with a strictly negative growth rate: this does not correspond to a bifurcation.
  We also show the dephasing $\Delta \phi = 2 \phi_1 - \phi_2$ between the amplitudes, which undergoes a pitchfork bifurcation; the direction of motion of the pattern is set by $\Delta \phi$.
  We have set $\alpha = \beta = \gamma = \delta = \num{1}$, $\epsilon = \num{+1}$ and $\mu_2 = \num{-0.1}$.
  }
\end{figure}

In Extended Data Fig.~\ref{linear_operator_spectrum_fingering}, we show the spectrum of the operator $L$ obtained by linearizing \eqref{amplitude_equation_interface_growth} around its steady-state (see Eq.~\eqref{jacobian_complex} for the definition of $L$, with the replacement $z_a \to A_a$). At the transition between a static solution (representing a static pattern) and a traveling wave solution (representing a traveling pattern), we observe the coalescence of the Goldstone mode with a damped mode at an exceptional point (red circle in the figure).
We also note that the presence of exceptional points away from zero in the spectrum of $L$ does not mark a phase transition (bifurcation).

\methodsection{Generalized PT symmetry and dynamical systems}
\label{generalized_PT_symmetry}

In this section, we show how the transition from the aligned (or antialigned) phase to the chiral phase can be seen as a spontaneous PT-symmetry breaking.

Let us first review (generalized) PT-symmetry in the context of linear operators. 
We refer the reader to Ref.~\cite{Mostafazadeh2002,Mostafazadeh2002b,Mostafazadeh2002c,Bender2002,Bender2010,Mostafazadeh2015} for details and proofs, and for a translation of the same concepts in the language of pseudo-Hermitian operators.
Consider an antiunitary operator $X$ (i.e., there is a linear operator $M_X$ such that $X \psi = M_X \overline{\psi}$) such as $X^2 = \Id$ (where $\Id$ is the identity matrix).
The operator $X$ can be seen as a generalization of the product $PT$ of parity $P$ and time-reversal $T$ (as it has the same properties), and is therefore called a (generalized) PT symmetry.
This generalized PT symmetry might have nothing to do with the physical parity and time-reversal operations: only the relevant algebraic structure is kept.
As $X^2 = \Id$, it is always possible to find a basis in which the antiunitary $X$ is represented by complex conjugation alone, by a reduction to Wigner normal form~\cite{Weigert2003,Wigner1960}.

Following Ref.~\cite{Mostafazadeh2015}, we say that a complex square matrix $H$ is $X$-symmetric when $[H,X] = 0$. 
The $X$-symmetry is called exact (or unbroken) when in addition, there is a complete set of eigenvectors $\psi_n$ of $H$ satisfying $X \psi_n = \psi_n$. Else, it is called inexact (or spontaneously broken).
It can be shown that the spectrum of $H$ is real if and only if there is an antiunitary $X$ such as $H$ has an exact $X$-symmetry.
In contrast with the case where PT symmetry is spontaneously broken ($H$ has an inexact $X$-symmetry), we say that PT symmetry is explicitly broken when $H$ is \emph{not} $X$-symmetric.

In a non-linear dynamical system $\partial_t \psi = f(\psi)$, the condition of PT-symmetry reads $X f(\psi) = f(X \psi)$, see e.g. Ref.~\cite{Konotop2016}.

Consider a harmonic solution of the form $\psi(t) = \ee^{s t} \psi_0$ where $s = \sigma + \ii \Omega$ is a complex growth rate (we are mostly interested in the case where $s = \ii \Omega$ is purely imaginary).
It satisfies $\partial_t \psi(t) = s \psi(t)$ (whether the system is PT-symmetric or not; this crucial property hinges on the fact that the motion is harmonic, and that (at least the relevant part of) the order parameter is written as a complex number, see Refs.~\cite{Golubitsky1985,Marques2013,Haragus2011}).
Further assuming that $f$ is $U(1)$-equivariant (with $f(\ee^{\ii \theta} \psi) = \ee^{\ii \theta} f(\psi)$), we find $f(\psi(t)) = \ee^{s t} f(\psi_0)$, leading to $s \psi_0 = f(\psi_0)$.
Using PT-symmetry, we deduce $\overline{s} X \psi_0 = X f(\psi_0) = f(X \psi_0)$. 
Hence, $X \psi_0 = \psi_0$ implies that $s$ is real.
Conversely, given a solution $\psi_0$ with a complex growth rate $s \not\in \RR$ that is not purely real, we find that $X \psi_0$ is another solution with complex growth rate $\overline{s}$.
(When $f(\psi) = H(\psi) \psi$ where $H(\ee^{\ii \theta} \psi) = H(\psi)$ is $U(1)$-invariant, the properties above can be reformulated in terms of the non-linear eigenvalue problem $s \psi_0 = H(\psi_0) \psi_0$, in which the analogy with standard PT-symmetry is more obvious. This is however not required.)

The procedure described above can be carried out for the systems in the main text, provided that we consider the complex order parameters $z_{a} = v_a^x + \ii v_a^y$. 
In this representation, a $SO(2) \simeq U(1)$ rotation corresponds to the multiplication by a phase $z_a \to \ee^{\ii \theta} z_a$ while mirror symmetry corresponds to complex conjugation $z_a \to \overline{z_a}$.
Hence, systems in both class I and II correspond to $U(1)$-equivariant dynamical systems.
Besides, systems in class I are PT-symmetric (with $X$ being complex conjugation), while systems in class II are not PT-symmetric (the parameters in $\alpha^*$ and $\beta^*$ explicitly break PT-symmetry).
Note that here, the physical parity (mirror symmetry) corresponds to the PT-symmetry.

The same strategy should allow for multiple frequencies, provided that they correspond to independent components of the order parameter, by generalizing the equation $f(\psi_0) = s \psi_0$ to $f(\psi_0) = \text{diag}(s_1, s_2, \dots) \psi_0$.
As an example, consider the case of directional interface growth presented in the section \methodsref{non_reciprocal_pattern_formation}{Non-reciprocal pattern-forming instabilities}.
The two complex amplitudes $A_1$ and $A_2$ do not transform in the same way under $U(1)$ (because they correspond to different wavevectors). 
Hence, harmonic solutions now satisfy $f(\psi_0) = \text{diag}(s_1, s_2)  \psi_0$ where $\psi_0 = (A_1, A_2)$ and $s_a = \sigma_a + \ii \Omega_a$ are the corresponding complex growth rates.
With this change, the parity-breaking transition from static patterns to traveling patterns can again be seen as a spontaneous PT-symmetry breaking.
Similarly, in the $O(3)$-symmetric system discussed in the SI Sec. VIII, one would need to decompose the 3D vectors $\vec{v}_a$ into in-plane and out-of-plane components.

\methodsection{List of supplementary movies}
\label{list_of_SI_movies}

\begin{enumerate}[wide,label={$\bullet$ SI Movie~\arabic*:},ref=\arabic*]
    \item \label{movie_vicsek} See Methods section \methodsref{molecular_dynamics_vicsek}{Molecular dynamics simulations of the Vicsek model}.
    \item \label{movie_order_parameter_phases} Evolution of the order parameter in the time-dependent phases (chiral, swap, chiral+swap) computed from the dynamical system Eq.~\eqref{general_SO2_invariant_main}.
    \item \label{movie_robots} See Methods section \methodsref{experimental_demonstration}{Demonstration with programmable robots}.
    \item \label{movie_patterns} This shows the pattern formation at fixed density when the incompressibility constraint is not enforced. See Methods section \methodsref{simulations_continuum_equations}{Simulations of the continuum equations in exceptional point induced pattern formation}.
    \item \label{movie_patterns_incompressible} This shows the pattern formation with the incompressibility constraint enforced. See SI Sec.~\siref{incompressibility_constraint}{XI}.
\end{enumerate}
 
\setcounter{section}{0}

\clearpage

\onecolumngrid
\renewenvironment{widetext}{}{}

\begin{center}{\bf SUPPLEMENTARY INFORMATION}\end{center} 
\setcounter{equation}{0}
\setcounter{figure}{0}
\renewcommand{\thefigure}{S\arabic{figure}}
\renewcommand{\theequation}{S\arabic{equation}}
\renewcommand{\theHfigure}{Supplement.\thefigure}
\renewcommand{\theHequation}{Supplement.\theequation}

In this Supplemental Information (SI), we provide details of the various analyses performed in this work. It is organized as follows: 
In Secs.~\ref{sec_hydro}--\ref{sec_excitation_supp}, we analyze a non-reciprocal flocking model. Section~\ref{sec_hydro} (page~\pageref{sec_hydro}) provides details of the coarse-graining procedure of a microsopic model of non-reciprocally interacting self-propelled particles (Eq. (2) in the main text), to derive hydrodynamic equations.
Sections~\ref{sec_mean_field} (page~\pageref{sec_mean_field}) and~\ref{sec_excitation_supp} (page~\pageref{sec_excitation_supp}) analyze the mean-field theory of the derived hydrodynamics and their excitation spectrum. Key features such as an exceptional transition from a (anti)flocking to time-dependent chiral phase, as well as the exceptional-point enforced pattern formation, are explicitly shown to arise analytically. 

In Sec.~\ref{sec_detailed_balance} (page~\pageref{sec_detailed_balance}), we point out that the non-reciprocity directly implies the breaking of detailed balance. 

In Sec.~\ref{two_agents} (page~\pageref{two_agents}), we show analytically that the time-dependent chiral phase is absent for the case of two (partially) non-reciprocal agents, hence demonstrating that the many-body character of non-reciprocal matter is a key ingredient of time-dependent phases.

In Sec.~\ref{sec_non_reciprocal_kuramoto} (page~\pageref{sec_non_reciprocal_kuramoto}), we provide details of non-reciprocal synchronization. We show how the mean-field dynamics performed in Methods reproduces direct simulation of microscopic model.
We give numerical evidences that the interplay between noise (or disorder) and many-body interaction are responsible for the emergence of time-dependent ordered phases, reminiscent of an ordered-to-disordered transition in frustrated many-body systems. 

In Sec.~\ref{arrhenius} (page~\pageref{arrhenius}), we show that noise can destroy the chiral phase by randomly flipping the chirality over time, but that this process is exponentially suppressed by many-body effects when the number of agents increases.

In Sec.~\ref{sec_O3} (page~\pageref{sec_O3}), we perform an analysis on a non-reciprocal O(3)-symmetric system, providing another example of an exceptional transition. 

In Sec.~\ref{lasers} (page~\pageref{lasers}), we discuss an example of single-body exceptional transition in laser physics and its destabilisation by the noise.

In Sec.~\ref{sec_multiple_populations} (page~\pageref{sec_multiple_populations}), we discuss systems with multiple populations.

Finally, in Sec.~\ref{incompressibility_constraint} (page~\pageref{incompressibility_constraint}), we analyze the effect of the incompressibility condition on pattern formation in the non-linear dynamics of the fluid.

\section{Microscopic and hydrodynamic descriptions of non-reciprocal multi-components active fluids}
\label{sec_hydro}

In this section, we describe a microscopic model of active self-propelled particles inspired by the Vicsek model~\cite{Vicsek1995}, in which several populations of aligning self-propelled particles interact. The coupling between individuals belonging to different populations is not necessarily reciprocal. This model is defined by Eq.~\eqref{eom_discrete}.
Using the methods described in references~\cite{Dean1996,Bertin2006,Bertin2009,Farrell2012,Yllanes2017} and summarized in the reviews \cite{Marchetti2013,Chate2019}, we perform a coarse-graining of this microscopic model to obtain a set of hydrodynamic equations generalizing the Toner-Tu equations \cite{Toner1995,Toner1998}, which is the basis of the analysis in the main text.
The main results are Eq.~\eqref{density_eom_si} and Eq.~\eqref{eq_less_hideous}, which are respectively the hydrodynamic equations for the densities of the active particles and for their polarization fields.
These equations describe an arbitrary number of species. 
In the main text, we focus on the case of two species described by Eq.~\eqref{eom_with_gradients_two_populations_si}.
In this SI, we denote the polarization field $\vec P^a$, which is called $\vec v_a$ in the main text. 

\medskip

Several methods of deriving continuum hydrodynamic equations from microscopics have been applied to active matter, going from (i) approaches based on the Fokker–Planck (Smoluchowski) equation for the hydrodynamic variables~\cite{Dean1996,Farrell2012,Marchetti2013}, to (ii) kinetic theory approaches based on the Boltzmann equation~\cite{Bertin2006,Bertin2009,Peshkov2014}, or (iii) directly from the Chapman-Kolmogorov equation~\cite{Ihle2011} (in increasing order of complexity).
Although coarse-graining microscopic models provides invaluable qualitative insights on the behaviour of the system, even current state-of-the-art coarse-graining procedures only provide a qualitative agreement, at best semi-quantitative, with the microscopic starting point~\cite{Chate2019,Mahault2019}. With this in mind, we use the easiest coarse-graining method (i) along with several simplifying approximations (see section~\ref{sec coarse-graining}). This procedure has the benefit of simplicity and allows to highlight the key features of a non-reciprocal multi-component fluid.
However, the correspondence between the resulting hydrodynamic equations and the microscopic model is only qualitative, in the sense that the values of the coefficients might be approximate.

\medskip

\subsection{Microscopic particle-based model}
\label{sec_microscopic_model}

Let us consider $N_{\text{pop}} $ populations $a = 1, \dots, N_{\text {pop}}$ of $N_{a}$ active particles moving in a plane.
Each particle is described by a position $r_i^{a}$ and an angle $\theta_i^a$, with $i = 1, \dots, N_{a}$.
The dynamics of the population is described by the set of equations
\begin{subequations}
\label{eom_discrete}
\begin{align}
	\dot{r}_i^a(t) &= v_0^a \hat{n}[\theta_i^a(t)] \label{eom_discrete_r} \\
	\dot{\theta}_i^a(t) &= \eta_i^a(t) + \sum_{b} \sum_{j=1}^{N_b} J_{i j}^{a b} \sin[ \theta_j^b(t) - \theta_i^a(t) \label{eom_discrete_theta} ]
\end{align}
\end{subequations}
where we have defined
\begin{equation}
	\hat{n}(\theta) = \begin{pmatrix}
		\cos(\theta) \\ \sin(\theta)
	\end{pmatrix}
\end{equation}
and where $\eta_i^a(t)$ are Gaussian white noises with $\braket{\eta_i^a(t)} = 0$ and
\begin{equation}
	\braket{\eta_{i}^a(t) \eta_{j}^b(t')} = 2 \eta \delta_{i j} \delta^{a,b} \delta(t-t').
\end{equation}
(This means that $\eta_{i}^a(t)$ scales as $\sqrt{\eta}$, contrary to what the notation might suggest.)
We set
\begin{equation}
	J_{i j}^{a b} = J^{a b} H(R_0 - \lVert r - r' \rVert).
\end{equation}
where $H$ is the Heaviside step function.
In the derivation of the hydrodynamic model, we will simplify the analysis by replacing the Heaviside step functions by Dirac distributions.

\subsection{Coarse-graining of the microscopic model to hydrodynamic equations}
\label{sec coarse-graining}

It will be convenient to write the equations of motion of section~\ref{sec_microscopic_model} in a slightly more general form as
\begin{subequations}
\label{eom_general}
\begin{align}
	\dot{r}_i^a &= A^a_{r}(r_i^a, \theta_i^a) + \sum_b \sum_{j=1}^{N_b} B^{a b}_{r}(r_i^a, \theta_i^a, r_j^b, \theta_j^b) \\
	\dot{\theta}_i^a &= A^a_{\theta}(r_i^a, \theta_i^a) + \sum_b \sum_{j=1}^{N_b} B^{a b}_{\theta}(r_i^a, \theta_i^a, r_j^b, \theta_j^b) + \eta_i^a(t).
\end{align}
\end{subequations}

For our equations \eqref{eom_discrete}, we have
\begin{equation}
	\label{A_for_vicsek_eom}
	A^a_{r}(r,\theta) = v_0^a n(\theta)
	\qquad
	\text{and}
	\qquad
	A^a_{\theta}(r,\theta) = 0
\end{equation}
while
\begin{equation}
	\label{B_for_vicsek_eom}
	B^{a b}_{r}(r,\theta,r',\theta') = 0
	\qquad
	\text{and}
	\qquad
	B^{a b}_{\theta}(r,\theta,r',\theta') =  J^{a b} H(R_0 - \lVert r_i - r_j \rVert)  \sin(\theta' - \theta).
\end{equation}

In order to obtain hydrodynamic equations, we first define the (stochastic) single-particle distributions
\begin{equation}
	c^{a}(r, \theta, t) = \frac{1}{N_a} \sum_{i = 1}^{N_a} \delta(r - r_i^a(t)) \delta(\theta - \theta_i^a(t)).
\end{equation}
They are the sum of the individual densities
\begin{equation}
	c_i^a(r, \theta, t) = \delta(r - r_i^a(t)) \; \delta(\theta - \theta_i^a(t)).
\end{equation}

We follow the procedure of Ref.~\cite{Dean1996} to obtain a Langevin equation for this quantity. 
To do so, let us first compute the time derivative of the individual densities
\begin{equation}
	\frac{\partial}{\partial t}\left[ c_i^a(r, \theta, t)  \right] = \frac{\partial}{\partial t}\left[\delta(r - r_i^a(t)) \right]  \delta(\theta - \theta_i^a(t)) + \delta(r - r_i^a(t)) \frac{\partial}{\partial t}\left[ \delta(\theta - \theta_i^a(t)) \right]
\end{equation}
so using Itô lemma,
\begin{equation}
\begin{split}
	\frac{\partial}{\partial t}\left[ c_i^a(r, \theta, t)  \right] = 
	&[- (\nabla_r \delta)(r - r_i^a(t)) \cdot \dot{r}_i^a(t)] \; \delta(\theta - \theta_i^a(t)) \\
	+ \; &\delta(r - r_i^a(t)) \; [- (\nabla_\theta \delta)(\theta - \theta_i^a(t)) \cdot \dot{\theta}_i^a(t) + \eta (\nabla_\theta^2 \delta)(\theta - \theta_i^a(t))].
\end{split}
\end{equation}
There is no diffusive term in the position equation, because there is no noise in the corresponding equation of motion~\eqref{eom_discrete_r}.

Let us now consider an arbitrary function $(r, \theta) \mapsto f(r, \theta)$. 
We indeed have
\begin{equation}
	f(r_i^a(t), \theta_i^a(t)) = \int \dd r  \dd \theta \, c_i^a(r, \theta, t) f(r, \theta).
\end{equation}
Hence, 
\begin{equation}
	\label{derivative_f}
	\frac{\dd }{\dd t}\left[ f(r_i^a(t), \theta_i^a(t)) \right] = \int \dd r  \dd \theta \, \frac{\partial}{\partial t}\left[ c_i^a(r, \theta, t)  \right] f(r, \theta)
\end{equation}
so we also have
\begin{equation}
\begin{split}
	\frac{\dd }{\dd t}\left[ f(r_i^a(t), \theta_i^a(t)) \right] = \int \dd r  \dd \theta \, &[- (\nabla_r \delta)(r - r_i^a(t)) \cdot \dot{r}_i^a(t)] \; \delta(\theta - \theta_i^a(t)) \\
	+ \; &\delta(r - r_i^a(t)) \; [- (\nabla_\theta \delta)(\theta - \theta_i^a(t)) \cdot \dot{\theta}_i^a(t) + \eta (\nabla_\theta^2 \delta)(\theta - \theta_i^a(t))] f(r, \theta).
\end{split}
\end{equation}
By integration by part and replacing the Dirac distributions with $c_i^a(r, \theta, t)$, we obtain
\begin{equation}
	\frac{\dd }{\dd t}\left[ f(r_i^a(t), \theta_i^a(t)) \right] =  \int \dd r  \dd \theta \, \left[(\nabla_r f) \cdot \dot{r}_i^a(t) + (\nabla_\theta f)(r, \theta) \cdot \dot{\theta}_i^a(t) + \eta (\nabla_\theta^2 f)(r, \theta) \right] c_i^a(r, \theta, t).
\end{equation}

Replacing the time derivatives with their values given by the equations of motion \eqref{eom_general} yields
\begin{equation}
\begin{split}
	\frac{\dd }{\dd t}\left[ f(r_i^a(t), \theta_i^a(t)) \right] = \int \dd r  \dd \theta \, (\nabla_r f) \cdot \left(
	 	A^a_{r}(r_i^a, \theta_i^a) + \sum_b \sum_{j=1}^{N_b} B^{a b}_{r}(r_i^a, \theta_i^a, r_j^b, \theta_j^b)
	\right) c_i^a(r, \theta, t) \\
	+ (\nabla_\theta f)(r, \theta) \cdot \left( 
		A^a_{\theta}(r_i^a, \theta_i^a) + \sum_b \sum_{j=1}^{N_b} B^{a b}_{\theta}(r_i^a, \theta_i^a, r_j^b, \theta_j^b) + \eta_i^a(t)
	\right) c_i^a(r, \theta, t) \\
	+ \eta (\nabla_\theta^2 f)(r, \theta)  c_i^a(r, \theta, t).
\end{split}
\end{equation}
After integration by parts,
\begin{equation}
\begin{split}
	\frac{\dd }{\dd t}\left[ f(r_i^a(t), \theta_i^a(t)) \right] = 
	\int \dd r  \dd \theta \, f(r, \theta) \Bigg(
		- \nabla_r \cdot \left[ \left( 
			A^a_{r}(r_i^a, \theta_i^a) + \sum_b \sum_{j=1}^{N_b} B^{a b}_{r}(r_i^a, \theta_i^a, r_j^b, \theta_j^b)
		\right) c_i^a(r, \theta, t) \right] \\
		- \nabla_\theta \cdot \left[ \left( 
			A^a_{\theta}(r_i^a, \theta_i^a) + \sum_b \sum_{j=1}^{N_b} B^{a b}_{\theta}(r_i^a, \theta_i^a, r_j^b, \theta_j^b)
		\right) c_i^a(r, \theta, t) \right] \\
		- \nabla_\theta \cdot \left[ \eta_i^a(t)c_i^a(r, \theta, t) \right] 
		+  \eta \nabla_\theta^2 c_i^a(r, \theta, t) 
	\Bigg).
\end{split}
\end{equation}
Comparing with \eqref{derivative_f}, we obtain
\begin{equation}
\begin{split}
	\frac{\partial}{\partial t}\left[ c_i^a(r, \theta, t)  \right] = 
	- \nabla_r \cdot \left[ \left( 
			A^a_{r}(r_i^a, \theta_i^a) + \sum_b \sum_{j=1}^{N_b} B^{a b}_{r}(r_i^a, \theta_i^a, r_j^b, \theta_j^b)
		\right) c_i^a(r, \theta, t) \right] \\
		- \nabla_\theta \cdot \left[ \left( 
			A^a_{\theta}(r_i^a, \theta_i^a) + \sum_b \sum_{j=1}^{N_b} B^{a b}_{\theta}(r_i^a, \theta_i^a, r_j^b, \theta_j^b)
		\right) c_i^a(r, \theta, t) \right] \\
		- \nabla_\theta \cdot \left[ \eta_i^a(t)c_i^a(r, \theta, t) \right] 
		+  \eta \nabla_\theta^2 c_i^a(r, \theta, t) 
\end{split}
\end{equation}

Summing over $i=1,\dots,N_a$ and replacing $r_i^a \to r$ and $\theta_i^a \to \theta$ as allowed by the Dirac distributions gives 
\begin{equation}
\begin{split}
	\frac{\partial}{\partial t}\left[ c^a(r, \theta, t)  \right] = 
	- \nabla_r \cdot \left[ \left( 
			A^a_{r}(r, \theta) + \sum_b \sum_{j=1}^{N_b} B^{a b}_{r}(r, \theta, r_j^b, \theta_j^b)
		\right) c^a(r, \theta, t) \right] \\
		- \nabla_\theta \cdot \left[ \left( 
			A^a_{\theta}(r, \theta) + \sum_b \sum_{j=1}^{N_b} B^{a b}_{\theta}(r, \theta, r_j^b, \theta_j^b)
		\right) c^a(r, \theta, t) \right] \\
		- \nabla_\theta \cdot \left[ \sum_{i} \eta_i^a(t) c_i^a(r, \theta, t) \right] 
		+  \eta \nabla_\theta^2 c^a(r, \theta, t) 
\end{split}
\end{equation}

Using again that for an arbitrary function $f$
\begin{equation}
	\int \dd r' \dd \theta' f(r', \theta', z) c_j^b(r', \theta') = f(r_j^b, \theta_j^b, z)
\end{equation}
\begin{equation}
\begin{split}
	\frac{\partial}{\partial t}\left[ c^a(r, \theta, t)  \right] = 
	- \nabla_r      \cdot \left[  \left( A^a_{r}(r, \theta)  + \sum_b \int  \dd r' \dd \theta' B^{a b}_{r}(r, \theta, r', \theta') c^b(r', \theta', t) \right) c^a(r, \theta, t) \right] \\
	- \nabla_\theta \cdot \left[  \left( A^a_{\theta}(r, \theta) + \sum_b \int \dd r' \dd \theta'  B^{a b}_{\theta}(r, \theta, r', \theta')   c^b(r', \theta', t)  \right) c^a(r, \theta, t) \right] \\
	- \nabla_\theta \cdot \left[\sum_{i} \eta_i^a(t)c_i^a(r, \theta, t) \right] +  \eta \nabla_\theta^2 c^a(r, \theta, t).
\end{split}
\end{equation}

The random contribution can then be handled to obtain a Markovian stochastic equation of motion following Ref.~\cite{Dean1996}.
This derivation suggests that noise $\eta(t)$ entering the equation of motion $\frac{\partial}{\partial t} \left[ c^a(r, \theta, t)  \right] = \text{(deterministic part)} + \eta(t)$ 
(i) is multiplicative in the density and (ii) has a correlation function of the form $\braket{\eta(t,r) \eta(0,0)} \propto \delta(t) \nabla^2 \delta(r)$, similar to fluids at thermal equilibrium (like model A in Ref.~\cite{Forster1976,Forster1977}).
In the Toner-Tu model~\cite{Toner1995,Toner1998,Toner2012}, this correlation function is usually assumed to be (i) not multiplicative in the density and (ii) with a correlation function of the form $\braket{\eta(t,r) \eta(0,0)} \propto \delta(t) \delta(r)$ (without Laplacian); in Ref.~\cite{Chen2016}, it is argued that this form is chosen because of the lack of linear momentum conservation. 
We refer to Refs.~\cite{Bertin2013,Marchetti2013,Mahault2019} for discussions.

Its noise-averaged version (where we use the same symbols for simplicity) is obtained by removing the noise and reads
\begin{equation}
\begin{split}
	\frac{\partial}{\partial t}\left[ c^a(r, \theta, t)  \right] = 
	- \nabla_r      \cdot \left[  \left( A^a_{r}(r, \theta)  + \sum_b \int  \dd r' \dd \theta' B^{a b}_{r}(r, \theta, r', \theta') c^b(r', \theta', t) \right) c^a(r, \theta, t) \right] \\
	- \nabla_\theta \cdot \left[  \left( A^a_{\theta}(r, \theta) + \sum_b \int \dd r' \dd \theta'  B^{a b}_{\theta}(r, \theta, r', \theta')   c^b(r', \theta', t)  \right) c^a(r, \theta, t) \right] \\
	+  \eta \nabla_\theta^2 c^a(r, \theta, t).
\end{split}
\end{equation}

We now replace the $A$ and $B$'s with equations \eqref{A_for_vicsek_eom} and \eqref{B_for_vicsek_eom} to get  
\begin{equation}
\begin{split}
	\frac{\partial}{\partial t}\left[ c^a(r, \theta, t)  \right] = 
	&- \nabla_r      \cdot \left[  v_0^a n(\theta) c^a(r, \theta, t) \right] \\
	&- \nabla_\theta \cdot \left[  \sum_b \int \dd r' \dd \theta' J^{a b} H(R_0 - \lVert r - r' \rVert)  \sin(\theta' - \theta) c^a(r, \theta, t) c^b(r', \theta', t)  \right] \\
	&+  \eta \nabla_\theta^2 c^a(r, \theta, t).
\end{split}
\end{equation}
that can be reorganized as
\begin{equation}
\begin{split}
	(\partial_t + v_0^a n(\theta) \cdot \nabla_r) c^a(r, \theta, t) = \eta \nabla_\theta^2 c^a(r, \theta, t) \\
	- \sum_b J^{a b} \nabla_\theta \cdot \left[ \int \dd r' \dd \theta'  H(R_0 - \lVert r - r' \rVert)  \sin(\theta' - \theta) c^a(r, \theta, t) c^b(r', \theta', t)  \right].
\end{split}
\end{equation}

To simplify this equation, we replace $H(R_0 - \lVert r - r' \rVert)$ by $2 \pi R_0^2 \delta(r - r')$, see e.g. Refs.~\cite{Marchetti2013,Yllanes2017} (the $2 \pi$ is here to simplify notations later), so that we obtain
\begin{equation}
\label{averaged_langevin_density_equation}
\begin{split}
	(\partial_t + v_0^a n(\theta) \cdot \nabla_r) c^a(r, \theta, t) = \eta \nabla_\theta^2 c^a(r, \theta, t)
	- \sum_b 2 \pi R_0^2 J^{a b} \nabla_\theta \cdot \left[ \int \dd \theta'   \sin(\theta' - \theta) c^a(r, \theta, t) c^b(r, \theta', t)  \right].
\end{split}
\end{equation}

Let us now define the angular moments 
\begin{equation}
	f_n^a(r, t) = \int \dd \theta \; \ee^{\ii n \theta} c^a(r, \theta, t).
\end{equation}
so that
\begin{equation}
	\label{expension_angular_moments}
	c^a(r, \theta, t) = \frac{1}{2\pi} \sum_n \ee^{-\ii n \theta} f_n^a(r, t)
\end{equation}
Note that by reality
\begin{equation}
	f_{-n}^a(r, t) = \overline{f_{n}^a(r, t)}
\end{equation}
where the overline represents complex conjugation.

We also define
\begin{equation}
	\partial_z = \partial_x - \ii \partial_y
	\qquad
	\text{and}
	\qquad
	\partial_{\bar{z}} = \partial_x + \ii \partial_y.
\end{equation}
Then
\begin{equation}
	n(\theta) \cdot \nabla_r = \cos(\theta) \partial_x + \sin(\theta) \partial_y = \frac{1}{2} \left[ \ee^{- \ii \theta} \partial_{z} + \ee^{\ii \theta} \partial_{\bar{z}} \right].
\end{equation}

Using the expansion \eqref{expension_angular_moments} into the equation \eqref{averaged_langevin_density_equation}, we get
\begin{equation}
\begin{split}
	\sum_n \ee^{-\ii n \theta} \partial_t f_n^a(r, t) 
	+ \frac{ v_0^a}{2} \sum_n \ee^{-\ii n \theta} \ee^{- \ii \theta} \partial_{z} f_n^a(r, t) 
	+ \frac{ v_0^a}{2} \sum_n \ee^{-\ii n \theta} \ee^{\ii \theta} \partial_{\bar{z}} f_n^a(r, t) 
	=
	\eta \sum_n \nabla_\theta^2 \ee^{-\ii n \theta} f_n^a(r, t) \\
	- \sum_b R_0^2 J^{a b} \nabla_\theta \cdot \left[ \int \dd \theta' 
	\frac{1}{2 \ii} \left[ \ee^{\ii (\theta' - \theta)} - \ee^{-\ii (\theta' - \theta)} \right]
	\sum_n \ee^{-\ii n \theta} f_{n}^a(r, t)
	\sum_n' \ee^{-\ii n' \theta'} f_{n'}^b(r, t)
	\right].
\end{split}
\end{equation}
i.e.
\begin{equation}
\begin{split}
	\sum_n \ee^{-\ii n \theta} \partial_t f_n^a(r, t) 
	+ \frac{ v_0^a}{2} \sum_n \ee^{-\ii (n+1) \theta} \partial_{z} f_n^a(r, t) 
	+ \frac{ v_0^a}{2} \sum_n \ee^{-\ii (n-1) \theta} \partial_{\bar{z}} f_n^a(r, t) 
	=
	\eta \sum_n (-\ii n)^2 \ee^{-\ii n \theta} f_n^a(r, t) \\
	- \sum_b R_0^2 J^{a b} \nabla_\theta \cdot \left[ \int \dd \theta' 
	\frac{1}{2 \ii} \left[
	\sum_{n, n'} \ee^{-\ii (n+1) \theta} \ee^{-\ii (n'-1) \theta'} f_{n}^a(r, t) f_{n'}^b(r, t)
	 - 
	\sum_{n, n'} \ee^{-\ii (n-1) \theta} \ee^{-\ii (n'+1) \theta'} f_{n}^a(r, t) f_{n'}^b(r, t)
	\right]
	\right].
\end{split}
\end{equation}
After reindexation,
\begin{equation}
\begin{split}
	\sum_n \ee^{-\ii n \theta} \partial_t f_n^a(r, t) 
	+ \frac{ v_0^a}{2} \sum_n \ee^{-\ii n \theta} \partial_{z} f_{n-1}^a(r, t) 
	+ \frac{ v_0^a}{2} \sum_n \ee^{-\ii n \theta} \partial_{\bar{z}} f_{n+1}^a(r, t) 
	=
	\eta \sum_n (-\ii n)^2 \ee^{-\ii n \theta} f_n^a(r, t) \\
	- \sum_b R_0^2 J^{a b} \nabla_\theta \cdot \left[ \int \dd \theta' 
	\frac{1}{2 \ii} \left[
	\sum_{n, n'} \ee^{-\ii n \theta} \ee^{-\ii n' \theta'} f_{n-1}^a(r, t) f_{n'+1}^b(r, t)
	 - 
	\sum_{n, n'} \ee^{-\ii n \theta} \ee^{-\ii n' \theta'} f_{n+1}^a(r, t) f_{n'-1}^b(r, t)
	\right]
	\right].
\end{split}
\end{equation}
Integrating over $\theta'$ gives $\delta_{n', 0}$ which removes the corresponding sum, and after applying the last derivative we obtain
\begin{equation}
\begin{split}
	\sum_n \ee^{-\ii n \theta} \partial_t f_n^a(r, t) 
	+ \frac{v_0^a}{2} \sum_n \ee^{-\ii n \theta} \partial_{z} f_{n-1}^a(r, t) 
	+ \frac{v_0^a}{2} \sum_n \ee^{-\ii n \theta} \partial_{\bar{z}} f_{n+1}^a(r, t) 
	=
	\eta \sum_n (-\ii n)^2 \ee^{-\ii n \theta} f_n^a(r, t) \\
	- \sum_b R_0^2 J^{a b} \left[
	\frac{1}{2 \ii}
	\sum_{n} (-\ii n) \ee^{-\ii n \theta} \left[ f_{n-1}^a(r, t) f_{1}^b(r, t) - f_{n+1}^a(r, t) f_{-1}^b(r, t) \right]
	\right].
\end{split}
\end{equation}
Finally, division by $\sum_n \ee^{-\ii n \theta}$ produces
\begin{equation}
\begin{split}
	\partial_t f_n^a(r, t) 
	+ \frac{v_0^a}{2} \partial_{z} f_{n-1}^a(r, t) 
	+ \frac{v_0^a}{2} \partial_{\bar{z}} f_{n+1}^a(r, t) 
	=
	\eta (-\ii n)^2 f_n^a(r, t) \\
	- \sum_b R_0^2 J^{a b}
	\frac{1}{2 \ii} (-\ii n)
	\left[ f_{n-1}^a(r, t) f_{1}^b(r, t) - f_{n+1}^a(r, t) f_{-1}^b(r, t) \right]
\end{split}
\end{equation}
and 
\begin{equation}
\begin{split}
	\partial_t f_n^a(r, t) 
	+ \frac{v_0^a}{2} \left[ \partial_{z} f_{n-1}^a(r, t)  + \partial_{\bar{z}} f_{n+1}^a(r, t)  \right]
	=
	- \eta n^2 f_n^a(r, t) \\
	+ \sum_b \frac{R_0^2 J^{a b}}{2} n \; \left[ f_{n-1}^a(r, t) f_{1}^b(r, t) - f_{n+1}^a(r, t) f_{-1}^b(r, t) \right]
\end{split}
\end{equation}

Hence, using the expansion \eqref{expension_angular_moments} into the equation \eqref{averaged_langevin_density_equation} finally yields
\begin{equation}
	\partial_t f_{n}^a + \frac{v_0^a}{2} \left( \partial_z f_{n-1}^a + \partial_{\bar{z}} f_{n+1}^a \right)
	= -n^2 \eta f_{n}^a + \frac{1}{2} \sum_b J_{a b} R_0^2 n \left[ f_{n-1}^a f_{1}^b - f_{n+1}^a f_{-1}^b \right].
\end{equation}

For $n=0,1,2$ we get
\begin{subequations}
\label{moments_equations_small_n}
\begin{align}
	\partial_t f_0^a + \frac{v_0^a}{2} \left( \partial_z \overline{f_{1}^a} + \partial_{\bar{z}} f_{1}^a \right) &= 0  \label{moments_equation_0} \\
	\partial_t f_1^a + \frac{v_0^a}{2} \left( \partial_z f_{0}^a + \partial_{\bar{z}} f_{2}^a \right)
	&= - \eta f_1^a + \frac{1}{2} \sum_b J_{a b} R_0^2 \left[ f_{0}^a f_{1}^b - f_{2}^a \overline{f_{1}^b} \right] \label{moments_equation_1} \\
	\partial_t f_{2}^a + \frac{v_0^a}{2} \left( \partial_z f_{1}^a + \partial_{\bar{z}} f_{3}^a \right)
	&= - 4 \eta f_{2}^a + \sum_b J_{a b} R_0^2 \left[ f_{1}^a f_{1}^b - f_{3}^a \overline{f_{1}^b} \right] \label{moments_equation_2}
\end{align}
\end{subequations}

Following \cite{Bertin2006,Bertin2009,Marchetti2013,Chate2019}, we close the hierarchy of moment equations by considering the last equation with the assumptions $f_3^a = 0$ and $\partial_t f_2^a = 0$, giving
\begin{equation}
\label{replacement_closure}
f_{2}^a = \frac{1}{4 \eta} \left[ -\frac{v_0^a}{2} \left( \partial_z f_{1}^a  \right) + \sum_b J_{a b} R_0^2 f_{1}^a f_{1}^b \right]
\end{equation}
The replacement \eqref{replacement_closure} used in equation \eqref{moments_equation_1} gives
\begin{equation}
\label{moments_equation_1_closed}
\begin{split}
	\partial_t f_1^a + \frac{v_0^a}{2} \partial_z f_{0}^a - \frac{(v_0^a)^2}{16 \eta} (\partial_{\bar{z}} \partial_z f_{1}^a) + \sum_b \frac{J_{a b} R_0^2 v_0^a}{8 \eta} \partial_{\bar{z}} (f_{1}^a f_{1}^b) \\
	= - \eta f_1^a + \sum_b \frac{J_{a b} R_0^2}{2} f_{0}^a f_{1}^b + 
	 \sum_b \frac{J_{a b} R_0^2 v_0^a}{16 \eta} \overline{f_{1}^b} (\partial_z f_{1}^a) - \sum_{b,c} \frac{J_{a b} J_{a c} R_0^4}{8 \eta} f_{1}^a \overline{f_{1}^b} f_{1}^c 
\end{split}
\end{equation}

We identify the density $\rho_a$ and polarization $\vec P^a=(P^a_x,P^a_y)^{\mathsf T}$ as
\begin{equation}
	f_0^a = \rho^a
	\qquad
	\text{and}
	\qquad
	f_1^a = P_x^a - \ii P_y^a.
\end{equation}
We note that the polarization $\vec{P}_a$ is called $\vec{v}_a$ in the main text. 

Equation~\eqref{moments_equation_0} gives
\begin{equation}
	\label{density_eom_si}
	\partial_t \rho^a + v_0^a \div(\vec P^a) = 0. 
\end{equation}

Neglecting spatial derivative terms $\partial_z$ and $\partial_{\bar{z}}$ for now, equation \eqref{moments_equation_1_closed} yields
\begin{equation}
\label{eq_without_gradients_hideous}
\begin{split}
	\partial_t \begin{pmatrix} P_x^a \\ P_y^a \end{pmatrix}
	= &- \eta \begin{pmatrix} P_x^a \\ P_y^a \end{pmatrix}
	+ \sum_b \frac{J_{a b} R_0^2}{2} \rho^a \begin{pmatrix} P_x^b \\ P_y^b \end{pmatrix} \\
	&- \frac{R_0^4}{8 \eta} \sum_{b,c} J_{a b} J_{a c} \begin{pmatrix}
		P_{x}^a P_{x}^b P_{x}^c + P_{y}^a P_{y}^b P_{x}^c - P_{y}^a P_{x}^b P_{y}^c + P_{x}^a P_{y}^b P_{y}^c \\
		P_{y}^a P_{y}^b P_{y}^c + P_{x}^a P_{x}^b P_{y}^c - P_{x}^a P_{y}^b P_{x}^c + P_{y}^a P_{x}^b P_{x}^c
	\end{pmatrix}
	+ \mathcal{O}(\nabla)
\end{split}
\end{equation}

Let us define the notation $(\vec x^*)_\mu = \epsilon_{\mu \nu} (\vec x)_\nu$ where $\epsilon_{\mu \nu}$ is the Levi-Civita symbol so we can rewrite the sum in the last term in Eq.~\eqref{eq_without_gradients_hideous} as $\vec P^a \braket{\vec P^b, \vec P^c} + \vec P^{a *} \braket{\vec P^{b *}, \vec P^c}$ where $\braket{\cdot, \cdot}$ is the standard Euclidean scalar product and write
\begin{equation}
	\partial_t \vec P^a
	= - \eta \vec P^a
	+ \sum_b \frac{J_{a b} R_0^2}{2} \rho^a \vec P^b
	- \frac{R_0^4}{8 \eta} \sum_{b,c} J_{a b} J_{a c} \left[ \vec P^a \braket{\vec P^b, \vec P^c} + \vec P^{a *} \braket{\vec P^{b *}, \vec P^c} \right]
	+ \mathcal{O}(\nabla).
\end{equation}

As $\braket{\vec P^*, \vec Q} = - \braket{\vec P, \vec Q^*}$, the term $\braket{\vec P^{b *}, \vec P^c}$ is antisymmetric in the exchange $b \leftrightarrow c$ while $J_{a b} J_{a c}$ is symmetric, so after the sum is applied this term is removed and we get
\begin{equation}
\label{eq_without_gradients_less_hideous}
	\partial_t \vec P^a
	= - \eta \vec P^a
	+ \sum_b \frac{J_{a b} R_0^2}{2} \rho^a \vec P^b
	- \frac{R_0^4}{8 \eta} \sum_{b,c} \vec P^a \braket{J_{a b} \vec P^b, J_{a c} \vec P^c}
	+ \mathcal{O}(\nabla).
\end{equation}

\medskip

We now explore the gradient contributions to $\partial_t f_1^a$, namely
\begin{equation}
\begin{split}
	 - \frac{v_0^a}{2} \partial_z f_{0}^a + \frac{(v_0^a)^2}{16 \eta} (\partial_{\bar{z}} \partial_z f_{1}^a) 
	 + \sum_b \frac{J_{a b} R_0^2 v_0^a}{16 \eta} \left[\overline{f_{1}^b} (\partial_z f_{1}^a) - 2 \partial_{\bar{z}} (f_{1}^a f_{1}^b) \right].
\end{split}
\end{equation}
The simple terms are
\begin{align}
	\partial_z f_{0}^a &\to \vec{\grad} \rho^a \\
	(\partial_{\bar{z}} \partial_z f_{1}^a) &\to \nabla^2 \vec{P}^a.
\end{align}
The last term is a mess, but the following should hold
\begin{align}
	\overline{f_1^b} (\partial_z f_1^a) &\to (\vec{P}^b \cdot \vec{\grad}) \vec{P}^a + (\vec{P}^{b *} \cdot \vec{\grad}) \vec{P}^{a *} \\
	\partial_{\bar{z}} (f_{1}^a f_{1}^b) &\to (\vec{P}^a \cdot \vec{\grad}) \vec{P}^b + \vec{P}^b \div(\vec{P}^a)
	- (\vec{P}^{a*} \cdot \vec{\grad}) \vec{P}^{b *}
	- \vec{P}^{b*} \div(\vec{P}^{a*})
\end{align}
where the $\grad$ are written in letters for clarity, but will soon be replaced by $\nabla$.

Hence, the hydrodynamic equation finally reads
\begin{equation}
\label{eq_less_hideous}
\begin{split}
	\partial_t \vec P^a
	= - \eta \vec P^a
	+ \sum_b j_{a b} \rho^a \vec P^b
	- \frac{1}{2 \eta} \sum_{b,c} \vec P^a (j_{a b} \vec P^b \cdot j_{a c} \vec P^c) 
	- \frac{v_0^a}{2} \nabla \rho^a + D_{a} \nabla^2 \vec{P}^a \\
	 + \sum_b \lambda_{a b} \left[
	 (\vec{P}^b \cdot \nabla) \vec{P}^a + (\vec{P}^{b *} \cdot \nabla) \vec{P}^{a *}
	 - 2 \left[
	 (\vec{P}^a \cdot \nabla) \vec{P}^b + \vec{P}^b \div(\vec{P}^a)
	- (\vec{P}^{a*} \cdot \nabla) \vec{P}^{b *}
	- \vec{P}^{b*} \div(\vec{P}^{a*})
	  \right]
	 \right].
\end{split}
\end{equation}
where we have defined
\begin{equation}
	\label{coefficients_hydro}
	j_{a b} = \frac{R_0^2}{2} \, J_{a b}
	\qquad
	D_{a} = \frac{(v_0^a)^2}{16 \eta}
	\qquad
	\lambda_{a b} = \frac{v_0^a j_{a b}}{8 \eta}
\end{equation}

In the stability analysis of section~\ref{sec_excitation_supp}, we will set $\lambda_{a b} = v_0^{a} v_0^{b}$.
We refer the reader to Ref.~\cite{Marchetti2013} for a discussion on this point in the case of a single population.

We also note that the relations between the numerous terms in Eq.~\eqref{eq_less_hideous} are merely a consequence of the particular derivation used here.
In generic non-reciprocal binary fluids, these terms might have unrelated coefficients. Nonetheless, we will focus on Eq.~\eqref{eom_with_gradients_two_populations_si} for simplicity.

\medskip

In some cases (such as when there is only one kind of active particle), Eq.~\eqref{eq_less_hideous} will be simplified by using the following identities for two vectors fields $u$ and $v$
\begin{subequations}
\label{identities_symmetrized_diff_stars}
\begin{align}
	(u^* \cdot \nabla) v^* + (v^* \cdot \nabla) u^* = \grad(u \cdot v) - u \div(v) - v \div(u) \\
	u^* \div(v^*) + v^* \div(u^*) = \grad(u \cdot v) - (u \cdot \nabla) v - (v \cdot \nabla) u.
\end{align}
\end{subequations}
They can also be written into a more symmetric but less useful way as
\begin{subequations}	
\begin{align}
\grad(u \cdot v) &= (u^* \cdot \nabla) v^* + (v^* \cdot \nabla) u^* + u \div(v) + v \div(u)  \\
\grad(u \cdot v) &= (u \cdot \nabla) v + (v \cdot \nabla) u + u^* \div(v^*) + v^* \div(u^*).
\end{align}
\end{subequations}

\subsubsection{Hydrodynamic equations for a single population}

In this subsection, we first specialize equation \eqref{eq_less_hideous} to the case of a single population to recover the standard Toner-Tu equations.

Starting from \eqref{eq_less_hideous}, we use $\vec P^{*} \cdot \vec P = 0$ and the identity
\begin{equation}
	\label{remove_stars_single_population}
	(\vec P^* \cdot \nabla) \vec P^* + 2 \left[ (\vec P^* \cdot \nabla) \vec P^* +  \vec P^* (\nabla \cdot \vec P^*) \right] = 5 \nabla(\vec P^2/2) - 3 \vec P (\nabla \cdot \vec P) - 2 (\vec P \cdot \nabla) \vec P.
\end{equation}
obtained from \eqref{identities_symmetrized_diff_stars} to get
\begin{equation}
\begin{split}
	\partial_t \vec P + \lambda_1 (\vec{P} \cdot \nabla) \vec{P}
	+ \lambda_2 \vec{P} \div(\vec{P}) 
	+ \lambda_3 \nabla(\vec{P}^2)
	=
	-\left[\alpha(\rho) + \beta \; \lVert \vec{P} \rVert^2 \right] \vec{P}
	- \frac{v_0}{2} \nabla \rho 
	+ D \nabla^2 \vec{P}
\end{split}
\end{equation}
where
\begin{equation}
	\alpha(\rho) = \eta - j \rho 
	\qquad
	\beta = \frac{j^2}{2 \eta}
	\qquad
	D = \frac{(v_0)^2}{16 \eta}
	\qquad
	\lambda_0 = \frac{j v_0}{8 \eta} 
	\qquad
	j = \frac{J R_0^2}{2}
\end{equation}
and $\lambda_1 = 3 \lambda_0$, $\lambda_2 = 5 \lambda_0$, $\lambda_3 = - 5/2 \lambda_0$.

\subsubsection{Hydrodynamic equations for two population}

We now specialize to the case where there are only two populations $a,b,c = A,B$ (here the capital letters $A$ and $B$ refer to the two populations and are not abstract indices), which is the situation analyzed in the main text.

A special case of this situation was derived and analyzed in Ref.~\cite{Yllanes2017}, with which our results agree.

We set $a=A$ for simplicity, and remove all spatial derivatives.
In this case, the hideous sum in equation \eqref{eq_without_gradients_hideous} reads
\begin{equation}
	J_{A A} J_{A A}
	\begin{pmatrix}
		P_{x}^A P_{x}^A P_{x}^A + P_{y}^A P_{y}^A P_{x}^A  \\
		P_{y}^A P_{y}^A P_{y}^A + P_{x}^A P_{x}^A P_{y}^A 
	\end{pmatrix}
	+
	2 J_{A A} J_{A B}
	\begin{pmatrix}
		P_{x}^A P_{x}^A P_{x}^B + P_{x}^A P_{y}^A P_{y}^B  \\
		P_{y}^A P_{y}^A P_{y}^B + P_{y}^A P_{x}^A P_{x}^B
	\end{pmatrix}
	+
	J_{A B} J_{A B}
	\begin{pmatrix}
		P_{x}^A P_{x}^B P_{x}^B + P_{x}^A P_{y}^B P_{y}^B \\
		P_{y}^A P_{y}^B P_{y}^B + P_{y}^A P_{x}^B P_{x}^B
	\end{pmatrix}
\end{equation}
Factoring out the polarization, one recognizes
\begin{equation}
\left[
	J_{A A} J_{A A} \lVert \vec P^A \rVert^2
	+
	2 J_{A A} J_{A B} \braket{\vec P^A, \vec P^B}
	+
	J_{A B} J_{A B} \lVert \vec P^B \rVert^2
\right]
	\begin{pmatrix}
	P_{x}^A \\
	P_{y}^A
	\end{pmatrix}
\end{equation}
i.e. (all quantities are real)
\begin{equation}
\left[
	\lVert J_{A A} \vec P^A \rVert^2
	+
	2 \braket{J_{A A} \vec P^A, J_{A B} \vec P^B}
	+
	\lVert J_{A B} \vec P^B \rVert^2
\right]
	\vec P^A
	= \lVert J_{A A} \vec P^A + J_{A B} \vec P^B \rVert^2 \;
	\vec P^A
\end{equation}

Using again $j_{a b} = (R_0^2/2) J_{a b}$, we obtain
\begin{equation}
	\partial_t \vec P^A
	= - \eta \vec P^A
	+ j_{A A} \rho^A \vec P^A
	+ j_{A B} \rho^A \vec P^B
	- \frac{1}{2 \eta} \lVert j_{A A} \vec P^A + j_{A B} \vec P^B \rVert^2 \; \vec P^A
	+ \mathcal{O}(\nabla)
\end{equation}
and a similar equation for $\vec P^B$ is obtained by permuting the indices.

Including the gradient terms, \eqref{eq_less_hideous} becomes for $a=A$ 
\begin{equation}
\begin{split}
	\label{eom_with_gradients_two_populations_si}
	\partial_t \vec P^A
	= 
	\left[
	j_{A A} \rho^A
	- \eta
	- \frac{1}{2 \eta}  \lVert j_{A A} P^A + j_{A B} \vec P^B \rVert^2
	\right]
	 \vec P^A
	+ j_{A B} \rho^A \vec P^B \\
- \frac{v_0^A}{2} \nabla \rho^A + D_A \nabla^2 \vec{P}^A \\
	+ \lambda_{A A} \left[
	 5/2 \nabla(\vec{P}^A \cdot \vec{P}^A) 
	 - 3 (\vec{P}^A \cdot \nabla) \vec{P}^A
	 - 5 \vec{P}^A \div(\vec{P}^A)
	 \right] \\
	 + \lambda_{A B} \left[
		     (\vec{P}^B \cdot \nabla) \vec{P}^A 
		 - 2 (\vec{P}^A \cdot \nabla) \vec{P}^B
		 - 2 \vec{P}^B \div(\vec{P}^A)
		 +   (\vec{P}^{B*} \cdot \nabla) \vec{P}^{A*}
		 + 2 (\vec{P}^{A*} \cdot \nabla) \vec{P}^{B*}
		 + 2 \vec{P}^{B*} \div(\vec{P}^{A*})
	 \right]
\end{split}
\end{equation}
where we already have used equation \eqref{remove_stars_single_population} to simplify the $AA$ terms, and where we have defined
\begin{equation}
	\label{eq_lambda0_derived}
	j_{a b} = \frac{R_0^2}{2} \, J_{a b}
	\qquad
	D_{a} = \frac{(v_0^a)^2}{16 \eta}
	\qquad
	\lambda_{a b} = \frac{v_0^a j_{a b}}{8 \eta}
\end{equation}
In the stability analysis of section~\ref{sec_excitation_supp}, we will set $\lambda_{a b} = v_0^{a} v_0^{b}$, see discussion above.

The equation for $a=B$ is obtained in the same way.

\section{Mean-field phase diagram in the steady state}
\label{sec_mean_field}

In this section, to grasp the influence of non-reciprocal interaction to the many-body state, we perform a mean-field approximation to the hydrodynamic theory derived in section~\ref{sec_hydro}, thereby neglecting the gradient terms in Eq.~\eqref{eom_with_gradients_two_populations_si}:
\begin{equation}
\label{TonerTuMF_supp}
\partial_t 
\begin{pmatrix}
\vec P^A \\
\vec P^B \\
\end{pmatrix}
=-\hat W[\vec P^A,\vec P^B]
\begin{pmatrix}
\vec P^A \\
\vec P^B \\
\end{pmatrix}
\end{equation}
where
\begin{equation}
	\hat W[\vec P^A,\vec P^B] =
\begin{pmatrix}
W^{AA}[\vec P^A,\vec P^B] & W^{AB} \\
W^{BA} & W^{BB}
[\vec P^A,\vec P^B]
\end{pmatrix}
= \begin{pmatrix}
\eta - j_{AA}\rho^A
+\frac{1}{2\eta}
\absvec{\vec {Q} ^A(t)}^2
& -j_{AB}\rho^A \\
-j_{BA}\rho^B & \eta - j_{BB}\rho^B
+\frac{1}{2\eta}
\absvec{\vec {Q} ^B(t)}^2
\end{pmatrix}
\end{equation}
and
\begin{subequations}
\begin{align}
\vec {Q} ^A(t)  = j_{AA}\vec P^A(t) +j_{AB}\vec P^B(t) , 
\\
\vec {Q} ^B(t)  = j_{BA}\vec P^A(t) +j_{BB}\vec P^B(t) .
\end{align}
\label{underbarP}
\end{subequations} 
The matrix $\hat W$ is in general non-Hermitian, i.e. $\hat W \ne \hat W^\dagger$.
We are especially interested in cases where the non-reciprocal interaction is pronounced enough that the inter-species couplings have opposite signs ($j_{AB} j_{BA}<0$). 
In such a situation, there are no configuration that can make both species satisfied. This situation shares conceptual similarities with the geometrical frustration present in systems ranging from (spin) glasses~\cite{Wannier1950,Toulouse1977,Vannimenus1977,Sherrington1975,Edwards1975,Binder1986,Nisoli2013} to ice, liquid crystals and colloidal systems~\cite{Sadoc1999,Turner2010,OrtizAmbriz2019}, which occurs when the interactions between different entities, such as spins or atoms, have competing effects (like for three spins with antiferromagnetic couplings on the vertices on a triangle).
The dynamical frustration present here has a different origin: instead of coming from multiple competing interactions, it arises from each individual non-reciprocal interaction.
It gives rise to a time-dependent phase where the direction of flocking continuously changes over time, see Fig. \ref{subfigure_bifurcation_ep}a. We call this the \enquote{chiral phase}.

In this section, we first give an analytic argument that illustrates how a spontaneous symmetry breaking from a flocking or anti-flocking phase (similar to the antiferromagnetic phase) to the chiral phase may occur by increasing the non-reciprocity of the coupling strength. We show that this phase transition is marked by a so-called exceptional points~\cite{Kato1984} of the matrix $\hat W$, which are the points where two of the eigenvectors of $\hat W$ \textit{coalesce}, and discuss its relation to PT symmetry breaking. 
This mechanism, unique to out-of-equilibrium systems, originates from the non-Hermitian structure of the matrix $\hat W$ that controls the dynamics. As such, it is a generic feature of non-reciprocal fluids, as we have illustrated from a general theory in the main text. 
Using the relation between the phase transition and the exceptional points of $\hat W$, we determine the phase boundaries in terms of the microscopic coupling strength $j_{ab}$ and compare them with numerics. 
We also show how chiral phase interpolates the flocking and the antiflocking phase (see Fig. \ref{subfigure_bifurcation_ep}b). 

In addition, we also find from direct numerical simulations of the mean-field equation (Eq.~\eqref{TonerTuMF_supp}) that another time-dependent phase appears in the phase diagram, which we call the "swap phase", see Fig.~\ref{subfigure_bifurcation_ep}c and SI Movie 2. 
The swap phase exhibits a time oscillation in the \textit{amplitude} of the macroscopic polarization (in contrast to the chiral phase exhibiting oscillations in their direction of the orientation), which is again triggered by the dynamical frustration.
Further, we find an interesting regime where these two oscillations coexist with \textit{different frequencies}, which its time dependence of the polarization field, as a result, becomes \textit{quasiperiodic}, see Figs. 2f and 2g in the main text and SI Movie 2. 
Discussions on the origin of these phases from the point of view of fluctuation modes are provided in section~\ref{sec_excitation_supp}.

We note that the mean-field approximation employed in this section assumes that the system reaches a uniform state.
This is not always true; we discuss finite momentum instabilities and pattern formation~\cite{Cross1993} in section.~\ref{sec_excitation_supp}.

\subsection{Emergence of the chiral phase by PT symmetry breaking}
\label{sec emergence of chiral phase}

\begin{figure*}
  \centering
  \includegraphics{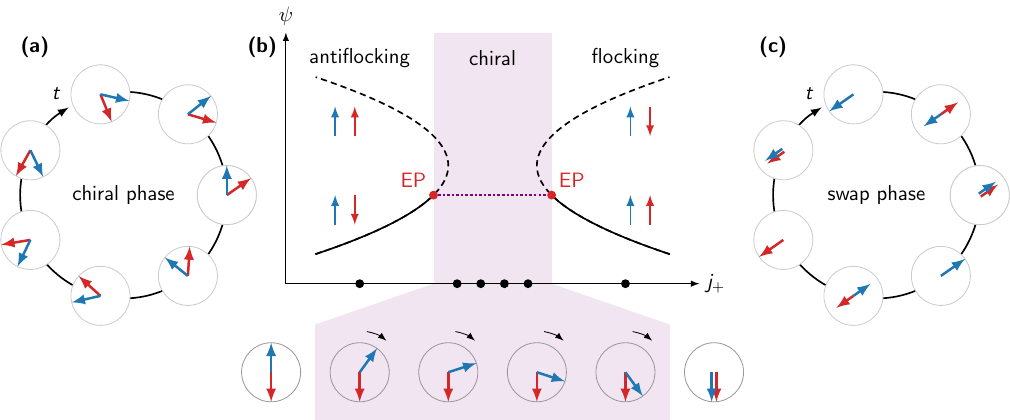}
  \caption{\label{subfigure_bifurcation_ep}\strong{Bifurcation and phases.}
  (a) Temporal diagram of the chiral phase. 
  (b) Schematic bifurcation diagram of the system along a line at fixed $j_{-}$, for different values of $j_{+}$.
  (c) Temporal diagram of the swap phase.
  In the chiral phase (panel a), both populations rotate in the same direction, with the same angular velocity. 
  The direction of rotation is chosen at random.
  The angle $\Delta \phi^{A B}$ between the velocities of the two populations continuously interpolates between the flocking (where it vanishes) and antiflocking phases (where it is maximal, i.e. equal to \ang{180}) along a line connecting them (panel b). 
  In the swap phase (panel c), the velocities oscillate along a fixed direction.
  }
\end{figure*}

Before attempting to directly solve the full nonlinear equation \eqref{TonerTuMF_supp}, here we provide an argument based on the non-Hermiticity and the symmetry of the matrix $\hat W$, that explains how a spontaneous breaking of time translation symmetry may emerge in this system. 
In particular, we show that, in addition to the uniform flocking and the antiflocking phase, the chiral phase, where the direction of the orientation continuously changes over time, can emerge as a steady state solution as a result of the non-Hermitian nature of the matrix $\hat W$ (see also Fig.~\ref{subfigure_bifurcation_ep}).
We also discuss its relation to the spontaneous PT symmetry breaking (with a generalized PT operator), often discussed in the context of open quantum mechanics \cite{Bender1998,Bender2007,Konotop2016,Mostafazadeh2002,Bender2002}.  
Since the details of the $\vec P^A(t),\vec P^B(t)$ dependence on the matrix $\hat W$ is essentially irrelevant to this discussion, the emergence of the chiral phase is a generic feature of  non-reciprocally interacting fluids, as we have shown by the general formalism presented in the main text.  

\subsubsection{Flocking and antiflocking phase}

Let us first look for the conventional, time-independent solutions, by assuming that the polarization eventually converges to a constant, i.e., $\vec P^{A(B)}(t)=\vec P^{A(B)}_0={\rm const.}$
Although, at a glance, it seems possible for the relative angle between $\vec P^A_0$ and $\vec P^B_0$ to take any value, we shall see in the following that it is only possible to be parallel or antiparallel to each other under this assumption, which we call the flocking and antiflocking phase, respectively. 

In the time-independent steady state, the mean-field equation~\eqref{TonerTuMF_supp} gives
\begin{eqnarray}
{\bf 0}&=&
\hat W_0
\left(\begin{array}{c}
\vec P^A_0 \\
\vec P^B_0
\end{array}\right)
=
\left(\begin{array}{cc}
W^{AA}_0 & W^{AB}_0 \\
W^{BA}_0 & W^{BB}_0
\end{array}\right)
\left(\begin{array}{c}
\vec P^A_0 \\
\vec P^B_0
\end{array}\right)
\nonumber\\
&=&
\left(\begin{array}{cc}
\eta - j_{AA}\rho^A
+\frac{1}{2\eta}
\absvec{\vec {Q} ^A_0} ^2
& -j_{AB}\rho^A \\
-j_{BA}\rho^B & \eta - j_{BB}\rho^B
+\frac{1}{2\eta}
\absvec{\vec {Q} ^B_0}^2
\end{array}\right)
\left(\begin{array}{c}
\vec P^A_0 \\
\vec P^B_0
\end{array}\right),
\label{TonerTuMFsteady}
\end{eqnarray}
with
\begin{subequations}
\begin{align}
\vec {Q}^A_0  = j_{AA}\vec P^A_0 +j_{AB}\vec P^B_0 , 
\\
\vec {Q}^B_0  = j_{BA}\vec P^A_0 +j_{BB}\vec P^B_0.
\end{align}
\label{underbarPsteady}
\end{subequations} 
Diagonalizing this matrix $\hat W_0$ in Eq.~\eqref{TonerTuMFsteady} gives 
\begin{eqnarray}
{\bf 0}= \left(\begin{array}{cc}
\Gamma_- & 0 \\
0 & \Gamma_+
\end{array}\right)
\left(\begin{array}{c}
\vec P^-_0 \\
\vec P^+_0
\end{array}\right).
\label{diagW0_flock}
\end{eqnarray}
Here, the eigenvalues $\Gamma_\pm$ roughly correspond to the decay rate of the eigenmodes, where their explicit expressions are given by 
\begin{eqnarray}
\Gamma_\pm &=& \frac{1}{2}[W^{AA}_0+W^{BB}_0\pm \sqrt{\Lambda_0}], \\
\Lambda_0 &=& (W^{AA}_0-W^{BB}_0)^2 + 4 W^{AB}_0W^{BA}_0,
\label{Lambda0_supp}
\end{eqnarray}
with the corresponding eigenmodes,
\begin{subequations}
\begin{align}
{\bf u_-} = 
\left(\begin{array}{c}
\frac{\sqrt{\Lambda_0}-(W^{AA}_0-W^{BB}_0)}{2}  \\
-W^{BA}_0
\end{array}\right), \\
{\bf u_+} = 
\left(\begin{array}{c}
W^{AB}_0 \\
\frac{\sqrt{\Lambda_0}-(W^{AA}_0-W^{BB}_0)}{2}  
\end{array}\right).
\end{align}
\label{eigenvectors}
\end{subequations}
The order parameter of the flocking phase is transformed accordingly into
\begin{widetext}
\begin{eqnarray}
\left(\begin{array}{c}
\vec P^-_0 \\
\vec P^+_0
\end{array}\right)
=\hat U_0
\left(\begin{array}{c}
\vec P^A_0 \\
\vec P^B_0
\end{array}\right)
=
\frac{1}{{\rm det} \hat U_0^{-1}}
\left(\begin{array}{cc}
\frac{\sqrt{\Lambda_0}-(W^{AA}_0-W^{BB}_0)}{2} 
& -W_0^{AB}
\\
W^{BA}_0 & \frac{\sqrt{\Lambda_0}-(W^{AA}_0-W^{BB}_0)}{2} 
\end{array}\right)
\left(\begin{array}{c}
\vec P^A_0 \\
\vec P^B_0
\end{array}\right),
\end{eqnarray}
\end{widetext}
where $\hat U_0^{-1}=(\bf u_-,\bf u_+)$, or inversely, 
\begin{eqnarray}
\left(\begin{array}{c}
\vec P^A_0 \\
\vec P^B_0
\end{array}\right)
=
\left(\begin{array}{cc}
[U_0^{-1}]^{A-} & [U_0^{-1}]^{A+} \\
{}[U_0^{-1}]^{B-} & [U_0^{-1}]^{B+} 
\end{array}\right)
\left(\begin{array}{c}
\vec P^-_0 \\
\vec P^+_0
\end{array}\right)
=\left(\begin{array}{cc}
\frac{\sqrt{\Lambda_0}-(W^{AA}_0-W^{BB}_0)}{2} 
& W^{AB}_0
\\
-W^{BA}_0 & \frac{\sqrt{\Lambda_0}-(W^{AA}_0-W^{BB}_0)}{2} 
\end{array}\right)
\left(\begin{array}{c}
\vec P^-_0 \\
\vec P^+_0
\end{array}\right).
\end{eqnarray}

It can be shown from Eq.~\eqref{diagW0_flock} that in addition to a trivial solution $\vec P^-_0=\vec P^+
_0=0$ that corresponds to a disordered phase, nontrivial solutions with $(\vec P_0^-,\vec P_0^+)^{\mathsf T}\ne 0$ 
can always be classified into two types~\cite{Hanai2019}: solutions that satisfy
$(\vec P^-_0 \ne 0,\vec P^+_0=0)$ and $(\vec P^-_0 = 0,\vec P^+_0\ne 0)$, 
which we call "$-$" and "$+$" solutions, respectively. 
This is readily seen as follows. 
Let us assume that $\vec P^{-(+)}_0\ne 0$. 
Then, it is necessary for the eigenvalue $\Gamma_{-(+)}$ to vanish in order to satisfy the first (second) line of Eq.~\eqref{diagW0_flock}.
In such case, since the eigenvalue of "$+(-)$" is finite $\Gamma_{+(-)}\ne 0$ as long as $\Gamma_-\ne\Gamma_+$, $\vec P^{+(-)}_0$ necessarily vanishes because of the second (first) line of Eq.~\eqref{diagW0_flock}. 
Thus, $\vec P^-_0$ and $\vec P^+_0$ cannot be nonzero simultaneously, letting us classify the solutions into two types.

The above property has a direct consequence that the polarization field of $A$ and $B$ agents can only be either parallel or antiparallel in the uniform steady state, which we call the flocking and antiflocking phase, respectively.
For example, for the "$-$" solution, polarization fields are given by
\begin{eqnarray}
\vec P_0^A &=&  [U_0^{-1}]^{A-} \vec P_0^-
= \frac{\sqrt{\Lambda_0}-(W^{AA}_0-W^{BB}_0)}{2} 
\vec P_0^-, 
\label{vecPA0} \\
\vec P_0^B &=& 
[U_0^{-1}]^{B-} \vec P_0^-
=-W^{BA}_0 \vec P_0^-,
\label{vecPB0}
\end{eqnarray}
explicitly showing that $\vec P^A_0$ and $\vec P^B_0$ are either parallel or antiparallel to each other, depending on the relative sign between $[U_0^{-1}]^{A-}$ and $[U_0^{-1}]^{B-}$.

As mentioned earlier, the eigenvalues $\Gamma_\pm$ roughly corresponds to the decay rate of the corresponding modes.
The condition $\Gamma_{-(+)}=0$ for the "$-(+)$" solution can be regarded as the defining property of a steady state.
Assuming $\Lambda_0>0$ (that assures $\Gamma_\pm$ to be real), 
the "$-(+)$" solution is likely to be (un)stable since $0<\Gamma_+(\Gamma_-<0)$, implying a positive (negative) decay rate of the "$+(-)$" mode, where we have used the relation  $\Gamma_-<\Gamma_+$.
This strongly suggest that the "$-$" solution is the solution that would be realized. 
Indeed, as shown in section~\ref{subsec_chiral} from a stability analysis, it can proven that "$+$" solution is always unstable, limiting the possible stable solution to the "$-$" solution.

It is important to emphasize that, for the flocking or antiflocking phase to be realized, it is necessary for $\Lambda_0$ to be positive since it requires $\sqrt{\Lambda_0}$ to be real such that the state can satisfy the relation $\Gamma_-=0$ ($\Gamma_+=0$) for "$-(+)$" solution.
This condition is assured to be satisfied when the sign of the inter-species coupling is the same,  i.e.,  $j_{AB}j_{BA}>0$ (or equivalently $W^{AB}_0W^{BA}_0>0$), which includes the reciprocal case, since the first term of Eq.~\eqref{Lambda0_supp}, $(W^{AA}_0-W^{BB}_0)^2$, is non-negative.

However, when the inter-species couplings have opposite sign (i.e.  $j_{AB}j_{BA}<0$ or $W^{AB}_0 W^{BA}_0<0$), $\Lambda_0$ may become negative, hence the eigenvalues can turn \textit{imaginary} implying the existence of a phase transition to a time-oscillating phase.
As we show in the following section, the system indeed exhibits a phase transition to an phase where the direction of the orientation continuously oscillates in time, which we call the chiral phase. 
The phase transition is  driven by the non-Hermitian nature of the matrix $\hat W$, which can be seen from the observation that the phase transition point, $\Lambda_0=0$ is the point where the two eigenvectors $\bm u_\pm$ (see Eq.~\eqref{eigenvectors}) coalesce; This is the so-called exceptional point \cite{Kato1984}.
We will further show in section~\ref{sec_PTsymmetry} that this transition can be regarded as a spontaneous PT symmetry breaking discussed in the field of non-Hermitian quantum mechanics. 

\subsubsection{Chiral phase}

Below, we look for solutions with an oscillating polarization field, described by the ansatz,  
\begin{subequations}
\begin{align}
\vec P^A (t)
=R^A
\left(\begin{array}{c}
\cos(\Omega t +\phi^A) \\
\sin(\Omega t +\phi^A)
\end{array}\right), \\
\vec P^B (t)
=
R^B
\left(\begin{array}{c}
\cos(\Omega t +\phi^B) \\
\sin(\Omega t +\phi^B)
\end{array}\right),
\end{align}
\label{chiral_ansatz}
\end{subequations}
with $\Omega$ being the frequency of the oscillation and $R^A,R^B(>0)$ the amplitude of the polarization fields of $A$ and $B$ species, respectively. 
This solution exhibits a "chiral" motion, in the sense that the direction of the orientation continuously evolves in time (while the amplitude of the polarization remains fixed) implying a collective chiral motion of agents, which is exactly what is observed in our microscopic Vicsek model simulation (Fig. 2d in the main text and SI Movie 1).

We note that, for solutions of the form~\eqref{chiral_ansatz}, the $O(2)$ symmetry of the mean-field system assures that $\hat W=\hat W_0$ does not depend on time.
This can be directly checked from the observation that the amplitudes of the vectors $\vec {Q} ^A(t), \vec {Q} ^B(t)$ given by Eq.~\eqref{underbarP} which shows that the magnitude of the nonlinearity in $W^{AA}_0$ and $W^{BB}_0$ are time-independent.

We show below that the ansatz \eqref{chiral_ansatz} satisfies, 
\begin{widetext}
\begin{eqnarray}
\Omega &=& \Omega_\pm 
=\pm \frac{1}{2}\sqrt{|\Lambda_0|}, \\
\Delta\phi^{AB} 
&=&
\phi^A-\phi^B 
= \begin{cases} 
\arccos\left[
+ \sqrt{
\left|\frac{W_0^2}{W^{AB}_0W^{BA}_0}\right|} 
\right]
= 
\arccos\left[
+\sqrt{
\left|1-\frac{|\Lambda_0|}{|W^{AB}_0W^{BA}_0|}\right|} 
\right] & (W_0 W_0^{AB}<0) \\
\arccos\left[
- \sqrt{
\left|\frac{W_0^2}{W^{AB}_0W^{BA}_0}\right|} 
\right]
= 
\arccos\left[
-\sqrt{
\left|1-\frac{|\Lambda_0|}{|W^{AB}_0W^{BA}_0|}\right|} 
\right] & 
(W_0 W_0^{AB}>0)
\end{cases}
\label{Deltaphi_supp}
\end{eqnarray}
\end{widetext}
with
\begin{eqnarray}
W^{AA}_0 = - W^{BB}_0 \equiv W_0, 
	\qquad
R^B =
\sqrt{\left|\frac{W_0^{BA}}{W_0^{AB}}\right|} R^A,
\end{eqnarray}
and importantly, $\Lambda_0<0$.
The last condition is satisfied only if $W^{AB}_0 W^{BA}_0<0$, i.e. non-reciprocal coupling with opposite signs.

Two comments are in order. 
Firstly, the fact that we find two solutions, $\Omega=\Omega_+>0$ and $\Omega=\Omega_-<0$, indicates the occurrence of a spontaneous chiral ($Z_2$) symmetry breaking to a left and right-handed phase, respectively.
Secondly, the exceptional point $\Lambda_0=0$ with $W_0>0(<0)$ is a continuous transition point to the (anti)flocking phase, with $\Delta\phi^{AB}=0(=\pi)$. 
Noting that $W_0$ may switch its sign inside the chiral phase, this implies that the chiral flocking phase lies in between the flocking and antiflocking phase (See Fig. \ref{subfigure_bifurcation_ep}b.), which is indeed the case (see also Figs. 2f and g).

\begin{widetext}
The mean-field equation \eqref{TonerTuMF_supp} we wish to solve with the ansatz \eqref{chiral_ansatz} takes the form
\begin{align}
-\Omega R^A  \sin (\Omega t + \phi^A) 
&= - W^{AA}_0 R^A \cos (\Omega t + \phi^A) - W^{AB}_0 R^B \cos(\Omega t+\phi^B) ,
\label{chiral mean-field A1} \\
\Omega R^A \cos (\Omega t +\phi^A) 
&= -W^{AA}_0 R^A \sin (\Omega t + \phi^A) - W^{AB}_0 R^B \sin(\Omega t + \phi^B) ,
\label{chiral mean-field A2}\\
-\Omega R^B  \sin (\Omega t + \phi^B) 
&= - W^{BA}_0 R^A \cos(\Omega t+\phi^A) - W^{BB}_0 R^B \cos (\Omega t + \phi^B),
\label{chiral mean-field B1}\\
\Omega R^B \cos (\Omega t +\phi^B) 
&=-W^{BA}_0 R^A \sin(\Omega t + \phi^A) - W^{BB}_0 R^B \sin(\Omega t + \phi^B).
\label{chiral mean-field B2}
\end{align}
Note that Eqs.~\eqref{chiral mean-field A1} and \eqref{chiral mean-field A2} (Eqs.~\eqref{chiral mean-field B1} and \eqref{chiral mean-field B2}) are equivalent.
These can be factorized as
\begin{align}
\tilde {R}^{a} 
\cos(\Omega t + \tilde{\phi}^a) &= 0, 
\label{Rtacos} \\
\tilde{R}^{a} 
\sin(\Omega t + \tilde{\phi}^a) &= 0,
\label{Rtasin} 
\end{align}
where $a=A,B$,
\begin{align*}
(\tilde{R}^A)^2 
&= (-W^{AA}_0 R^A\cos \phi^A - W^{AB}_0 R^B \cos \phi^B
+\Omega R^A \sin \phi^A )^2 
+ (W^{AA}_0 R^A \sin\phi^A + W^{AB}_0 R^B \sin\phi^B
+ \Omega R^A\cos \phi^A)^2
\\
(\tilde{R}^B )^2 
&= (-W^{BA}_0 R^A\cos \phi^A - W^{BB}_0 R^B \cos \phi^B
+\Omega R^B \sin \phi^B  )^2 
+ (W^{BA}_0 R^A \sin\phi^A + W^{BB}_0 R^B \sin\phi^B 
+ \Omega R^B\cos \phi^B)^2  
\end{align*}
and $\tilde\phi^A,\tilde\phi^B$ are real constant numbers determined from the parameters $W_0^{ab},R^a,\phi^a$. 
Equations \eqref{Rtacos} and \eqref{Rtasin} are satisfied at arbitrary $t$ when 
$\tilde R^A
=\tilde R^B=0$. 
\end{widetext}

Let us first determine $\Omega$ by solving $\tilde R^A = 0$. 
This gives, 
\begin{eqnarray}
\Omega &=& \frac{R^B}{R^A} W^{AB}_0
\sin\Delta\phi^{AB}
\pm i (W^{AA}_0 + W^{AB}_0 \frac{R^B}{R^A} \cos\Delta\phi^{AB}). 
\label{Omega_1}
\end{eqnarray}
Since we require the frequency $\Omega$ to be real, we demand the imaginary part of Eq.~\eqref{Omega_1} to vanish, 
\begin{eqnarray}
\Delta\phi^{AB} 
=\arccos \Big[-\frac{W^{AA}_0}{W^{AB}_0}\frac{R^A}{R^B}
\Big].
\label{phipmA}
\end{eqnarray}
Plugging this back into Eq.~\eqref{Omega_1}, we get 
\begin{eqnarray}
\label{chiral_omega_from_params}
\Omega&=&
\Omega_{\pm} = \frac{R^B}{R^A} W^{AB}_0
\sin\Delta\phi^{AB}
=
\pm\frac{R^B}{R^A}W^{AB}_0\sqrt{1-\left(\frac{R^A}{R^B}\frac{W^{AA}_0}{W^{AB}_0}\right)^2}.
\label{OmegapmA}
\end{eqnarray}

We can similarly compute for $\Omega$ and $\Delta\phi^{AB}$ by solving $\tilde R^B=0$, where we get
\begin{eqnarray}
\Delta\phi^{AB}
&=&
\arccos \Big[-\frac{W^{BB}_0}{W^{BA}_0}\frac{R^B}{R^A}\Big], 
\label{phipmB}
\\
\Omega
&=&\Omega_{\pm} = - \frac{R^A}{R^B} W^{BA}_0
\sin\Delta\phi^{AB}
=\mp \frac{R^A}{R^B}W^{BA}_0
\sqrt{1-\left(\frac{R^B}{R^A}\frac{W^{BB}_0}{W^{BA}_0}\right)^2}.
\label{OmegapmB}
\end{eqnarray}
The solution sets given by Eqs.~\eqref{phipmA}, \eqref{OmegapmA} and by Eqs.~\eqref{phipmB}, \eqref{OmegapmB} should be identical. 
Noting that $W^{AB}_0 W^{BA}_0<0$ and thus $\Omega_\pm$ in Eqs.~\eqref{OmegapmA} and \eqref{OmegapmB} have the same sign, we get the relation
\begin{widetext}
\begin{eqnarray}
\frac{W^{AA}_0}{W^{AB}_0}\frac{R^A}{R^B} 
&=& \frac{W^{BB}_0}{W^{BA}_0}\frac{R^B}{R^A}, \\
\frac{R^B}{R^A}W^{AB}_0\sqrt{1-\left(\frac{R^A}{R^B}\frac{W^{AA}_0}{W^{AB}_0}\right)^2}
&=&
-\frac{R^A}{R^B}W^{BA}_0
\sqrt{1-\left(\frac{R^B}{R^A}\frac{W^{BB}_0}{W^{BA}_0}\right)^2}.
\end{eqnarray}
\end{widetext}
Solving the above yields, 
\begin{eqnarray}
W^{AA}_0 &=& -W^{BB}_0 \equiv W_0,\\
R^A &=& \sqrt{-\frac{W^{AB}_0}{W^{BA}_0}}R^B 
= \sqrt{\left|\frac{W^{AB}_0}{W^{BA}_0}\right|}R^B,
\end{eqnarray}
giving,
\begin{eqnarray}
\Omega_\pm &=& \pm \sqrt{|W^{AB}_0 W^{BA}_0| - W_0^2}=\pm\frac{\sqrt{|\Lambda_0|}}{2}, 
\end{eqnarray}
and 
\begin{eqnarray}
\Delta\phi^{AB}
&=&
\arccos\left[\sqrt{\frac{W_0^2}{|W^{AB}_0W^{BA}_0|}}\right],
\end{eqnarray}
for $W_0 W_0^{AB}<0$, and for $W_0W_0^{AB}>0$,
\begin{eqnarray}
\Delta\phi^{AB}
&=&
\arccos\left[-\sqrt{\frac{W_0^2}{|W^{AB}_0W^{BA}_0|}}\right].
\end{eqnarray}
Hence, we arrive at the relations \eqref{chiral_ansatz}-\eqref{Deltaphi_supp}.

\subsubsection{The (anti)flocking to chiral phase transition as spontaneous PT symmetry breaking}
\label{sec_PTsymmetry}

In this subsection, we show that our system is PT symmetric (in the sense defined below) and the (anti)flocking-to-chiral phase transition can be regarded as an instance of spontaneous PT symmetry breaking, often discussed in the context of non-Hermitian quantum mechanics~\cite{Bender1998,Bender2007,Konotop2016,Mostafazadeh2002,Mostafazadeh2002b,Mostafazadeh2002c,Bender2002,Mostafazadeh2015,Bender2010}.
In the following, we consider the operations executed by the operators ${\mathcal P}$ and ${\mathcal T}$, which are the \textit{generalized} parity and time-reversal operator, respectively (i.e., they are not necessarily related to the physical parity and time-reversal operations). Here, ${\mathcal P}$ is defined to be a generic Hermitian and unitary operator and ${\mathcal T}$ is a generic antiunitary operator, expressible as $K$ times a unitary matrix (where $K$ is a complex conjugation) that satisfy
\begin{eqnarray}
{\mathcal P}^2={\mathcal T}^2=1,
	\qquad
[{\mathcal P},{\mathcal T}]=0, 
\label{generic PT}
\end{eqnarray}
as would the conventional parity and time-reversal operators. 

The system is said to be PT symmetric if we can find a ${\mathcal P}{\mathcal T}$ operator that commutes with the matrix $\hat W_0$ that controls the dynamics:
\begin{eqnarray}
[{\mathcal P}{\mathcal T},\hat W_0]=0.
\label{PT symmetric Hamiltonian}
\end{eqnarray}
The PT symmetry of a PT symmetric system is said to be unbroken if any eigenstate of the matrix $\hat W_0$ is simultaneously an eigenstate of the ${\mathcal P}{\mathcal T}$ operator. 
Otherwise, the PT symmetry is said to be spontaneously broken \cite{Konotop2016}.

We argue below that (i) our system is PT symmetric, (ii) the (anti)flocking phase is a PT unbroken phase, and (iii) the chiral phase is a PT broken phase. Hence, the (anti)flocking-to-chiral phase transition is an instance of spontaneous PT symmetry breaking. 

Our system is PT symmetric since we can find  operators ${\mathcal P}$ and ${\mathcal T}$ that satisfy Eq.~\eqref{PT symmetric Hamiltonian}. 
To see this explicitly, following Ref.~\cite{Mostafazadeh2002}, we express the matrix $\hat W_0$ and the operators ${\mathcal P}$ and ${\mathcal T}$ in terms of Pauli matrices,
\begin{eqnarray}
\hat W_0 &=& w_0^0 \bm 1 + \bm w_0\cdot \bm \sigma 
\\
{\mathcal P} &=& p^0 \bm 1 + \bm p\cdot \bm \sigma, \\
{\mathcal T} &=& K\sigma_2( t^0 \bm 1 + \bm t\cdot \bm \sigma),
\end{eqnarray}
where $\bm \sigma=(\sigma_1,\sigma_2,\sigma_3)^{\mathsf T}$ is a vector composed of Pauli matrices. 
From their definitions introduced above, the vectors $\bm p$ and $\bm t$ need to be real with
\begin{eqnarray}
p^0=t^0=0, \bm p\cdot\bm p=\bm t\cdot \bm t=1,
\bm p\cdot \bm t=0.
\end{eqnarray}
Further decomposing the vector $\bm w_0$ into real and imaginary part as $\bm w_0=\bm w_0^R + i\bm w_0^I$, we find~\cite{Mostafazadeh2002}
\begin{eqnarray}
{\mathcal P}{\mathcal T}\hat W_0
({\mathcal P}{\mathcal T})^{-1}-\hat W_0
=2\bm\sigma\cdot \bm F
-2i\bm \sigma\cdot \bm G
\label{PT criterion}
\end{eqnarray}
where
\begin{eqnarray}
\bm F&=&(\bm w_0^R\cdot\bm p)\bm p+(\bm w_0^R\cdot\bm t)\bm t-\bm w_0^R, 
\label{Fvector}\\
\bm G&=&(\bm w_0^I\cdot\bm p)\bm p+(\bm w_0^I\cdot \bm t)\bm t.
\label{Gvector}
\end{eqnarray}
The existence of $\bm p$ and $\bm t$ that makes the right-hand side of Eq.~\eqref{PT criterion} vanish means that the system is PT symmetric. 

We can indeed find such $\bm p$ and $\bm t$ for our system, by using the property that $\hat W_0$ is a real matrix that restrict $w_0^0$ to be real and the real vectors $\bm w_0^R,\bm w_0^I$ to take the form $\bm w_0^R=(w_0^1,0,w_0^3)^{\mathsf T}$ and $\bm w_0^I=(0,w_0^2,0)^{\mathsf T}$. Thus, 
\begin{eqnarray}
\bm w_0^R\cdot \bm w_0^I = 0.
\label{wRwI orthogonal}
\end{eqnarray}
By choosing 
\begin{eqnarray}
\bm p=\frac{\bm w_0^R}{|\bm w_0^R|},
\end{eqnarray}
since $\bm p\cdot\bm t=0$, we get $\bm w_0^R\cdot \bm t=0$. Plugging these into Eq.~\eqref{Fvector} yields $\bm F=0$.
Further, since $\bm p\propto \bm w_0^R$, Eq.~\eqref{wRwI orthogonal} gives
\begin{eqnarray}
\bm p\cdot \bm w_0^I = 0,
\end{eqnarray}
leading to $\bm G=(\bm w_0^I\cdot \bm t)\bm t$. 
Since $\bm w_0^I$ and $\bm t$ are in a plane orthogonal to $\bm p$, we can always find $\bm t$ that are also orthogonal to $\bm w_0^I$. Choosing such a vector $\bm t$, i.e.,
\begin{eqnarray}
\bm t
=\frac{\bm w_0^R\times\bm w_0^I}
{|\bm w_0^R\times\bm w_0^I|},
\end{eqnarray}
we get $\bm G=0$ and therefore the right-hand side of Eq.~\eqref{PT criterion} vanishes. Thus, our system is PT symmetric.   

Now we argue that the flocking and antiflocking phases are in a PT unbroken phase while the chiral phase is in a PT broken phase. 
In a PT symmetric system, the eigenstates $\bm u_\pm$, defined as states that satisfies 
\begin{eqnarray}
\hat W_0\bm u_\pm = \Gamma_\pm \bm u_\pm,
\label{eigenstate PT}
\end{eqnarray}
also satisfies
\begin{eqnarray}
\hat W_0({\mathcal P}{\mathcal T}  \bm u_\pm)
=\Gamma_\pm^*({\mathcal P}{\mathcal T}\bm u_\pm),
\label{eigenvalue PT unbroken}
\end{eqnarray}
showing that ${\mathcal P}{\mathcal T}\bm u_\pm$ is also an eigenstate of this system.
Here, we have operated ${\mathcal P}{\mathcal T}$ from the left and used Eq.~\eqref{PT symmetric Hamiltonian}.

In a PT unbroken phase, since ${\mathcal P}{\mathcal T}\bm u_\pm \propto\bm u_\pm$ and thus satisfies 
\begin{eqnarray}
\hat W_0  \bm u_\pm
=\Gamma_\pm^*\bm u_\pm,
\end{eqnarray} 
the eigenvalues of PT unbroken phase is real ($\Gamma_\pm = \Gamma_\pm^*$).
On the other hand, in a PT broken phase, ${\mathcal P}{\mathcal T}\bm u_{+(-)}$ is a distinct vector from $\bm u_{+(-)}$. Thus, the eigenstate with the eigenvalue $\Gamma_{+(-)}^*$ is a different state from that with $\Gamma_{+(-)}$.
Since there are at most two eigenvalues in our two component system, the two eigenvalues are complex conjugate of each other,
\begin{eqnarray}
\Gamma_+ = \Gamma_-^*.
\label{PT broken complex conjugate}
\end{eqnarray}

While the flocking and antiflocking  phase corresponds to real eigenvalues, the chiral phase corresponds to complex eigenvalues that are complex conjugate of each other, resulting in an oscillation in time. Hence, the former is in a PT unbroken phase while the latter is in a PT broken phase, marking the phase transition point as a PT symmetry breaking point.

\subsection{Mean-field phase diagram}

So far, we have observed how the flocking/antiflocking phase may be destabilized into a chiral phase, without paying too much attention to the concrete form of the matrix $\hat W_0$.
Here, we directly compute  analytically the mean-field equation~\eqref{TonerTuMF_supp} to show that the flocking, antiflocking, and the chiral phase predicted from the above analysis indeed arise. We determine the phase diagram in terms of the microscopic coupling strengths $j_{ab} (a,b=A,B)$.

Below, we parameterize
\begin{eqnarray}
j_\pm&=&\frac{1}{2}(j_{AB}\pm j_{BA}), 
\label{jpm}
\end{eqnarray}
for our convenience, where $j_+$ and $j_-$ characterize the reciprocal and non-reciprocal component of the coupling, respectively.

\subsubsection{Ordered-to-disordered phase transition}

Firstly, we analyze the ordered-to-disordered phase transition point within mean-field approximation.
Starting from the ordered phase  (i.e. flocking and antiflocking phase), the order parameter $\vec P_0^a$ approaches zero as moving towards the phase boundary.
Thus, the ordered-to-disordered phase transition point should satisfy, \begin{eqnarray}
&&{\rm det}\hat W_0 (\vec P^A_0\rightarrow 0,\vec P^B_0\rightarrow 0) 
=\Gamma_-(\vec P^A_0\rightarrow 0,\vec P^B_0\rightarrow 0)
\Gamma_+(\vec P^A_0\rightarrow 0,\vec P^B_0\rightarrow 0) = 0.
\end{eqnarray}
This can be solved analytically with the result (For simplicity, we assume below  $\rho^A=\rho^B=\rho$.),
\begin{eqnarray}
j_+^c=\pm\frac{\sqrt{\eta^2-\eta(j_{AA}+j_{BB})\rho+(j_{AA}j_{BB}+j_-^2)\rho^2}}
{\rho}.
\label{orderdisordered}
\end{eqnarray}

Note that, as discussed earlier, the ordered phase should be described as the stable "$-$" solution and not the unstable "$+$" solution. 
While the "$-$" solution satisfies  $\Gamma_-=0,\Gamma_+>0$, the "$+$" solution satisfies $\Gamma_-<0,\Gamma_+=0$. Thus, the sign of the average $(\Gamma_++\Gamma_-)/2$ indicates which type the obtained solution is.
Since
\begin{eqnarray}
&&\Gamma_\pm(\vec P^A_0\rightarrow 0,\vec P^B_0\rightarrow 0) 
=
-\frac{1}{2}[(j_{AA}+j_{BB})\rho-2\eta \pm\sqrt{\Lambda_0}],
\end{eqnarray}
the average of the two eigenvalues are given by
\begin{eqnarray}
\frac{\Gamma_+ + \Gamma_-}{2} 
= -\frac{1}{2}\big[
(j_{AA}+j_{BB})\rho-2\eta
\big].
\end{eqnarray}
Thus, Eq.~\eqref{orderdisordered} is valid only when
\begin{eqnarray}
\eta>(j_{AA}+j_{BB})\rho
\end{eqnarray}
is satisfied such that it describes the destabilization towards the stable "$-$" solution. 
We have drawn this ordered-to-disordered phase boundary in Fig. 2a in the main text (black line), giving an excellent agreement with the numerical result. 

\subsubsection{Exceptional point}

We next determine the exceptional point that marks the exceptional transition point from the (anti)flocking to a chiral phase for a given parameter set $(\rho^A,\rho^B,j_{AA},j_{BB})$. 
At the transition point, the following relations are satisfied:
\begin{eqnarray}
\Gamma_- &=& \bar W_0 - \frac{\sqrt\Lambda_0}{2} = 0, 
\label{steadyCEP}\\
\Lambda_0 &=&(\Delta W_0)^2 + 4 W_0^{AB}W_0^{BA} = 0,
\label{Lambda0CEP}\\
R^A &=& \left|\frac{\sqrt\Lambda_0-\Delta W_0}{2W_0^{BA}}\right|R^B,
\label{amplitudeCEP} 
\end{eqnarray} 
where $R^{a}=\absvec{\vec P^a_0}(>0)$, 
\begin{eqnarray}
\bar W_0 
&=& \frac{W_0^{AA}+W_0^{BB}}{2}
=  -\frac{1}{2}\big[
(j_{AA}+j_{BB})\rho-2\eta
-\frac{1}{2\eta}
(\absvec{\vec {Q} ^A_0}^2
+\absvec{\vec {Q} ^B_0}^2)
\big] ,
\\
\Delta W_0  
&=& \frac{W_0^{AA}-W_0^{BB}}{2} 
=  -\frac{1}{2}\big[
(j_{AA}-j_{BB})\rho-2\eta
-\frac{1}{2\eta}
(\absvec{\vec {Q} ^A_0}^2-\absvec{\vec {Q} ^B_0}^2)
\big], 
\end{eqnarray}
with
\begin{eqnarray}
\absvec{\vec {Q} ^A_0}^2
&=&\absvec{ j_{AA}\vec P_0^A+j_{AB}\vec P_0^B}^2, \\
\absvec{\vec {Q} ^B_0}^2
&=&\absvec{j_{BA}\vec P_0^A+j_{BB}\vec P_0^B}^2.
\end{eqnarray}
Here, Eq.~\eqref{steadyCEP} is the steady state condition, Eq.~\eqref{Lambda0CEP} is condition for the exceptional point, and Eq.~\eqref{amplitudeCEP} is derived from Eqs.~\eqref{vecPA0} and \eqref{vecPB0}. 
Eq. \eqref{Lambda0CEP} shows that non-reciprocal interaction with $W_0^{AB}W_0^{BA}<0$ (hence $j_{AB}j_{BA}<0$) is necessary for the chiral flocking phase to appear.
Below, we assume without loss of generality that the interaction are non-reciprocal with opposite sign, i.e. $j_{AB}j_{BA},W_0^{AB}W_0^{BA}<0$.

From Eqs.~\eqref{steadyCEP} and \eqref{Lambda0CEP},
\begin{eqnarray}
\Delta W_0 = \pm 2\sqrt{|W_0^{AB}W_0^{BA}|}.
\label{Lambda0CEPb}
\end{eqnarray}
Plugging this into Eq.~\eqref{amplitudeCEP}, we get
\begin{eqnarray}
R_B=\sqrt{\left|\frac{W_0^{BA}}{W_0^{AB}}\right|}R_A
=\sqrt{\left|\frac{j_{BA}\rho^B}{j_{AB}\rho^A}\right|}R_A. 
\label{RARB}
\end{eqnarray}

For the flocking phase, 
\begin{eqnarray}
\absvec{\vec {Q} ^A_0}^2
&=&(j_{AA}R_A+j_{AB}R_B)^2, \\
\absvec{\vec {Q} ^B_0}^2&=&(j_{BA}R_A+j_{BB}R_B)^2,
\end{eqnarray}
while for the  antiflocking phase, 
\begin{eqnarray}
\absvec{\vec {Q} ^A_0}^2&=&(j_{AA}R_A-j_{AB}R_B)^2, \\
\absvec{\vec {Q} ^B_0}^2&=&(j_{BA}R_A-j_{BB}R_B)^2.
\end{eqnarray}
We solve the coupled equations
\begin{eqnarray}
\bar W_0 &=& 0,
\label{W0=0}
\end{eqnarray}
Eq.~\eqref{Lambda0CEPb}, and Eq.~\eqref{RARB} for $R_A,R_B$ and the critical value $j_+^{EP}$ for a given $j_-$. 

We can solve the above equations analytically in the case $j_{AA}=j_{BB}(\equiv j)$, as we perform in the following.
We only consider here the flocking-to-chiral phase transition point, as the antiflocking-chiral transition point can be computed in a similar way. 
Using Eq.~\eqref{RARB} to eliminate $R_B$, Eq.~\eqref{Lambda0CEPb} yields the relation 
\begin{eqnarray}
R_A^2
=\frac{4\eta j_{AB}\sqrt{-j_{AB}j_{BA}}\rho}
{2 j j_{AB}(j_{AB}-j_{BA})\sqrt{-j_{BA}/j_{AB}}
+j^2(j_{AB}+j_{BA})-j_{AB}j_{BA}(j_{AB}+j_{BA})}.
\end{eqnarray}
We substitute this to Eq.~\eqref{W0=0} to get, 
\begin{equation}
\begin{split}
(j^2 +j_{AB} j_{BA})\big[j
(j_{AB}+j_{BA})
+ \sqrt{-j_{AB}j_{BA}}
(j_{AB}-j_{BA})
\big]
\rho \\
= \big[
(j^2 -j_{AB}j_{BA}) (j_{AB}+j_{BA})
+ 2j \sqrt{-j_{AB}j_{BA}}
(j_{AB}-j_{BA})
\big ]
\eta.
\end{split}
\end{equation}
This can be rewritten in terms of $j_\pm$ introduced in Eq.~\eqref{jpm}, 
\begin{eqnarray}
\label{equation_chiral_phase_visek_jpm}
(j^2+j_+^2-j_-^2)
(j j_+ + j_- \sqrt{j_-^2-j_+^2})\rho
=
(2j j_-\sqrt{j_-^2 - j_+^2}
+ j_+)
(j j_+ + j_- \sqrt{j_-^2-j_+^2})\eta,
\end{eqnarray}
which can be organized into a cubic equation in terms of $j_+^2$ as,
\begin{eqnarray}
a j_+^6 + b j_+^4 + c j_+^2 +d = 0
\end{eqnarray}
with
\begin{eqnarray}
a &=& (\eta+j\rho)^2 + j_-^2 \rho^2, \\
b &=& 2j^4 \rho^2 - 2\eta^2(j^2-j_-^2)
-8\eta j j_-^2 - 3 j_-^4\rho^2, \\
c &=& \eta^2 j^4 + 6\eta^2 j^2 j_-^2 + \eta^2 j_-^4 
-2\eta j^5 \rho - 4\eta j^3 j_-^2 \rho + 10\eta j j_-^4 \rho + j^6 \rho ^2 - j^4 j_-^2 \rho^2 -3j^2j_-^4 \rho ^2 +3j_-^6 \rho^2, \\
d&=&
j_-^4
[-4\eta^2 j^2(1+j\rho)-j^4\rho^2
-4\eta j\rho j_-^2 
+2 j^2\rho^2 j_-^2
-\rho^2 j_-^4].
\end{eqnarray}
This can be solved exactly  using Cardano's formula. Its real solution is plotted as a red line in Figs. 2b and 2c in the main text, giving an excellent agreement with the (anti)flocking/chiral phase transition lines obtained numerically.

\section{Excitation spectrum and stability analysis}
\label{sec_excitation_supp}

In this section, we provide a linear analysis on the fluctuations around the mean-field solution obtained in section~\ref{sec_mean_field} to study the  properties of excitations,  the stability of the phases described above,  as well as the nature of the phase transitions between them.
We confirm that there is a wide range of parameters where the flocking, antiflocking, and chiral phases are stable, as summarized in the phase diagram in Fig. 3a in the main text. 

We show from this analysis that the chiral phase emerges by the \textit{coalescence} of the collective eigenmodes in the transverse channel, which is a fundamentally different mechanism from the conventional phase transition scenarios~\cite{Hanai2020}. 
We further show that the emergence of the swap phase can be understood as the instability of the flocking/antiflocking phase against the global longitudinal fluctuations. 
Based on these results, we argue that it leads to the appearance of chiral-swap mixed phase exhibiting quasiperiodic oscillation in time, and explain the occurrence of tetracritical points with reduced codimension in the phase diagram. 
We also describe how the combination of exceptional points and the convective terms enforces the occurrence of the pattern-forming instabilities at the (anti)flocking-to-chiral phase transition.

\subsection{General considerations}
\label{linear_stability_general}

We assume the existence of a homogeneous solution $\psi(t) = (\rho_a(t), P_a(t))_a$ to the equation \eqref{eq_less_hideous}, which therfore also satisfies the mean-field Eq.~\eqref{TonerTuMF_supp}.
This equation is of the form
\begin{equation}
	\label{eom_ss_calW}
	\partial_t \psi(t,r) = f(\psi(t,r), \nabla \psi(t,r), \dots)
\end{equation}
The linear stability and the fate of small excitations (e.g. waves) on top of the steady-state are ruled by the linearized equation of motion for small perturbations $\delta \psi = \psi - \psi_{\text{ss}}$ on top of a steady-state $\psi_{\text{ss}}$. 
At first order in $\delta \psi$, the perturbations are described by the equation
\begin{equation}
	\partial_t \delta \psi(t,r) 
	=  f(\psi(t,r), \nabla \psi(t,r), \dots) -  f(\psi_{\text{ss}}(t), \nabla \psi_{\text{ss}}(t) \equiv 0, \dots)
	\simeq \hat{\mathcal{L}}(t) \delta \psi(t,r) + \mathcal{O}(\delta \psi^2)
\end{equation}
where $\hat{\mathcal{L}}(t)$ is a linear (differential) operator, that may depend on time through the steady-state $\psi_{\text{ss}}(t)$.
As this operator is linear, we use the Fourier transform to block-diagonalize the differential operators in momentum space, where we are left with a family of linear equations of the form
\begin{equation}
	\label{linear_system_stability_waves}
	\partial_t \delta \psi(t,k) = \mathcal{L}(t,k) \delta \psi(t,k)
\end{equation}
where $\mathcal{L}(t,k)$ are finite matrices.
In terms of the perturbations $\delta P^a_\mu$ and $\delta \rho^a$ of the polarization fields and density fields,
\begin{subequations}
\label{linear_system_stability_waves_explicit}
\begin{align}
	\partial_t \delta P^a_\mu &= [\mathcal{L}_{P P}]^{a b}_{\mu \nu} \delta P^b_\nu + [\mathcal{L}_{P \rho}]^{a b}_{\mu} \delta \rho^b \\
	\partial_t \delta \rho^a &= [\mathcal{L}_{\rho \rho}]^{a b} \delta \rho^b + [\mathcal{L}_{\rho P}]^{a b}_{\nu} \delta P^b_\nu.
\end{align}
\end{subequations}

Hence, the matrix elements of $\mathcal{L}$ are
\begin{equation}
\label{LPP_stability_operator}
\begin{split}
	[\mathcal{L}_{P P}]^{a b}_{\mu \nu} = 
	- D_{\theta} \delta_{a b} \delta_{\mu \nu}
	+ j_{a b} \rho^a \delta_{\mu \nu}
	- \frac{1}{2 D_{\theta}} \sum_{c,d,\rho} j_{a c} j_{a d} P^c_\rho P^d_\rho \delta_{a b} \delta_{\mu \nu}
	- \frac{1}{D_{\theta}} \sum_{c} j_{a b} j_{a c} P^a_\mu P^c_\nu  \\
	+ D_a (-k^2) \delta_{a b} \delta_{\mu \nu}  \\
	+ \sum_c \lambda_{a c} \Big[
		-   \ii k_\rho  P^c_\rho \delta_{a b} \delta_{\mu \nu} 
		+ 3  \ii k_\mu P^c_\nu  \delta_{a b} 
	    - 3  P^c_\mu \ii k_\nu  \delta_{a b} 
	\Big] \\
	+ \lambda_{a b} \Big[
		- 2  P^a_\mu \ii k_\nu  
		+ 2  \ii k_\mu P^a_\nu    
		- 2 P^a_\rho \ii k_\rho \delta_{\mu \nu} 
	\Big]
\end{split}
\end{equation}
and
\begin{align}
	[\mathcal{L}_{P \rho}]^{a b}_{\mu} &= \left[ (\sum_c j_{a c} P^c) - \frac{v_0^a}{2} \ii k \right] \delta_{a b} \\
	[\mathcal{L}_{\rho P}]^{a b}_{\mu} &= - \delta_{a b} v_0^b \ii k_\mu \\
	[\mathcal{L}_{\rho \rho}]^{a b} &= 0.
\end{align}

In the following, we neglect density fluctuations, setting $\delta \rho_a = 0$. Hence, we will only consider the stability of the polarization channel described by $\hat L=\mathcal{L}_{P P}$.
This can happen, for instance, when the system is incompressible (see Refs.~\cite{Chen2015,Chen2018} for an analysis of the consequences in the Toner-Tu model for a single population).
Here, we do not impose the divergence constraint $\div(\vec{v}_a) = 0$ that would usually arise from the continuity equation by integrating out high-frequency density fluctuations; instead, we assume that the continuity equation includes appropriate source terms so that density fluctuations can be traced out without producing constraints on the velocities. 
We refer to section~\ref{incompressibility_constraint} for a discussion of the situation in which the incompressibility constraint is imposed.

\subsection{Stability of the flocking/antiflocking phase}

We start with the stability analysis on the flocking and the antiflocking phase.
In these phases, the steady state solutions can be written as
\begin{eqnarray}
\vec P^{a}_0 = P^a_0 \; \vec{e}_\sslash. 
\end{eqnarray}
for $a=A,B$, where $P_0^a$ are real numbers and $\vec e_\sslash$ is a unit vector pointing at the flocking direction. Crucially $\vec{e}_\sslash$ is the same for $a=A$ and $a=B$. 
With this notation, $P^A_0 P^B_0 > 0 \, (<0)$ corresponds to the (anti)flocking phase.
It is convenient to decompose the fluctuations into longitudinal ($\sslash$) and transverse ($\perp$) components,  
\begin{eqnarray}
&&\delta \vec P^a(\bm r,t)
=\vec e_\sslash 
\delta P^a_\sslash(\bm r,t)
+\vec e_\perp \delta P^a_\perp (\bm r,t)
=P^a_0
\big[
\vec e_\sslash \delta n^a(\bm r,t)
+\vec e_\perp 
\delta \phi^a(\bm r,t)
\big],
\end{eqnarray}
where $\vec e_\perp$ is a unit vector perpendicular to the flocking direction $\vec e_\sslash$. In this way, $\delta n^a$ represents longitudinal fluctuations normalized by $P_0^a$, and $\delta\phi^a$ is the angle describing the deviation from the flocking direction.

The relevant part of the linearized version \eqref{linear_system_stability_waves} of the coupled Toner-Tu equations~\eqref{eom_with_gradients_two_populations_si} then reads, in Fourier space,
\begin{eqnarray}
\hat L(\bm k) 
d\vec y(\bm k)
= s(\bm k)
d\vec y(\bm k), 
\end{eqnarray}
where $\bm k=k_\sslash \vec e_\sslash+k_\perp \vec e_\perp$, $\hat L=\mathcal{L}_{P P}$ as defined in \eqref{LPP_stability_operator},
\begin{eqnarray}
d\vec y(\bm k)=
\left(\begin{array}{c}
\delta \phi^A(\bm k) \\
\delta \phi^B(\bm k) \\
\delta n^A(\bm k) \\ 
\delta n^B(\bm k) 
\end{array}\right),
\end{eqnarray}
and $\omega(\bm k)$ is the dispersion relation of a collective mode. 
We decompose the 4x4 matrix $\hat L(\bm k)$ controlling the linear excitations of the system as the block matrix
\begin{eqnarray}
\hat L(\bm k)
=\left(\begin{array}{cc}
\hat L_{\perp\perp}(\bm k) 
&\hat  L_{\perp\sslash}(\bm k) 
 \\
\hat L_{\sslash\perp}(\bm k) 
& \hat L_{\sslash\sslash}(\bm k) 
\end{array}\right),
\end{eqnarray}
where the blocks correspond to the transverse and longitudinal channels (blocks on the diagonal) and their coupling (off-diagonal). They have the explicit form
\begin{subequations}
\label{blocks_linear_op}
\begin{align}
\hat L_{\perp\perp}(\bm k)
 &=
-\left(\begin{array}{cc}
W^{AA}_0
& \frac{W^{AB}_0 P^B_0}{P^A_0}
 \\
\frac{W^{BA}_0 P^A_0}{P^B_0}
& W^{BB}_0
\end{array}\right)
-i\hat \lambda^{\perp\perp}k_\sslash
-
\left(\begin{array}{cc}
D_0^{A}  & 0 \\
0 & D_0^{B}
\end{array}\right)
\bm k^2,
\label{Lkperpperp}
\\
\hat L_{\perp\sslash}(\bm k) &=
i\hat \lambda^{\perp\sslash}k_\perp,
\\
\hat L_{\sslash\perp}(\bm k)
&=
-i\hat \lambda^{\sslash\perp}k_\perp,\\
\hat L_{\sslash\sslash}(\bm k) &=
-\left(\begin{array}{cc}
W^{AA}_0
& \frac{W^{AB}_0 P^B_0}{P^A_0}
 \\
\frac{W^{BA}_0 P^A_0}{P^B_0}
& W^{BB}_0
\end{array}\right)
-\left(\begin{array}{cc}
\frac{j^{AA}}{\eta}
P_0^A{Q}^A_0
&  
\frac{j^{AB}}{\eta}
P_0^B{Q}^A_0
\\
\frac{j^{BA}}{\eta}
P_0^A{Q}^B_0
& 
\frac{j^{BB}}{\eta}
P_0^B
{Q}^B_0
\end{array}\right)
-i\hat\lambda^{\sslash\sslash}k_\sslash
-
\left(\begin{array}{cc}
D_0^{A}  & 0 \\
0 & D_0^{B}
\end{array}\right)
\bm k^2,
\end{align}
\end{subequations}
where
\begin{subequations}
\begin{align}
{Q}^A_0  = j_{AA} P^A_0 +j_{AB} P^B_0 , 
\\
{Q}^B_0  = j_{BA}P^A_0 +j_{BB}P^B_0,
\end{align}
\label{underbarPsteadyexcitation}
\end{subequations} and 
$\hat\lambda^{\perp\perp},\hat\lambda^{\perp\sslash},\hat\lambda^{\sslash\perp}$,
and $\hat\lambda^{\sslash\sslash}$ are 2$\times$2 real matrices that originate from the convective terms, given by
\begin{eqnarray}
\hat \lambda^{\perp\perp}
=\hat \lambda^{\sslash\sslash}
&=&\begin{pmatrix}
3 P^A_0\lambda_{AA}+P^B_0 \lambda_{AB} 
& 2 P^B_0 \lambda_{AB} \\
2 P^A_0 \lambda_{BA}
& P^A_0\lambda_{BA}+3 P^B_0 \lambda_{BB}
\end{pmatrix}, \\
\hat \lambda^{\perp\sslash}
=
\hat \lambda^{\sslash\perp}
&=&\begin{pmatrix}
5 P^A_0 \lambda_{AA} + 3P^B_0 \lambda_{AB}
& 2 P^B_0 \lambda_{AB} \\
2 P^A_0 \lambda_{BA} 
& 3 P^A_0 \lambda_{BA} + 5 P^B_0 \lambda_{BB} 
\end{pmatrix}.
\end{eqnarray}
We use
\begin{equation}
	D_{a} = \frac{(v_0^a)^2}{16 \eta}
	\qquad
	\lambda_{a b} = v_0^a v_0^b
\end{equation}
in our analysis (see section~\ref{sec_hydro} for further discussion on this point).

\subsubsection{Computation of the stability regions of the phase diagram}

We determine the regions of stability of the (anti)flocking phases (Fig.~3a of the main text) by first solving numerically the mean-field dynamical system in Eq.~\eqref{TonerTuMF_supp} to obtain the quantities $R_{a} = \lVert \vec{P}_{a} \rVert$ ($a=A,B$) and $\Delta \phi^{A B} = \text{angle}(\vec{P}_A, \vec{P}_B)$.
This allows to identify the phase, and to obtain the matrix $\hat L(k)$ using Eq.~\eqref{LPP_stability_operator}. 
The eigenvalues $s_\alpha(k) = \sigma_\alpha(k) + \ii \omega_\alpha(k)$  of this matrix give the growth rates $\sigma_\alpha(k)$ and the frequencies $\omega_\alpha(k)$ of the perturbations. 
Because of the existence of a Goldstone mode, the largest growth rate is pinned to $\sigma=0$ at $k=0$. 
To evaluate the stability of the phase, we determine whether $k=0$ is a local maximum by computing the sign of the eigenvalues of the Hessian matrix of $\sigma(k)$ at $k = 0$, which is obtained by discretizing the second derivatives in momentum space.
The result is presented in Fig.~3a of the main text.
We have verified manually (by directly computing the growth rates as a function of $k$) that for a large range of parameters, an instability is present only if $k=0$ is not a local minimum. However, this hypothesis might fail in particular cases. 
This shortcoming will be addressed in future works by determining the stability regions directly from the full momentum dependent growth rates.
The largest growth rates as a function of the wavevector in Fig.~4b are directly obtained by diagonalizing $\hat L(k)$ at each point.

\subsubsection{Goldstone's theorem and destabilization towards the chiral phase}
\label{subsec_chiral}

Since the flocking and antiflocking phase are spontaneous symmetry broken phases,  Goldstone's theorem assures that at least one gapless eigenmode (i.e. the Goldstone mode) exists \cite{Altland2010}. This can be shown explicitly as follows. Using the steady state mean-field equation \eqref{TonerTuMFsteady}, or 
\begin{eqnarray}
W^{AA}_0
&=&
-\frac{W^{AB}_0P^B_0}{P^A_0}
=\frac{j^{AB}\rho^A P^B_0}{P^A_0},
\label{WAAjAA}
\\
W^{BB}_0
&=&
-\frac{W^{BA}_0P^A_0}{P^B_0}
=\frac{j^{BA}\rho^BP^A_0}{P^B_0},
\label{WBBjBB}
\end{eqnarray}
the transverse-transverse  block of the dynamical matrix $\hat L$ at $\bm k=0$ can be simplified to 
\begin{eqnarray}
\nonumber\\
\hat L_{\perp\perp}(\bm k={\bf 0})=
-\left(\begin{array}{cc}
W^{AA}_0
& -W^{AA}_0
 \\
-W^{BB}_0
& W^{BB}_0
\end{array}\right).
\end{eqnarray}
Then, it is readily shown, by noting that the transverse and the longitudinal fluctuations decouple at $\bm k={\bf 0}$ since $\hat L_{\perp\sslash}(\bm k={\bf 0})=L_{\sslash\perp}(\bm k={\bf 0})=0$, that the vector 
\begin{eqnarray}
d\vec y(\bm k={\bf 0})=
\left(\begin{array}{c}
\delta \phi^A({\bf 0}) \\
\delta \phi^B({\bf 0}) \\
\delta n^A({\bf 0})  \\
\delta n^B({\bf 0}) 
\end{array}\right)
=
\left(\begin{array}{c}
1 \\
1 \\
0  \\
0
\end{array}\right),
\label{inphase}
\end{eqnarray}
which corresponds to the global in-phase rotation of the direction of the flocks, is a gapless mode: 
\begin{eqnarray}
s(\bm k={\bf 0})=0.
\label{Goldstone mode spectrum}
\end{eqnarray} 

We also find another eigenmode
\begin{eqnarray}
d\vec y(\bm k={\bf 0})=
\left(\begin{array}{c}
\delta \phi^A({\bf 0}) \\
\delta \phi^B({\bf 0}) \\
\delta n^A({\bf 0})  \\
\delta n^B({\bf 0}) 
\end{array}\right)
=
\left(\begin{array}{c}
1 \\
-W_0^{BB}/W_0^{AA} \\
0  \\
0
\end{array}\right),
\label{gappedmodetransverse}
\end{eqnarray}
associated with a gapped spectrum
\begin{eqnarray}
s(\bm k={\bf 0})=-
(W_0^{AA}+W_0^{BB})
=\sigma_{\perp\perp},
\label{gapped spectrum}
\end{eqnarray}
where $\sigma_{\perp\perp}$ is a growth rate of this mode.
Since the "$-$" solution satisfies (note that $\Lambda_0>0$ in the flocking and antiflocking phase)
\begin{eqnarray}
0=\Gamma_- =\frac{1}{2}(-\sigma_{\perp\perp}-\sqrt{\Lambda_0}),
\end{eqnarray}
we get 
\begin{eqnarray}
\sigma_{\perp\perp}=-\sqrt{\Lambda_0}<0,
\label{decay rate gapped -}
\end{eqnarray}
meaning that that the "$-$" solution is stable, at least against the global transverse fluctuations involving a change in the relative flocking angle between the two species.
In contrast, the "$+$" solution that satisfies, 
\begin{eqnarray}
0=\Gamma_+ =\frac{1}{2}(-\sigma_{\perp\perp}+\sqrt{\Lambda_0}),
\end{eqnarray}
gives a positive growth rate of the mode, 
\begin{eqnarray}
\sigma_{\perp\perp}=\sqrt{\Lambda_0}>0,
\end{eqnarray}
thus is always unstable. 
This tells us that the only possible stable solution of the flocking and antiflocking phase is the "$-$" solution. 

We remark that the decay rate (at $\bm k=0$) of the two collective transverse modes obtained here, i.e. $\gamma_{\perp\perp}=-\sigma_{\perp\perp} = 0$ for the Goldstone mode and $\gamma_{\perp\perp}=\sqrt{\Lambda_0}$ for the gapped mode (See Eqs.~\eqref{Goldstone mode spectrum}, \eqref{gapped spectrum} and \eqref{decay rate gapped -}.), are identical to the eigenvalues of the matrix $\hat W_0$~\eqref{TonerTuMFsteady}, 
\begin{eqnarray}
\Gamma_- = 0, 
\qquad 
\Gamma_+ = \sqrt{\Lambda_0}. 
\end{eqnarray}
Since the exceptional point of $\hat W_0$ (which satisfies $\Lambda_0=0$) marks the phase transition point from the (anti)flocking phase to the chiral phase (as shown in Sec. \ref{sec emergence of chiral phase}), 
the above property shows that this point is simultaneously an exceptional point of $\hat L_{\perp\perp}(\bm k=0)$.
At the transition point, 
similarly to the conventional phase transition, the damping gap $\gamma_{\perp\perp}\equiv-\sigma_{\perp\perp}=\sqrt{\Lambda_0}$ of the "$-$" solution vanishes signaling the instability in the transverse channel to the chiral phase. 
However, there is a fundamental difference; the eigenvector given in Eq.~\eqref{gappedmodetransverse} \textit{coalesces} with the Goldstone mode given by Eq.~\eqref{inphase} ~\cite{Hanai2020}.
This mechanism by which the (anti)flocking phase gets destabilized into the chiral phase by the coalescence of a transverse fluctuations mode with the Goldstone mode is a mechanism unique to non-equilibrium systems where the dynamics is controlled by non-Hermitian matrices, and is in sharp contrast with the conventional phase transition mechanism where the gapped mode simply softens at the critical point but are still orthogonal to other eigenmodes.

The above coincidence of the eigenvalues of $\hat W_0$ and the decay rate of the transverse modes obtained from $\hat L_{\perp\perp}(\bm k=0)$ is due to the property that $\hat W_0$ is invariant under the perturbation (where $\phi^a$ is the steady-state orientation angle of the polarization)
\begin{eqnarray}
\phi^A \rightarrow \phi^A+\delta\phi^A,
\qquad
\phi^B \rightarrow \phi^B+\delta\phi^B
\end{eqnarray}
in the linear order, i.e.
\begin{eqnarray}
\hat W_0(\phi^A+\delta\phi^A,\phi^B+\delta\phi^B)
=\hat W_0(\phi^A,\phi^B)
+O((\delta\phi^{A})^2,\delta\phi^{A}\delta\phi^{B},(\delta\phi^{B})^2).
\end{eqnarray}
It follows from this that  linearizing the mean-field equation~\eqref{TonerTuMF_supp} in the transverse channel gives
\begin{eqnarray}
\partial_t 
\begin{pmatrix}
\delta \phi^A \\
\delta \phi^B \\
\end{pmatrix}
=-\hat W_0
\begin{pmatrix}
\delta \phi^A \\
\delta \phi^B \\
\end{pmatrix}.
\end{eqnarray}
Therefore, the two collective transverse modes are computed as,
\begin{eqnarray}
s=-\Gamma_-, 
\qquad
s=-\Gamma_+,
\end{eqnarray}
which must be identical to the ones obtained from $\hat L_{\perp\perp}(\bm k=0)$. 

\subsubsection{Destabilization towards the swap phase and the emergence of the active time-quasicrystal}
\label{subsec_swap}

In this paragraph, we show that the longitudinal fluctuations can also get destabilized by the non-reciprocal interaction implying a phase transition to the swap phase (see SI Movie 2). 

The existence of two independent mechanisms of destabilization, with one in the transverse channel (that leads to the chiral phase) and one in the longitudinal channel (that leads to the swap phase), suggests the existence of a phase where both channels destabilize. 
As we show in Figs.~2b, 2c and 2g of the main text, a mixed chiral-swap phase indeed exists, as a result of the simultaneous occurrence of these two instabilities.

We focus on the uniform longitudinal fluctuations in this channel, described by the linear operator
\begin{eqnarray}
\hat L_{\sslash\sslash}(\bm k={\bf 0})&=&
-\left(\begin{array}{cc}
W^{AA}_0
& -W^{AA}_0
 \\
-W^{BB}_0 & W^{BB}_0
\end{array}\right)
-\left(\begin{array}{cc}
\frac{j^{AA}}{\eta}
P_0^A{Q}^A_0
&  
\frac{W^{AA}}{\rho^A \eta}
P_0^A{Q}^A_0
\\
\frac{W^{BB}}{\rho^B \eta}
P_0^B{Q}^B_0
& 
\frac{j^{BB}}{\eta}
P_0^B
{Q}^B_0
\end{array}\right),
\end{eqnarray}
where we have used the steady state condition \eqref{TonerTuMFsteady} or \eqref{WAAjAA}, \eqref{WBBjBB}. 
The collective eigenmodes are given by 
\begin{eqnarray}
s^{\sslash\sslash}_\pm({\bf 0})
=-\bigg[
\zeta_\sslash 
\pm \sqrt{\zeta_\sslash^2-\eta_\sslash}
\bigg],
\label{longitudinal gap}
\end{eqnarray}
where
\begin{eqnarray}
\eta_\sslash
&=&
\frac{4}{\eta}
\Big[
j^{AA}W^{BB}_0 P^A_0
{Q}^A_0
+j^{BB}W^{AA}_0 P^B_0
{Q}^B_0
+W^{AA}_0W^{BB}_0
(P^A_0
{Q}^A_0
+P^B_0
{Q}^B_0)
+\mu_\sslash
P^A_0
{Q}^A_0
P^B_0
{Q}^B_0
\Big],
\label{etaparallel}
\\
\mu_\sslash
&=&
\frac{1}{\eta}
\bigg[
j^{AA}j^{BB}
-\frac{W^{AA}_0W^{BB}_0}{\rho^A\rho^B}
\bigg]
=\frac{1}{\eta}
\bigg[
j^{AA}j^{BB}
-
\big(
j^{AA}-\frac{\eta}{\rho^A}
-\frac{(Q ^A_0)^2}{2\eta\rho^A}
\big)
\big(
j^{BB}-\frac{\eta}{\rho^B}
-\frac{(Q ^B_0)^2}{2\eta\rho^B}
\big)
\big)\bigg]
>0, \\
\zeta_\sslash
&=&W^{AA}_0+W^{BB}_0
+\frac{1}{\eta}
\big[
j^{AA}P^A_0 {Q}^A_0
+j^{BB}P^B_0 {Q}^B_0
\big].
\end{eqnarray}
The inequality $\mu_\sslash>0$ is shown under the assumption  $j^{AA},j^{BB}>0$.
Equation \eqref{longitudinal gap} shows that this mode is (un)stable when $\eta_\sslash>0(<0)$, because the largest growth rate of this channel is given by
\begin{eqnarray}
\sigma_{\sslash\sslash}
\equiv
{\rm max}\big[
{\rm Re}[s_-(\bm k={\bm 0)},
{\rm Re}[s_+(\bm k={\bm 0})]
\big]
=\sqrt{\zeta_\sslash^2-\eta_\sslash}
-|\zeta_\sslash|
<0(>0).
\end{eqnarray}

\begin{figure}
  \centering
  \includegraphics[width=8cm]{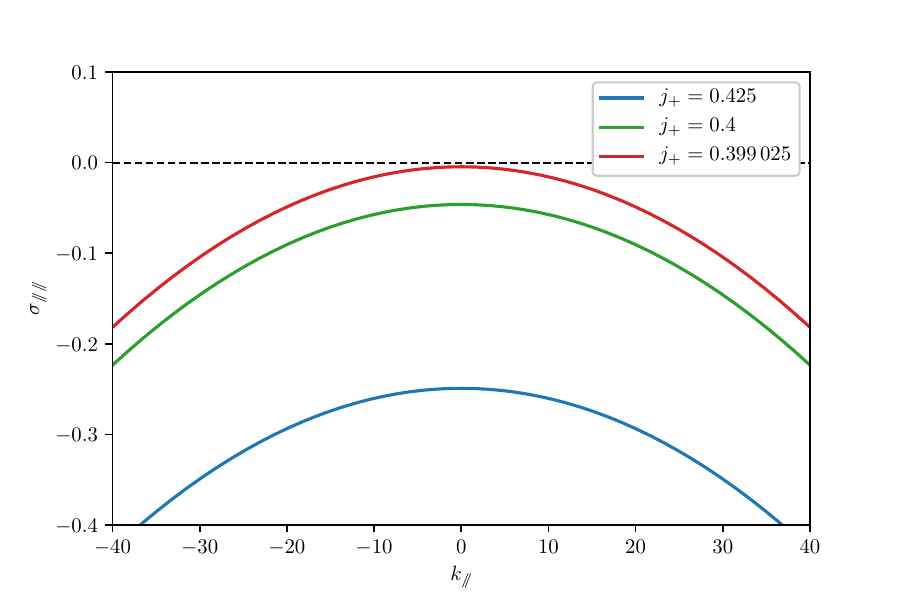}
  \caption{\label{si_figure_growth_rate_swap_transition}\strong{Growth rate for the longitudinal perturbations near the swap transition.}
  We have used $j_{A A} = j_{B B} = \num{1}$, $\rho_A = \rho_B = 1$, $j_- = \num{0.75}$, $\eta = \num{0.5}$, $v_0^A = \num{0.01}$ and $v_0^A = \num{0.03}$. Here $j_+^{\text{swap}} \approx \num{0.399}$ is the flocking-to-swap phase transition point. 
}
\end{figure}

In the case where the inter-species coupling has the same sign $j^{AB}j^{BA}>0$ (which includes the reciprocal case $j^{AB}=j^{BA}$), $\eta_\sslash$ is always positive and thus stable. 
This is seen as follows. 
When $j^{AB},j^{BA}>0(<0)$, the system is in a (anti)flocking phase at low enough noise strength, giving $P^A_0 P^B_0>0(<0)$. 
Then, from Eqs.~\eqref{WAAjAA} and \eqref{WBBjBB}, we get $W^{AA}_0,W^{BB}_0>0$.
Similarly, from the definition of ${Q}^a_0$ given by Eq.~\eqref{underbarPsteadyexcitation}, $P_0^{A(B)}$ and ${Q}^{A(B)}_0$ have the same signs irrespective of the sign of the inter-species coupling $j^{AB},j^{BA}$, i.e. $P_0^a {Q}^a_0>0$.
Since all the terms contributing to $\eta_\sslash$ in Eq.~\eqref{etaparallel} is thus positive as long as $j^{AB}$ and $j^{BA}$ has the same sign, the parameter  $\eta_{\sslash}$ is positive and thus stable 
($\sigma_{\sslash\sslash}<0$).

This restriction is lifted in the strong non-reciprocal case where the inter-species interaction have opposite signs $j^{AB}j^{BA}<0$.
In particular, $W^{AA}_0,W^{BB}_0,P_0^a{Q}^a_0$ may all become negative. 
Thus, as the non-reciprocal interaction increases,  $\eta_\sslash$ gets smaller and smaller until it approaches zero, or equivalently, $\sigma_{\sslash\sslash}=0$. 
Such softening of the  longitudinal mode implies a phase transition to a swap phase.

This scenario is confirmed in Fig.~\ref{si_figure_growth_rate_swap_transition}, which shows the growth rate $\sigma_{\sslash\sslash}$ of the collective mode in this channel, in the vicinity of the phase transition point to the swap phase. 
As expected, the negative growth rate $\sigma_{\sslash\sslash}$ in the flocking phase softens (i.e. $\sigma_{\sslash\sslash}={\rm Re} s_+(\bm k=0)$ approaches zero) at the flocking-to-swap phase transition point ($j_+=j_+^{\text{swap}}$).

\subsubsection{Tetracritical point}
\label{subsec_tetracritical}

\begin{figure}
  \centering
  \includegraphics[width=8cm]{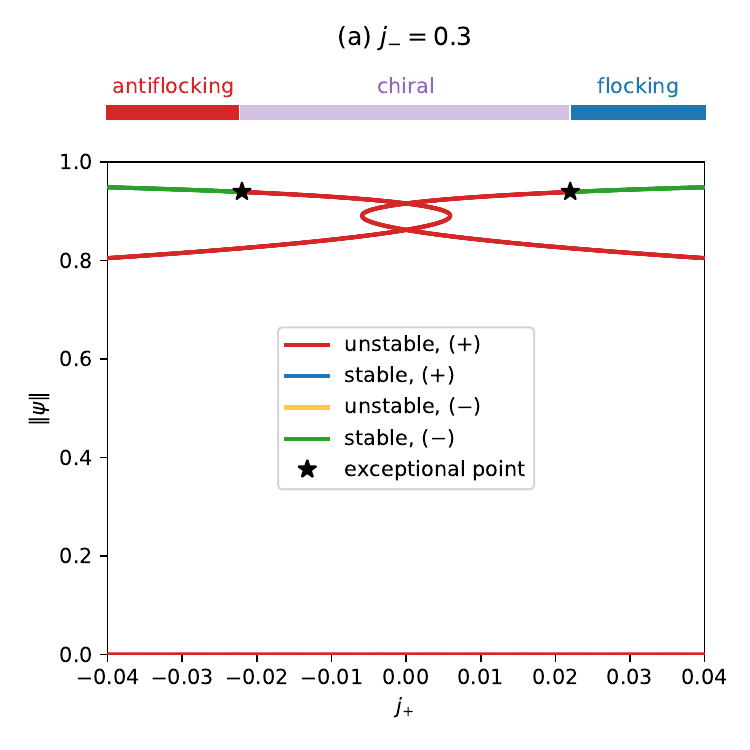}
  \includegraphics[width=8cm]{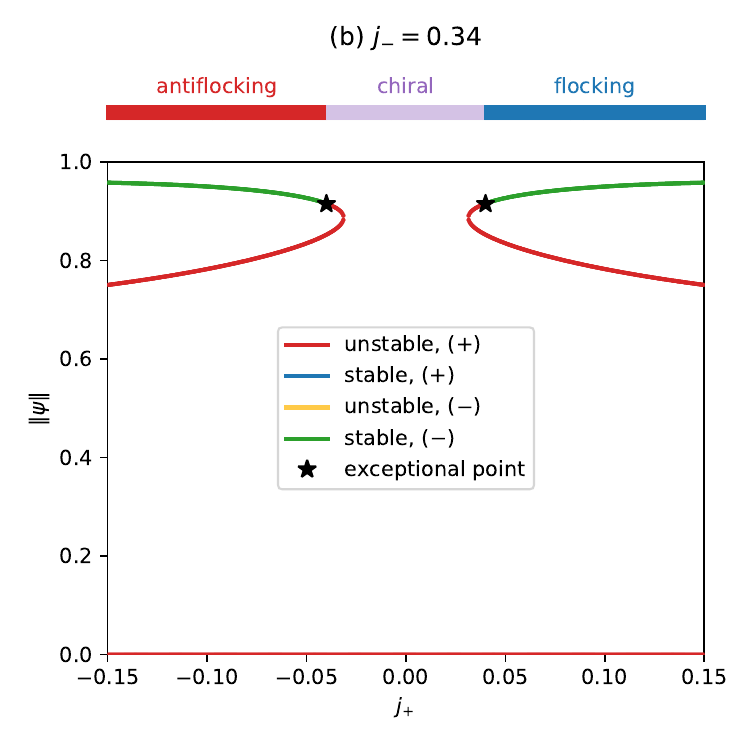}
  \includegraphics[width=8cm]{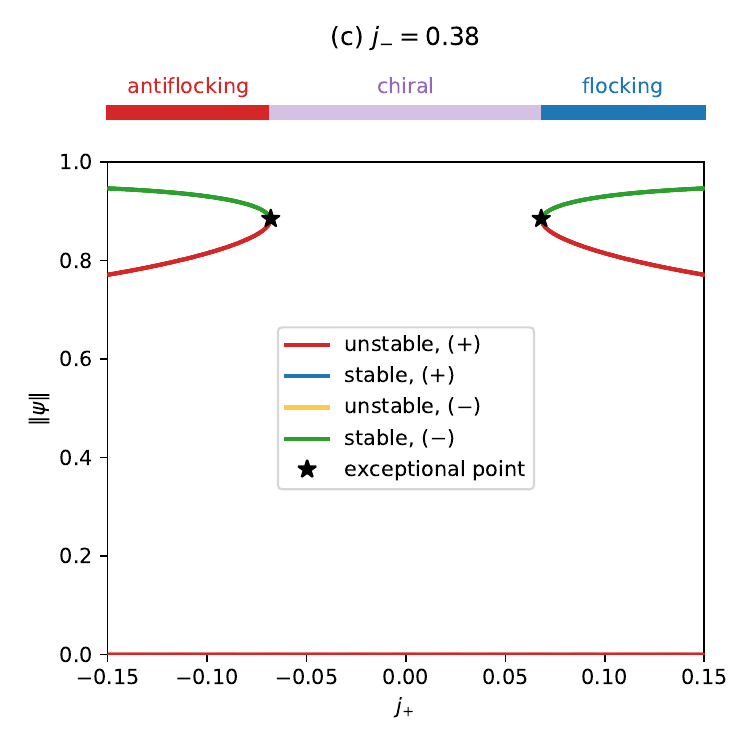}
  \includegraphics[width=8cm]{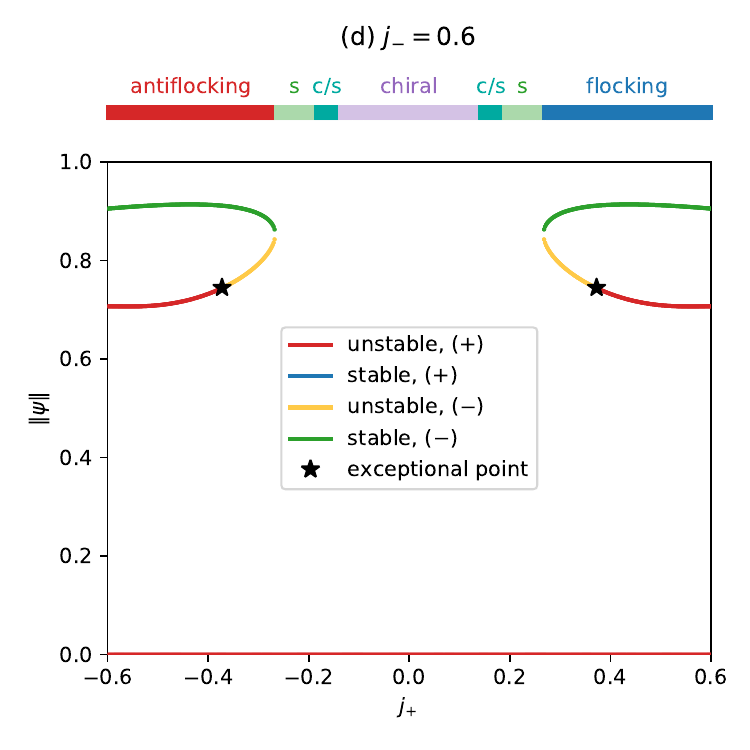}
  \caption{\label{bifurcation_diagrams}\strong{Bifurcation diagrams.}
  Unless otherwise specified, the parameters are the same as in Fig. 2c of the main text.
  An exceptional point (marked by a black star) separates the branches labelled $(+)$ and $(-)$. 
  The tetracritical points are found in the panel c and Fig.~2c in the main text at $(j_-^*,j_+^*) \approx (\pm 0.38,\pm 0.07)$. Note that the x-axis scales are different in the four panels.
  }
\label{fig_bifurcation}
\end{figure}

Here we argue that, from the properties shown above, a tetracritical point, which is the point where the four phases (i.e. the (anti)flocking, chiral, swap, and the chiral-swap phase) meet at a single point, naturally emerges at least within the mean-field approximation.
This is demonstrated in Fig. 2c in the main text, where the tetracritical points are marked by black dots.

The properties that we have shown so far can be
summarized as follows:
\begin{enumerate}
\item 
In the (anti)flocking phase, there exists two types of solutions, namely, the "$-$" and "$+$" solutions, but the "$+$" solution is always unstable. 
On the other hand, as long as the system is away from the exceptional point $\Lambda_0>0$, the "$-$" solution  is always stable against the global ($\bm k=0$) transverse channel fluctuations.

\item
Approaching the exceptional point $\Lambda_0=0$ (which is the point where the "$-$" and the "$+$" solution coalesces), however, the (anti)flocking phase  destabilizes in the transverse channel, signaling the phase transition to the chiral phase.

\item Independently from the above mechanism, "$-$" solution can also destabilize in the global longitudinal channel fluctuations, implying the phase transition into the swap phase.

\item In the parameter region between the chiral and swap phase, a mixed chiral-swap time quasicrystal phase emerges.
\end{enumerate}

The properties 1 and 2 show that the (anti)flocking-to-chiral phase transition always occurs at the many-body exceptional point.
As an example of such a situation, in Figs. \ref{fig_bifurcation}a and b, we have plotted the amplitude of the polarization $\absvec{\psi}$ (where $\psi=(\vec P^{A{\mathsf T}}_0,\vec P^{B{\mathsf T}}_0)^{\mathsf T}$) for both the stable and unstable solutions in the (anti)flocking phase with $j_-<j_-^*$.
As seen in the figure, the stable "$-$" solution and the unstable "$+$" solution coalesce at the exceptional point, which marks the phase transition point from the (anti)flocking phase to the chiral phase. 

On the other hand, property 3 shows that there are cases where the "$-$" solution of the (anti)flocking phase exhibits   phase transition to the swap phase. 
The property that this phase transition is associated with the destabilization of the longitudinal channel implies that the "susceptibility" of the amplitude of the polarization
(i.e. the sensitivity of the amplitude of the polarization against a parameter change)
diverges.
This is indeed seen in the region $j_->j_-^*$ in Fig. \ref{fig_bifurcation}d at the (anti)flocking-to-swap phase transition point, where the derivative of the amplitude $\absvec{\psi}$ in the stable branch of the "$-$" solution diverges, i.e.
$\left|\, \partial \absvec{\psi}/\partial j_+ \, \right|
\rightarrow \infty$ at the transition point. 
In this case, the exceptional point sits in the unstable branch of solutions. 

At $j_-=j_-^*$, these two types of transition points (i.e. the exceptional point and the diverging susceptibility point) merges at $j_+=j_+^*$ as shown in Fig. \ref{fig_bifurcation}c.
Since these are the signals of the transition to the chiral and swap phase, respectively, and keeping in mind that the chiral/swap time quasicrystal phase occurs in the region between the chiral and the swap phase (property 4), the point $(j_-^*,j_+^*)$ is nothing but the tetracritical point. 

\subsubsection{Exceptional point enforced pattern formation}
\label{subsec exceptional point instability}

The chiral and swap phases appear in mean-field theory as spatially uniform instabilities of the (anti)flocking phases. 
Now we show that the flocking and antiflocking phases generically exhibit a finite wavelength instability in the transverse fluctuation channel in the vicinity of the exceptional point (except in highly fine-tuned situations), implying the occurrence of pattern formation. This originates from the singular behavior of the mean-field operator at this point combined with the presence of convective terms.
At the exceptional point $\Lambda_0=0$ of the matrix $\hat W_0$,
\begin{equation}
0=\Gamma_-=\Gamma_+=\frac{1}{2}(W_0^{AA}+W_0^{BB}),
\end{equation}
and hence, $W^{AA}_0=-W^{BB}_0 (\equiv W_0)$.
As a consequence, the block $\hat L_{\perp\perp}(\bm k)$ corresponding to transverse fluctuations in Eq.~\eqref{blocks_linear_op} reduces to
\begin{eqnarray}
\hat L_{\perp\perp}(\bm k)
&=&
-i\left(\begin{array}{cc}
W_0 & -W_0
 \\
W_0 & -W_0
\end{array}\right)
-i\hat \lambda^{\perp\perp}k_\sslash
-
\left(\begin{array}{cc}
D_0^{A}  & 0 \\
0 & D_0^{B}
\end{array}\right)
\bm k^2.
\label{LperpperpEP}
\end{eqnarray}
The matrix $\hat L_{\perp\perp}(\bm k)$ has an exceptional point at $\bm k=0$, where the two collective modes given by Eqs.~\eqref{inphase} and \eqref{gappedmodetransverse}) coalesce~\cite{Hanai2020}.
Below, we restrict ourselves to $k_\perp=0$, where the transverse fluctuations decouple from the longitudinal mode such that the eigenvalues of $\hat L_{\perp\perp}(\bm k=k_\sslash \vec e_\sslash)$ describe the collective modes of the system in this channel (i.e. $\hat L_{\perp\sslash}(\bm k=k_\sslash \vec e_\sslash)
=\hat L_{\sslash\perp}(\bm k=k_\sslash \vec e_\sslash)=0$).
The complex growth rates ($s = \sigma + i \omega$ including the frequencies $\omega$ as imaginary parts) of the collective modes are
\begin{eqnarray}
s_\pm^{\perp\perp}(k_\sslash \vec e_\sslash)=
\pm i \sqrt{iC_0 k_\sslash
+O (k_\sslash^2)}
-i \lambda_0^{\perp\perp} k_\sslash
-D_0 k_\sslash^2.
\label{dynamicalinstabilityEP}
\end{eqnarray}
Here,
\begin{equation}
C_0 = -2W_0(\lambda_2^{\perp\perp}+\lambda_3^{\perp\perp})
\qquad
\text{and}
\qquad
D_0 = \frac{D_0^A+D_0^B}{2},
\end{equation}
where $\hat\lambda^{\perp\perp}$ is parameterized as
\begin{eqnarray}
\hat \lambda^{\perp\perp}=\lambda_0^{\perp\perp} \hat \sigma_0
+
\lambda_1^{\perp\perp} \hat \sigma_1
+
i\lambda_2^{\perp\perp} \hat \sigma_2
+
\lambda_3^{\perp\perp} \hat \sigma_3,
\end{eqnarray}
where $\hat\sigma_\alpha$ ($\alpha=0,1,2,3$) are the Pauli matrices, and $\lambda^{\perp\perp}_\alpha$ are real numbers since $\hat \lambda^{\perp\perp}$ is a real matrix.

The leading order contribution in respect to $k_\sslash$ has a singular form  $s_\pm^{\perp\perp}\sim\pm i\sqrt{iC_0 k_\sslash}$ as long as $C_0\ne 0$.
As a result, since the quantity inside the square root is purely imaginary,
at least one of the two modes is inevitably unstable, irrespective of the sign of $C_0$.
This shows that in the vicinity of a phase transition point from a (anti)flocking to the chiral phase, the uniform state is always destabilized by transverse fluctuations.  

This originates from the exceptional point physics in the presence of convective terms. Typically, the eigenvalues in the vicinity of the exceptional point behave as $s_\pm \sim \pm i\sqrt{\Delta}$, where $\Delta$ is a characteristic distance from the exceptional point. 
In our situation, recalling that $\hat L_{\perp\perp}(\bm k={\bf 0})$ is at an exceptional point, the finite momentum $\bm k$ contribution to $\hat L_{\perp\perp}(k_\sslash \vec e_\sslash)$ can be regarded as the contributions that makes $\hat L_{\perp\perp}(k_\sslash \vec e_\sslash)$  deviate from that point.
Because of the presence of the convective term $\hat \lambda^{\perp\perp}k_\sslash$, the leading order is $O(k_\sslash)$. As a result, $\Delta\sim i k_\sslash$, leading to $s_\pm\sim \pm i\sqrt{i k_\sslash}$, so at least one of the complex growth rates has a positive real part, implying an instability leading to pattern formation. The uniform flocking phase may be stabilized by moving away from the exceptional point.

\subsection{Floquet stability analysis of the chiral phase}

Here we provide a stability analysis on the chiral phase. 
As the mean-field solution in this phase depends on time, the method used to analyze the stability of the (anti)flocking phases cannot be directly applied.
Instead, we take advantage of the periodicity in time of the mean-field solution $\vec P^a(t)$ given by Eq.~\eqref{chiral_ansatz} we perform a Floquet stability analysis of the phase~\cite{Floquet1883,Lyapunov1907,Teschl2012} (see also Ref.~\cite{Barkley1996} for an example in the context of fluid mechanics).
We note that the transformation to the rotating frame does not completely eliminate the time dependence, due to the presence of convective terms.

\def\kk{\bm k}
The mean-field solution in the chiral phase ($\vec P^a(t)$ in Eq.~\eqref{chiral_ansatz}) is $T$-periodic in time with a period $T= 2\pi/\Omega$, where $\Omega$ is given by Eq.~\eqref{chiral_omega_from_params}.
Specializing the relevant part of Eq.~\eqref{linear_system_stability_waves} to this case, we obtain
\begin{equation}
\label{stability_floquet}
\partial_t \delta\vec P(t,\kk) = L(t, \kk) \delta\vec P(t, \kk)
\end{equation}
where crucially $L(t, \kk)$ is $T$-periodic, namely $L(t + T, \kk)=L(t, \kk)$.
This equation is formally solved by the evolution operator $U(t)$ such $\delta \vec P(t, \kk) = U(t) \delta \vec P(0, \kk)$ is a solution of \eqref{stability_floquet} for any initial condition $\delta \vec P(0, \kk)$ (in the context of ODEs, the evolution operator is called the principal fundamental matrix solution of the differential equation).
This evolution operator is the time-ordered exponential
\begin{equation}
	U(t, \kk) = T\exp\left( \int_0^t L(\kk,s) \dd s \right)
\end{equation}
which can also be written as the infinite series
\begin{equation}
	\label{evolution_operator_series}
	U(t, \kk) = \lim_{\delta t/t \to 0} \prod_{n=1}^{\lfloor t/\delta t \rfloor} \exp \left( n \delta t \, L(\kk, n \delta t) \right).
\end{equation}
Assuming the periodicity of $L(t) = L(t+T)$, the Floquet theorem \cite{Floquet1883,Teschl2012} implies that the evolution operator satisfies (pointwise in $\kk$)
\begin{equation}
	\label{evolution_operator_periodicity}
	U(t+nT) = U(t) [U(T)]^{n}
\end{equation}
The evolution operator evaluated after one period $U(T)$ is called the Floquet operator (also known as the monodromy matrix).
It is convenient to write it as $U(T) = \ee^{T L^{\text{eff}}}$ (where $L^{\text{eff}}$ is defined up to a choice of branch cut for the logarithm).
The property \eqref{evolution_operator_periodicity} allows to decompose the evolution operator as $U(t) = V(t) \ee^{t L^{\text{eff}}}$ where $V(t) = V(t+T)$ is periodic in time. 
This decomposition shows that crucially, the long-time behavior of the solutions of Eq.~\eqref{stability_floquet}, and in particular their stability, is fully controlled by the Floquet operator~$U(T)$.

Indeed, any solution of \eqref{stability_floquet} can be decomposed into fundamental solutions $\delta \vec{P}_\alpha(\kk,t)$, which are obtained from the eigenvalues $\lambda_\alpha(\kk)$ and eigenvectors $\delta \vec{P}^0_\alpha(\kk)$ of $U(T, \kk)$ as
\begin{equation}
\label{floquet_fundamental_solutions}
\delta \vec{P}_\alpha(\kk,t) = (\lambda_\alpha)^t \, \vec{p}_\alpha(t, \kk)
\end{equation}
where $\vec{p}_\alpha(t, \kk) = V(t) \delta \vec{P}^0_\alpha(\kk) = p_\alpha(t+T, \kk)$ is periodic in time.
The eigenvalues $\lambda_\alpha(\kk)$ of $U(T, \kk)$ are called Floquet multipliers (or characteristic multipliers), and they can be written as $\lambda_\alpha(\kk) = \ee^{s_\alpha({\kk})}$ where $s_\alpha({\kk})$ is called a Floquet exponent (the Floquet exponents are also the eigenvalues of $L^{\text{eff}}$, and they are only defined up to a phase).
The Floquet exponent $s_\alpha({\kk}) = \sigma_\alpha({\kk}) + \ii \omega_\alpha({\kk})$ can be decomposed into real and imaginary parts, which correspond to the growth rate $\sigma_\alpha({\kk})$ of the corresponding fundamental solution and its oscillation quasi-frequency $\omega_\alpha({\kk})$ (only defined up to multiples of $2 \pi/T)$.
A positive (negative) growth rate corresponds to an unstable (stable) solution. 
Equivalently, the solution is stable when the absolute value of the Floquet multiplier $|\lambda_\alpha|$ is smaller than  unity.

\medskip 

To determine the stability of the chiral phase, we first solve numerically the mean-field dynamical system in Eq.~\eqref{TonerTuMF_supp} to obtain the time-independent quantities $R_{a} = \lVert \vec{P}_{a} \rVert$ ($a=A,B$) and $\Delta \phi^{A B} = \text{angle}(\vec{P}_A, \vec{P}_B)$.
We obtain the time-dependent mean-field solution $\vec{P}_{a, \text{ss}}(t)$ as well as its period $T$ using Eqs.~\eqref{chiral_ansatz} and~\eqref{chiral_omega_from_params}. 
This allows us to compute the time-dependent matrix $L(t)$ from Eq.~\eqref{LPP_stability_operator}, where $\vec{P}_a$ is replaced by $\vec{P}_{a, \text{ss}}(t)$.
We then use a discretized version of Eq.~\eqref{evolution_operator_series} (where $\delta t$ is finite) to compute $U(T, \kk)$, which is diagonalized to determine the Floquet multipliers $\lambda_\alpha(\kk)$. 
A direct inspection of the spectra shows that as in the time-independent case, we always have $|\lambda_\alpha({\bm 0})| \leq 1$ in the chiral phase (because it is the mean-field solution), with one marginal eigenvalue pinned at $|\lambda| = 1$.
We focus on the multiplier $\lambda({\bm 0})$ with maximal absolute value, for which $|\lambda({\bm 0})| = 1$.
We use the same procedure as for the (anti)flocking phases to estimate the stability of the phase from the sign of the eigenvalues of the Hessian matrix of the function $\kk \mapsto |\lambda(\kk)|$ at $\kk = {\bm 0}$, and the same caveat applies. 
By carrying out this procedure, we find wide regions in parameter space where the chiral phase is stable, as shown in Fig.~3a of the main text.

\section{Non-reciprocity and broken detailed balance}
\label{sec_detailed_balance}
\def\WW{\mathbb{W}}

In this section, we show that the non-reciprocity in Eq.~(2) of the main text ($J_{m n} \neq J_{n m}$) implies the breaking of detailed balance in the corresponding Markov process.
The broken detailed balance (and hence, the lack of microscopic reversibility) implies that the system is out-of-equilibrium, allowing the appearance of oscillating states~\cite{Kondepudi2014,Zhabotinsky2007,Lan2012,Cao2015,Barato2015,delJunco2020}, see also Refs.~\cite{Yan2013,Wang2008}.

We consider a set of coupled Langevin equations
\begin{equation}
  \label{coupled_langevin}
  \partial_t X_m(t) = F_m(X) + \sigma_{m n} \eta_n(t)
\end{equation}
where $X_n$ are a set of random variables for $n=1, \dots, N$, and $\eta_n(t)$ are white noises with unit standard deviation.
This general form encompasses Eq.~(2) of the main text. (It can also describe the coarse-grained Eq.~(1) of the main text, provided that a noise is added to the right-hand side.)
The evolution of the probability distribution $p(t, x)$ of the random variables $X=(X_1, \dots, X_N)$ ruled by Eq.~\eqref{coupled_langevin} is described by the Fokker-Planck equation~\cite{vanKampen2007,Risken1989}
\begin{equation}
  \partial_t p = \hat{\WW} p
\end{equation}
where the operator $\hat{\WW}$ is defined by
\begin{equation}
  \hat{\WW} p = - \partial_m [ F_m(x) p(t,x) ] + D_{m n} \partial_m \partial_n p(t,x)
\end{equation}
for an arbitrary distribution $p$, where $\partial_m = \partial/\partial x_m$ and $D = \sigma \sigma^T/2$.
In the following, we shall assume that the diffusion tensor $D_{m n} = D_m \delta_{m n}$ is diagonal (as $D$ is real and symmetric, this can always be achieved through an appropriate change of variable $x_m \to J_{m m'} x_{m'}$).

Assume that the process defined by $\hat{\WW}$ has a stationary distribution $p_{\text{s}}(x)$ (such that $\hat{\WW} p_{\text{s}} = 0$).
We define a weighted scalar product on distributions by
\begin{equation}
  \braket{f,g}_{\text{s}} = \int \frac{1}{p_{\text{s}}(x)} \, f(x) g(x) \dd x.
\end{equation}
The operator $\hat{\WW}$ satisfies detailed balance (for $p_{\text{s}}$) if it is self-adjoint with respect to this scalar product~\cite{vanKampen2007}, namely if for any distributions $f$ and $g$, 
\begin{equation}
\label{detailed_balance}
\braket{f,\hat{\WW} g}_{\text{s}} = \braket{\hat{\WW} f,g}_{\text{s}}.
\end{equation}
A few tedious but straightforward manipulations show that for the Langevin Eq.~\eqref{coupled_langevin}, the detailed balance condition Eq.~\eqref{detailed_balance} is equivalent to~\cite{Lan2012}
\begin{equation}
  F_m = D_m \frac{\partial_m p_{\text{s}}}{p_{\text{s}}}
\end{equation}
for all $m=1,\dots,N$.
This equation implies (but is not necessarily equivalent to)
\begin{equation}
  \frac{\partial_n F_m}{D_m} = \frac{p_{\text{s}} (\partial_m \partial_n p_{\text{s}}) - (\partial_m p_{\text{s}}) (\partial_n p_{\text{s}}) }{p_{\text{s}}^2}
\end{equation}
which in turn implies that for all $m,n$, one has~\cite{Lan2012}
\begin{equation}
  D_n \partial_n F_m = D_m \partial_m F_n.
\end{equation}
For the system considered in the main text, Eq.~(2), we have $D_m = D_n = \eta^2/2$ for all $m,n$ and
\begin{equation}
  F_m(\theta) = \sum_{n} J_{m n} \sin(\theta_n - \theta_m).
\end{equation}
Hence (for $n \neq m$),
\begin{equation}
  \partial_n F_m(\theta) = J_{m n} \cos(\theta_n - \theta_m)
\end{equation}
and by permuting the indices, 
\begin{equation}
  \partial_m F_n(\theta) = J_{n m} \cos(\theta_m - \theta_n) = J_{n m} \cos(\theta_n - \theta_m)
\end{equation}
We conclude that $J_{m n} \neq J_{n m}$ implies that detailed balance is broken (in the situation when the noises can have different strengths, this condition would read $D_n J_{m n} \neq D_m J_{n m}$).

\section{Nonexistence of the chiral phase with two partially non-reciprocal agents}
\label{two_agents}

In this section, we analyze the evolution of two and only two non-reciprocal agents $A$ and $B$, that represent either two oscillators in the Kuramoto model (without a natural frequency), or two self-propelled agents in the Vicsek model.
We show analytically that stable chiral motion is only possible when the interactions are fully non-reciprocal: any amount of reciprocal interaction eventually leads to alignment, and the chiral behavior is a transient.

We assume that the agents always close enough to interact, and neglect the effect of noise.
The evolution of their angles $\theta_A$ and $\theta_B$ is then described by the equations
\begin{subequations}
\label{two_agents_system}
\begin{align}
  \partial_t \theta_A &= J_{A B} \sin(\theta_B - \theta_A) \\
  \partial_t \theta_B &= J_{B A} \sin(\theta_A - \theta_B).
\end{align}
\end{subequations}
It is convenient to define
\begin{equation}
	\bar{\theta} = \theta_A + \theta_B
	\quad
	\Delta \theta = \theta_A - \theta_B
	\quad
	J_{\pm} = J_{AB} \pm J_{BA}
\end{equation}
in terms of which the dynamical system in Eq.~\eqref{two_agents_system} becomes
\begin{subequations}
\label{two_agents_system_rotated}
\begin{align}
  \partial_t \bar{\theta} &= - J_- \sin(\Delta \theta) \label{two_agents_system_rotated_thetabar}  \\
  \partial_t \Delta \theta &= - J_+ \sin(\Delta \theta) \label{two_agents_system_rotated_Deltatheta}.
\end{align}
\end{subequations}
The coefficient $J_{+}$ represents the reciprocal part of the interactions, while $J_{-}$ represents the non-reciprocal part. 
When the reciprocal interactions vanish, $J_{+}=0$, then $\Delta \theta(t) = \Delta \theta(0)$ is a constant (equal to its initial value), and the average angle $\bar{\theta}(t)$ increases linearly.
This corresponds to a circular motion at a frequency $1/[J_- \sin(\Delta \theta(0))]$, whose characteristics are highly sensitive to the initial conditions.
In an exactly anti-reciprocal system (and in the absence of noise), this circular motion goes on forever.
However, in this very simple model, any amount of reciprocal interaction $J_{+} \neq 0$ leads to the eventual suppression of the circular motion, on a time scale of order $1/J_{+}$.
Indeed, when the reciprocal part of the interaction $J_{+}$ is nonzero, $\Delta \theta$ relaxes to either $0$ or $\pi$ depending on the sign of $J_{+}$ (the solution flows from the unstable fixed point to the stable one). When $J_{+}$ vanishes, the situation is marginal and $\Delta \theta$ remains constant.

We now solve the system \eqref{two_agents_system_rotated} explicitly.
In terms of $y = \tan (\Delta \theta/2)$, for which $y'(t) = - J_{+} y(t)$, Eq.~\eqref{two_agents_system_rotated_Deltatheta} describes an exponential relaxation on a time scale $1/J_{+}$, namely
\begin{equation}
	\tan\left( \frac{\Delta \theta(t)}{2} \right) = \tan\left( \frac{\Delta \theta(0)}{2} \right) \, \exp\left( - J_+ \, t \right).
\end{equation}
The evolution of the average angle $\bar{\theta}(t)$ is slaved to the evolution of $\Delta \theta$, as
\begin{equation}
	\bar{\theta}(t) = \bar{\theta}(0) - J_- \int_{0}^{t} \sin(\Delta \theta(t')) \dd t'.
\end{equation}
As a consequence, $\bar{\theta}(t)$ becomes approximately constant when $\Delta \theta(t)$ approaches $0$ or $\pi$.
More precisely, we find
\begin{equation}
	\bar{\theta}(t) = \bar{\theta}(0) - 2 \frac{J_-}{J_+} \big[ \operatorname{arccot}\left( y_0 \, \ee^{- t \, J_+} \right) - \operatorname{arccot}\left( y_0 \right) \big]
\end{equation}
where $y_0 = \tan\left( \Delta \theta(0)/2 \right)$.
This expression exhibits an indeterminate form as $J_{+} \to 0$, that can however be resolved and yields a linear behavior in time consistent with the previous discussion, 
\begin{equation}
	\bar{\theta}(t) = \bar{\theta}(0) - 2 J_- \, \frac{y_0 \, t}{1+y_0^2} = \bar{\theta}(0) - J_- \sin[\Delta \theta(0)] t.
\end{equation}

\section{Non-reciprocal Kuramoto model}
\label{sec_non_reciprocal_kuramoto}

In this section, we consider a non-reciprocal version of the Kuramoto model~\cite{Ott2008,Ott2009,Watanabe1993,Watanabe1994,Marvel2009,Pikovsky2011,Pikovsky2008,Hong2011,Hong2011b,Hong2012,Tyulkina2018,Montbrio2015} with two communities ($a=A,B$).
We first give numerical evidence that the Ott-Antonsen type mean-field analysis~\cite{Ott2008,Ott2009,Watanabe1993,Watanabe1994,Marvel2009,Pikovsky2011,Pikovsky2008,Hong2011,Hong2011b,Hong2012,Tyulkina2018,Montbrio2015} used in the Methods is consistent with direct simulations of the microscopic model Eq.~\eqref{kcomp_kuramoto_microscopic_si}.
We then show analytically that the time-dependent phases are absent when all oscillators have identical natural frequencies (pictorially, $\Delta_a = 0$) and there is no random noise. 
Perhaps counter-intuitively, noise/disorder is necessary for the occurrence of such time-dependent ordered phases (Sec.~\ref{lack_of_time_dependent_phases}).
We further argue that the interplay between noise/disorder (that continuously resets the evolution, thereby promoting the chase-and-runaway behavior) and the many-body interactions (that makes the state \enquote{rigid}) is responsible for the emergence of time-dependent ordered phases. We support this claim by computing the standard deviations of the order parameter (Sec.~\ref{sec_convergence}). 

The emergence of the time-dependent phases (such as the chiral phase) as a consequence of the interplay between disorder/noise and many-body interactions is strikingly similar to the phenomenon of order-by-disorder that often arises in frustrated many-body systems~\cite{Villain1980,Henley1989,Ionita2013}. 
There, the presence of thermal noise or disorder stabilizes an ordered phase even when the ground state is degenerate.
This analogy implies a surprising connection between geometric frustration and the dynamical frustration due to non-reciprocal interactions that may only arise in non-equilibrium systems. 

We consider the Kuramoto model~\cite{Kuramoto1984,Shinomoto1986,Sakaguchi1988,Acebron2005}
\begin{equation}
\label{kcomp_kuramoto_microscopic_si}
\partial_t \theta_m^a 
= \omega_m^a 
+ \sum_{b=A,B} \sum_{n=1}^{N_b}
J_{a b} \sin(\theta_n^b-\theta_m^a)
+\eta_a(t),
\end{equation}
where $\theta_m^a$ and $\omega_m^a$ are the phase and natural frequency of the $m$-th oscillator in the community $a$,  
$N_a$ is the number of $a$-species oscillators, $\eta_a(t)$ is a white noise acting on species $a$, and 
$J_{ab}$ is the coupling strength between the oscillators in the communities $a$ and $b$. 
The inter-community interaction can be non-reciprocal by allowing $J_{AB}\ne J_{BA}$. 
We assume that all-to-all connections between the oscillators. Unless otherwise specified, we assume that the natural frequencies of species $a$ follow the Lorentzian distribution,  
\begin{eqnarray}
g_{a}(\omega) = \frac{1}{\pi} 
\frac{\Delta_a}{(\omega - \omega_a)^2 + \Delta_a^2}.
\end{eqnarray}
(We will also consider the case where all the frequencies are equal.)
We are mainly interested in the behavior of the order parameters that characterizes the synchronization is given by,
\begin{equation}
z_a(t) \equiv r_a(t) \ee^{\ii\phi_a(t)} 
=\frac{1}{N_a}
\sum_{m=1}^{N_a} \ee^{\ii\theta_m^a(t)},
\end{equation}
which becomes finite when synchronization occurs. 
In the following, we restrict our attention to the $O(2)$ symmetric situation where $\omega_A = \omega_B=0$ (it is crucial that the detuning $\omega_A - \omega_B$ is set to zero, but we can restore a common oscillation frequency $\omega_A = \omega_B$ by a change of reference frame). The $SO(2)$ symmetric case where a detuning can exist (and in which PT-symmetry is explicitly broken) is discussed in the Methods.

\subsection{Time-dependent phases in the PT-symmetric Kuramoto model}
\label{sec PT Kuramoto numerics}
\def\kuramotoj{j}

Here, we compare the dynamics of the non-reciprocal Kuramoto model (without noise) with the mean-field dynamical system
\begin{equation}
	\label{kcomp_kuramoto_ott_antonsen}
	\partial_t z_{a} = (\ii \omega_a - \Delta_a) z_a + \frac{1}{2} \sum_{b} \kuramotoj_{a b} \left[ z_{b} - z_a^2 \, \overline{z_b} \right]
\end{equation}
describing the evolution of the complex order parameter $z_a(t)$ of each community $a$, and where $\overline{z_b}$ is the complex conjugate of $z_b$.
In the limit where $N_a \to \infty$, this mean-field reduction should be exact, see Refs.~\cite{Ott2008,Ott2009,Watanabe1993,Watanabe1994,Marvel2009,Pikovsky2011,Pikovsky2008,Hong2011,Hong2011b,Hong2012,Tyulkina2018,Montbrio2015,Bick2019}.

\begin{figure}
  \centering
  \includegraphics[width=8cm]{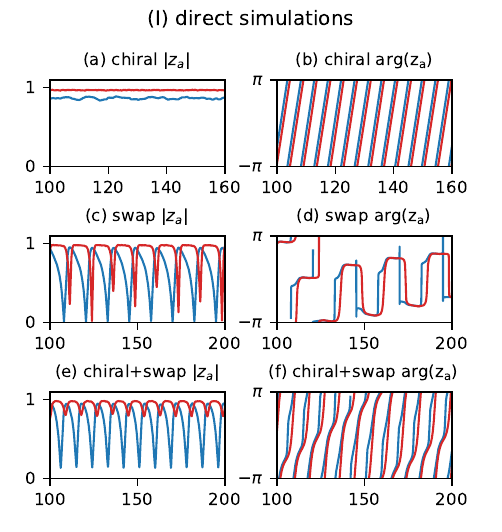}
  \hspace{1cm}
  \includegraphics[width=8cm]{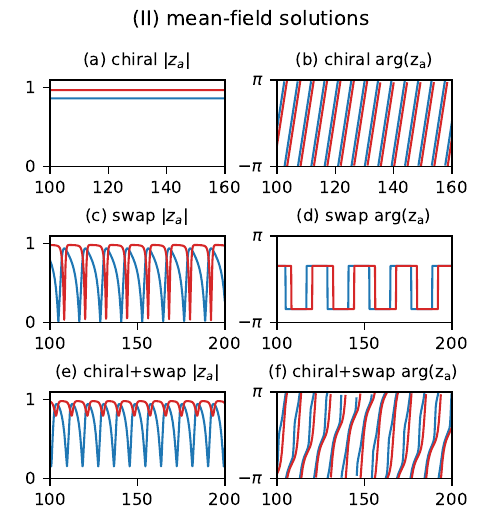}
  \caption{\label{kuramoto_phases_direct_simulations}\strong{Comparison of direct simulations of the Kuramoto model with mean-field solutions.}
  (A) Simulation of $N = \num{4000}$ particles (separated in two species) with Lorentzian frequency distributions with parameters 
  $\omega_{A} =  \omega_{B} = \num{0}$
  $\Delta_A = \Delta_B = \num{0.05}$, without noise.
  We have set $\kuramotoj_{A A} = \kuramotoj_{B B} = \num{1.0}$ and $\kuramotoj_{-} = \num{1.6}$, as well as
  (a) chiral $\kuramotoj_{+} = \num{0.1}$
  (b) swap $\kuramotoj_{+} = \num{0.7}$
  (c) chiral+swap $\kuramotoj_{+} = \num{0.4}$.
  The total duration is $T_{\text{sim}} = \num{200}$ with (a,c) $\delta t = \num{0.05}$ and (b) $\delta t = \num{0.025}$.
  (B) Solutions of the corresponding mean-field dynamical systems.
  The agreement is very good. The most striking difference between (A) and (B) is the existence of a slow drift in the swap phase in the direct simulations (not present in the mean-field analysis), that we attribute to finite-size effects.
  }
\end{figure}

The results of the comparison are presented in Fig.~\ref{kuramoto_phases_direct_simulations}, and show a good agreement between the microscopic simulations and the mean-field description.

\subsection{Lack of time-dependent phases without disorder or noise}
\label{lack_of_time_dependent_phases}

\begin{figure}
  \centering
  \includegraphics[width=12cm]{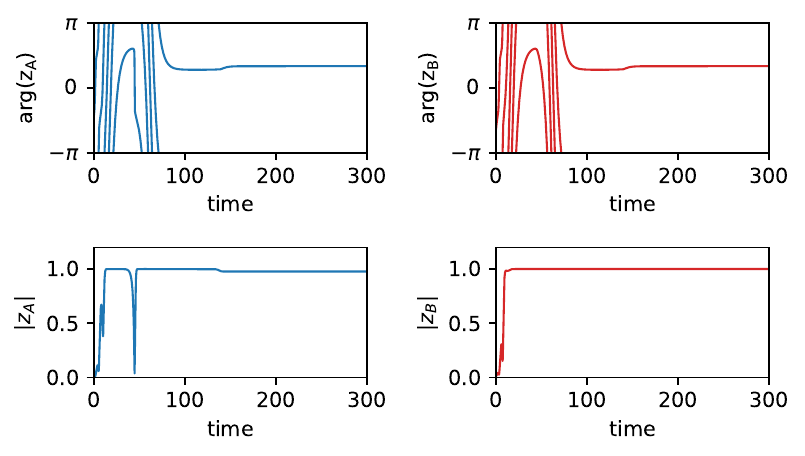}
  \caption{\label{si_figure_eventual_alignment}\strong{Eventual alignment in the absence of noise and/or frequency spread.}
  We compute the standard deviations of the order parameter in the chiral phase (the norms of $\vec{P}_{A}$ and $\vec{P}_{B}$, and their relative angle $\Delta \phi$) obtained by simulations of the microscopic model Eq.~\eqref{eom_discrete}.
  We used the same parameters as in Fig.~\ref{kuramoto_phases_direct_simulations}a, with the exception of the frequency distribution: here all the natural frequencies are equal to $\omega_{A} =  \omega_{B} = \num{0}$.
  In contrast with what Fig.~\ref{kuramoto_phases_direct_simulations}a, we observe here that the oscillators all eventually align.
  We have used $\delta t = \num{0.01}$.
  }
\end{figure}

In this section, we show that when all the oscillators have identical natural frequencies and there is no noise, the dynamics of two \emph{communities} of oscillators can be mapped to the dynamics of two oscillators.
As a consequence, the time-dependent phases (such as the chiral (travelling wave) and swap (periodic synchronization) phases) of the PT-symmetric Kuramoto model are absent in this case.
This explicitly shows that these time-dependent phases are driven by noise/disorder.

We note that this result hinges on the infinite-ranged coupling between the oscillators, and might not hold true with finite-ranged couplings. For instance, Eq.~\eqref{equation_chiral_phase_visek_jpm} in the non-reciprocal Vicsek model suggests the possibility of a chiral phase at zero disorder $\eta = 0$ for $|j_{-}| > |j|$ (although other instabilities not taken into account by this equation might also take place). Yet, we expect the range of parameters in which the chiral is stable to be enlarged by the noise even in this case (compare Fig.~2b and Fig.~2c in the main text).

In the absence of noise ($\eta_a = 0$), the steady-state with frequency $\Omega$ has an order parameter of the form
\begin{equation}
\label{Kuramoto steady state si}
    z_a(t) = z_a \ee^{\ii \Omega t}
    \quad
    \text{with}
    \quad
    z_a = r_a \ee^{\phi_a^0},
\end{equation}
that satisfies the self-consistency condition (derived in the Methods; note that here we start directly from the Kuramoto model Eq.~\eqref{kcomp_kuramoto_microscopic_si}, not from the Ott-Antonsen mean field description)
\begin{equation}
\label{za_si}
z_a = R_a e^{i\alpha_a}F_a[z_a,z_b]
\end{equation}
where
\begin{subequations}
\begin{align}
\label{Ra_si}
R_a \ee^{\ii\alpha_a} &= \sum_b \kuramotoj_{a b} z_b, \\
\label{Fa_si}
F_a[z_A,z_B] &= \int_{-\pi/2}^{\pi/2} \dd\theta \cos\theta \, \ee^{\ii\theta} g_a(R_a\sin\theta +\Omega) +\int_{-\pi}^\pi \dd\theta\int_{|x|>1} \dd x
\, \ee^{\ii\theta} g_a(R_a x + \Omega)
\frac{\sqrt{x^2-1}}{|x-\sin \theta|},
\end{align}
\end{subequations}
and  $\kuramotoj_{a b} = J_{a b} N_b$.
The first and the second term of Eq.~\eqref{Fa_si} are contributions from synchronized and unsynchronized oscillators, respectively. 

In the limit where the width of the distribution of natural frequencies vanish $\Delta_a\rightarrow 0^+$ (so all oscillators have the same natural frequency), we get
\begin{equation}
\label{Fa_delta_si}
F_a = \frac{e^{\ii \zeta_a}}{R_a}
\qquad
\text{with}
\qquad
\zeta_a = - \sin\Big(\frac{\Omega}{R_a}\Big),
\end{equation}
as long as the synchronization frequency is not very far away from resonance $|\Omega| < R_a$, where there are no contribution from the second term of Eq.~\eqref{Fa_si}.

Plugging Eq.~\eqref{Fa_delta_si} into  Eq.~\eqref{za_si} yields
\begin{equation}
z_a(t) = r_a e^{i\phi_a(t)}
= e^{i(\alpha_a+\zeta_a)}e^{i\Omega t}, 
\end{equation}
giving perfect synchronization $r_a = 1$ and $\phi_a(t) = \alpha_a + \zeta_a +\Omega t$.
The perfect synchronization $r_a=1$ implies that all the oscillators in the community $a$ satisfy $\theta_m^a(t)=\phi_a(t)$. 
It then follows (by substituting this in Eq.~\eqref{kcomp_kuramoto_microscopic_si})
that the dynamics (in the long time limit where the amplitude $r_a(t)$ has converged to unity) are described by 
\begin{align}
\label{kcomp_kuramoto_no_noise_si}
\partial_t \phi_A 
&= \kuramotoj_{A B} \sin(\phi_B-\phi_A), \\
\partial_t \phi_B 
&= \kuramotoj_{B A} \sin(\phi_A-\phi_B)
\end{align}
which are identical to the dynamical equations Eq.~\eqref{two_agents_system} for the case of two agents!
This is because, in the absence of noise and disorder, all the oscillators perfectly synchronize within each population, so each group of oscillators simply behaves as a single (macroscopic) oscillator characterized solely by its phase $\phi_a$.

As we show in Sec.~\ref{two_agents}, the state of the system in these phases eventually approaches a constant (i.e. $\Omega=0$) with $\phi_A-\phi_B = 0 (\pi)$ when $J_{AB}+J_{BA}>0 (<0),$ corresponding to the static, coherent ($\pi$-state) phase. 
This explicitly shows that the time-dependent phases that cannot occur without noise or disorder.

These conclusions are numerically confirmed in Fig.~\ref{si_figure_eventual_alignment} from direct numerical simulations of Eq.~\eqref{kcomp_kuramoto_microscopic_si} with $\eta_a=0$ and $\omega_m^a=0$. 
As expected from the arguments above, the phases of the two order parameters eventually align, even though the system would be deep in the chiral phase when a small width in the frequency distribution is introduced (compare with Fig.~\ref{kuramoto_phases_direct_simulations}a where $\Delta_A=\Delta_B=0.05$). 
The magnitude of the order parameters $|z_A|$ and $|z_B|$ approaches one, which is also consistent with the discussions above.
Hence, the existence of the time-dependent (chiral, swap, and chiral+swap) phases crucially hinges on a finite amount of disorder in the frequencies and/or on the presence of a random noise.
The noise/disorder alone would however simply suppress any order without the rigidity provided by many-body interactions, see Fig.~\ref{standard_deviations_sqrt_N} in Sec.~\ref{sec_convergence}.

The exactly non-reciprocal case $J_{AB}=-J_{BA}$ is a singular point, where a finite synchronization frequency $\Omega=\pm(J_{AB}+J_{BA})$ is obtained in the long-time limit.
However, even here, there is an important difference from the chiral phase: the relative phase $\phi_A-\phi_B$ can take an \emph{arbitrary} value set by the initial condition, while in the chiral phase they are determined intrinsically. (See the detailed analysis in Sec.~\ref{sec emergence of chiral phase} for the chiral phase in a non-reciprocal Vicsek model.)

The above analysis clearly shows that noise/disorder is a \emph{necessary} element for the time-dependent phases to emerge in this system. 
This can be intuitively understood as follows:
Let $J_{AB}>0$ and $J_{BA}<0$ with $J_{AB}>|J_{BA}|$.
As discussed earlier, the \emph{predator} group A oscillators would stop chasing the \emph{prey} group B oscillators once they \emph{caught} them (i.e. become aligned). 
(We emphasize that we are not describing a prey-predator model in the usual sense, and that the chase occurs on the circle in which the phases of the oscillators or angles of the self-propelled particles live, not in physical space.)
However, noise can kick this state out of this optimal state for the A component, restarting the chase again. 
Since noise occurs continuously, this (randomized) chase-and-runaway motion will continue even in the long-time limit.
This continuous resetting of the dynamics is reminiscent of the predator-prey population dynamics discussed in Ref.~\cite{McKane2005,Horsthemke1984,Gammaitoni1998}.

\subsection{Convergence of the chiral phase}
\label{sec_convergence}

\begin{figure}
  \centering
  \includegraphics[width=8cm]{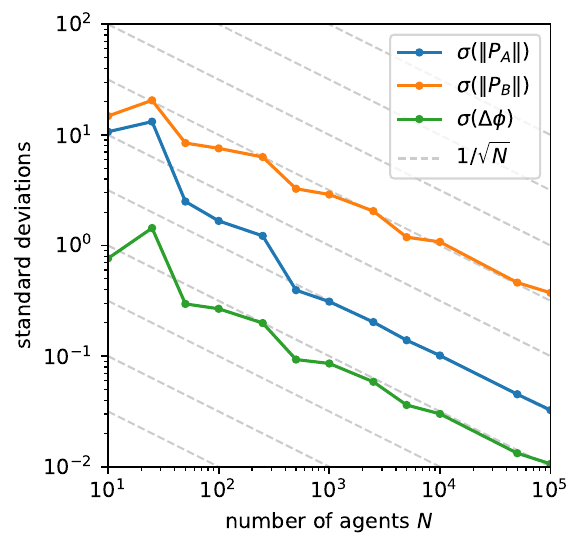}
  \caption{\label{standard_deviations_sqrt_N}\strong{Standard deviations of the order parameters in the chiral phase.}
  We compute the standard deviations of the order parameter in the chiral phase (the norms of $\vec{P}_{A}$ and $\vec{P}_{B}$, and their relative angle $\Delta \phi$) obtained by simulations of the microscopic model Eq.~\eqref{eom_discrete}.
  Here, there is a slight amount of reciprocal interaction between the two species. Without noise, the chiral phase does not exist as the populations eventually synchronize or antisynchronize.
  Noise restores the chiral behavior, at the price of fluctuations of the order parameter, but they decrease approximately as $1/\sqrt{N}$ with the number of agents $N$.
  The gray lines are equally spaced $1/\sqrt{N}$ curves and are meant as a guide to the eye (not a fit).
  In order to single out the influence of the number of agents from other, we set their velocities to zero and their interaction range to be infinite.
  We used $J_{AA} = J_{BB} = \num{1}$, $J_{A B} = \num{1}$, $J_{BA} = \num{-1.1}$, $\eta = \num{8e-2}$. 
  The total duration is $T_{\text{sim}}/\delta t = \num{4000}$ with $\delta t = \num{0.5}$, over which the standard deviation is computed.
  }
\end{figure}

The discussion in the previous subsection confirms that noise/disorder is necessary for the time-dependent phases to appear. However, there is an important caveat: the noise may simultaneously destroy the order. 
Here we argue that the many-body interaction effect plays an equally crucial role in the appearance of the time-dependent phases by making this phase rigid and stable against noise.

Our claim is supported in Fig.~\ref{standard_deviations_sqrt_N} for the chiral phase, where the evolution of the standard deviations\footnote{Note that the amplitude of order parameter contains information on the standard deviation of the individual oscillators. 
This is not what we analyze here. Instead, we discuss the standard deviation of the order parameter itself, which captures whether the phase is well-defined or not.} of the order parameters with $N$ is obtained from simulations (where $N=N_A+N_B$ is the total number of agents).
When the number of agents is still far from the thermodynamic limit ($N\lesssim 10$), the standard deviations are of $O(1)$ (or even larger) meaning that the fluctuations are larger than the order parameter itself, questioning the robustness of the order. 
However, as $N$ gets larger, the standard deviations decrease approximately as $1/\sqrt{N}$, signaling the stabilization of the chiral phase.
A similar scaling is observed in the case of the standard Kuramoto model, see Ref.~\cite{Acebron2005}.

\section{Many-body stabilization of the chiral phase against changes of the chirality}
\label{arrhenius}

We have seen in Sec.~\ref{lack_of_time_dependent_phases} that noise is crucial in the establishment of the chiral phase. 
In this paragraph, we show how noise can also destroy the chiral phase by randomly flipping the chirality over time. 
We also show that this process is exponentially suppressed when the number of agents increases.

This situation is very similar to Néel relaxation theory~\cite{Neel1950,Brown1963} which describes the mean transition time between the two equilibrium states available to the magnetization of single-domain ferromagnetic nanoparticles.

The Kuramoto and Vicsek models
without microscopic noise can be described by an accurate mean-field dynamical system, that (correctly) predicts the lack of a chiral phase in some circumstances.
The microscopic noise leads to the apparition of the chiral phase by a mechanism of constant resetting of the dynamics.
When only a few agents are present, the dynamics is effectively random and it is not clear that it can be correctly captured by a noisy mean-field equation of motion. When the number of agents increases, the effect of noise is reduced and we can expect to capture some aspects of the dynamics (such as the chirality flips) by a noisy dynamical system. As we shall see, the effect of the noise is exponentially reduced with the number of agents, leading to a deterministic dynamical system in the hydrodynamic limit.

\medskip

\begin{figure}
  \centering
  \includegraphics[width=8cm]{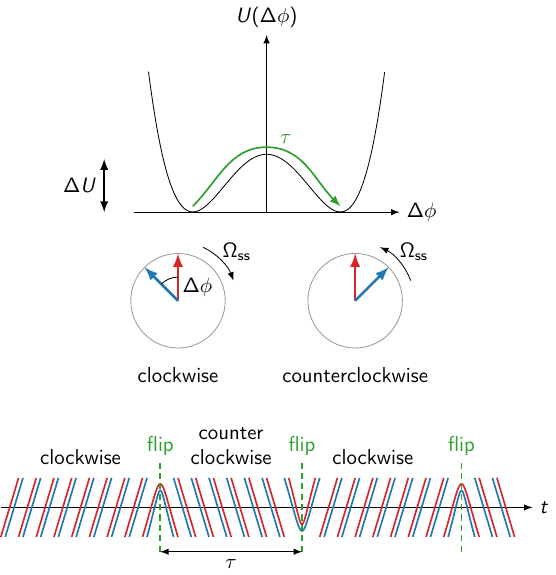}
  \caption{\label{si_figure_chiral_flip}\strong{Noise-activated chirality inversion.}
  }
\end{figure}

The chiral phase spontaneously breaks parity: in a deterministic dynamical system (without noise), the resulting motion can either be clockwise or counterclockwise, depending on the initial conditions.
In reality, some noise (thermal or not) is always present, and the mean-field dynamical system $\partial_t X = f(X)$ (e.g. Eq.~(1) of the main text) becomes a stochastic dynamical system $\partial_t X = f(X) + \xi(t)$ in which $\xi(t)$ is a random noise.
A large enough fluctuation can flip the system from clockwise to counterclockwise, as represented in Fig.~\ref{si_figure_chiral_flip}.
This leads to a statistical restoration of parity: the system undergoes a succession of clockwise and counterclockwise motions, that have an average lifetime $\tau$ (the time between two successive flips), which has to be compared with the duration of an experiment.

The chirality flips occur when the angle $\Delta \phi$ between the order parameters $\vec{v}_A$ and $\vec{v}_B$ changes sign (see Fig.~\ref{si_figure_chiral_flip}, and e.g. Eqs.~\eqref{chiral_ansatz}-\eqref{Deltaphi_supp} in the case of flocking).
Away from the phase transitions, we expect every mode to relax way faster than $\Delta \phi$ (except the global rotation of both $\vec{v}_A$ and $\vec{v}_B$, which can be ignored because of the symmetry).
Hence, we focus only on $\Delta \phi$ and write a reduced dynamical system for $\Delta \phi$, such as
\begin{equation}
    \label{eom_phi_only}
	\partial_t \Delta \phi = h(\Delta \phi) + \xi(t)
\end{equation}
Intuitively, $\Delta \phi$ essentially evolves in a double-well potential, in which the two minima $\pm \Delta \phi_\text{c}$ correspond to the clockwise and counterclockwise chiral phases. 
The variable $\Delta \phi(t)$ fluctuates because of the noise, and can sometimes overcome the potential barrier to go from one minimum to another.
(Despite this intuitive description in terms of a potential, we do not assume that the system is at equilibrium. We simply start from Eq.~\eqref{eom_phi_only}.)
This situation is known as the Kramers problem, and the characteristic time $\tau$ between two transitions follows an Arrhenius law of the form~\cite{Zwanzig2001,Hanggi1990}
\begin{equation}
	\tau = \tau_0 \exp\left(\frac{\Delta U}{\sigma_\xi^2} \right)
\end{equation}
in which $\sigma_\xi$ is the standard deviation of the Gaussian noise $\xi(t)$ with zero mean and with $\braket{\xi(t)\xi(0)} = 2 \sigma_\xi^2 \delta(t)$, and $\Delta U$ is the effective barrier height defined as
\begin{equation}
	\Delta U = - \int_{\pm \Delta \phi_\text{c}}^{0} h(\vartheta) \, \dd \vartheta.
\end{equation}
Here, we have ignored the additional complexity coming from the fact that $\Delta \phi$ is an angle, which does not qualitatively affect the argument.

In section \ref{sec_convergence}, we have seen from numerical simulations that the fluctuations of the order parameter decrease as $1/\sqrt{N}$ when the number of agents $N$ increases (when $N$ is large).
This can be understood as a consequence of the central limit theorem, according to which the average of $N$ independent and identically distributed variables with finite variances $\sigma^2$ converges to a normal distribution with variance $\sigma^2/N$.
Accordingly, we assume that 
\begin{equation}
	\sigma_\xi = \frac{\sigma_0}{\sqrt{N}}
\end{equation}
in which $\sigma_0$ is a constant representing the standard deviation of the noise acting on a single agent (in a thermal system, we would have $\sigma_0^2 \propto k_{\text{B}} T$).
The quantities $\tau_0$ and $\Delta U$ are properties of the mean-field dynamical system, so they do not depend on $N$.
Hence, we find that the escape time behaves as
\begin{equation}
	\tau \simeq \tau_0 \exp\left(N \, \frac{\Delta U}{\sigma_0^2} \right)
\end{equation}
when $N$ is large enough.
In other terms, the lifetime of the chiral phase increases exponentially with the number of agents, leading to a stable chiral phase in the hydrodynamic limit.

\section{Exceptional transition in a non-reciprocal O(3)-symmetric system}
\label{sec_O3}

In this section, we discuss an example of an exceptional transition in a different symmetry group, namely the group of three-dimensional rotations $O(3)$.
This supports the suggestion that exceptional transitions are a generic phenomenon not limited only to $O(2)$-equivariant systems (note, however, that $O(2)$ is indeed a subgroup of $O(3)$). 

The dynamical system
\begin{equation}
	\label{general_O3_invariant}
	\partial_t \vec{v}_a = \alpha_{a b} \vec{v}_b + \beta_{a b c d} \braket{\vec{v}_b, \vec{v}_c} \vec{v}_d
\end{equation}
in which $\vec{v}_a(t)$ is now an element of $\RR^3$ is $O(3)$-equivariant (with the standard diagonal action).
For concreteness, we use the same $\alpha_{a b}$ and $\beta_{a b c d}$ as in the flocking system, and write
\begin{equation}
	\label{flocking_like_O3}
	\partial_t \vec{v}_A
	= 
	\left[
	j_{A A} \rho_A
	- \eta
	- \frac{1}{2 \eta}  \lVert j_{A A} \vec{v}_A + j_{A B} \vec{v}_B \rVert^2
	\right]
	 \vec{v}_A
	+ j_{A B} \rho_A \vec{v}_B
\end{equation}
in which the vectors are now 3D. We define $j_{\pm} = [j_{AB} \pm j_{BA}]/2$.

\begin{figure}
  \centering
  \includegraphics[width=8cm]{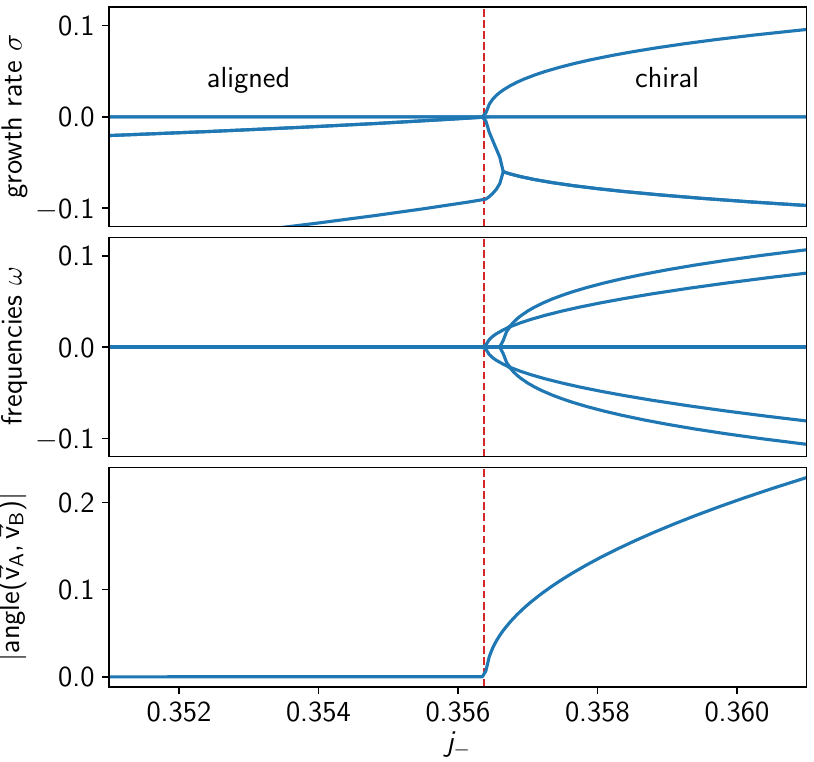}
  \caption{\label{linear_operator_spectrum_O3}\strong{Exceptional transition in a O(3)-symmetric system.}
  The spectrum of the Jacobian $L$ corresponding to Eq.~\eqref{flocking_like_O3} exhibits an exceptional point at the transition between static alignment and chiral motion in which the order parameters orbit in the same plane, with a fixed angle between them.
  The two most unstable distinct eigenvalues $\lambda_i = \sigma_i + \ii \omega_i$ of $L$ coalesce at $j_{-} \approx \num{0.36}$.
  In the aligned phase, they correspond to Goldstone modes (with $\lambda = 0$ in the whole phase) and massive (i.e., damped) modes. 
  This value coincides with the transition (marked by a red dashed line) from a constant solution to a traveling waves state in which both order parameters rotate in a common plane.
  (Note the presence of other exceptional points with finite eigenvalues (away from the red dashed line): they are irrelevant in this discussion.)
  We also show the absolute value of the angle between $\vec{v}_{A}$ and $\vec{v}_{B}$.
  We have set $j_{AA} = j_{BB} = \num{1}$, $\rho_A = \rho_B = \num{1}$, $\eta = \num{0.5}$, and $j_{+} = \num{0.05}$.
  }
\end{figure}

Direct simulations of this dynamical system show that it exhibits a time-independent phase in which the $O(3)$ symmetry is spontaneously broken: $\vec{v}_A$ and $\vec{v}_B$ acquire a finite value (and point in the same direction). This phase is destabilized by non-reciprocal interactions. We observe a transition to a chiral phase, in which the order parameters $\vec{v}_A(t)$ and $\vec{v}_B(t)$ rotate in a common plane (which plane, and in which direction is random). 
This transition is marked by an exceptional point at $s \equiv \sigma + \ii \omega = 0$ in the spectrum of the Jacobian $L$ of the dynamical system, as we show in Fig.~\ref{linear_operator_spectrum_O3}.

\section{Exceptional transition in coupled lasers}
\label{lasers}

In this section, we discuss an example of single-body exceptional transition in laser physics and its destabilisation by the noise.

\medskip

In exceptional point lasers, the main idea is to start with a linear system (e.g., coupled cavities) exhibiting an exceptional point, on top of which a lasing transition occurs. When the EP is close to the laser threshold, various interesting phenomena occur~\cite{Miri2019,Konotop2016,Feng2014,Peng2014,Brandstetter2014,Hodaei2014,Liertzer2012,Lumer2013,Hassan2015}.
Most of the initial literature focuses on linear models~\cite{Konotop2016}, with non-linearities playing essentially no role beyond determining the amplitude of the lasing emission~\cite{Liertzer2012}.
For instance, the EP identified in Ref.~\cite{Liertzer2012} occurs below the laser threshold (see Fig.~2 in Ref.~\cite{Liertzer2012}).
The dramatic effects of EP on the lasing transition are not related with the exceptional transitions analyzed in this paper (this can be understood as follows: the lasing transition, by which the $U(1)$ phase symmetry is spontaneously broken, should be compared to the flocking transition, or more generally the disorder-to-(anti)aligned transition).

\medskip

Further studies analyzed the non-linear behavior of lasers (after threshold), see e.g. Refs~\cite{Lumer2013,Hassan2015,Konotop2016}.
In Ref.~\cite{Hassan2015}, a secondary threshold was predicted and observed, which coincides with what we call exceptional transitions, as we now discuss.

\begin{figure}
  \centering
  \includegraphics[width=8cm]{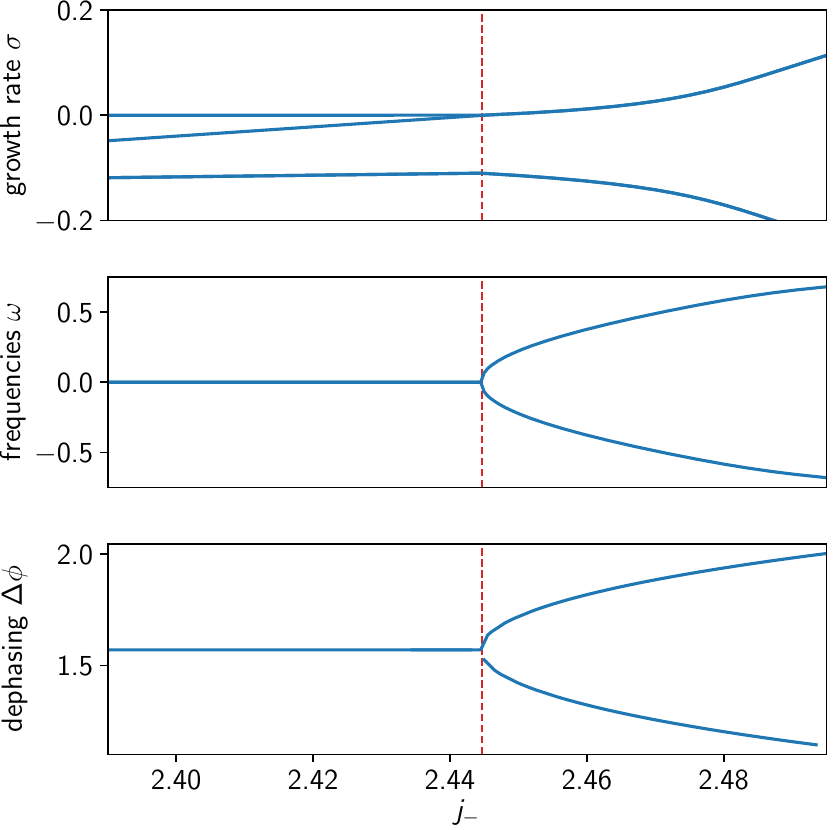}
  \caption{\label{linear_operator_spectrum_laser}\strong{Exceptional transition in a coupled laser cavities.}
  The spectrum of the Jacobian $L$ corresponding to Eq.~\eqref{laser_equations} exhibits an exceptional point at the transition between two different lasing phases, corresponding to the aligned and chiral phases.
  In the aligned lasing phase, the phases $\phi_1$ and $\phi_2$ of the complex amplitudes $A_1$ and $A_2$ described by Eq.~\eqref{laser_equations} are constant, while they increase or decrease at constant rate in the chiral phase (depending on the branch). The transition between these two phases is marked by a red dashed line.
  The two most unstable distinct eigenvalues $\lambda_i = \sigma_i + \ii \omega_i$ of $L$ coalesce at $g_0 \approx \num{2.44}$.
  In the aligned phase, they correspond to the Goldstone mode of the broken $U(1)$ symmetry (with $\lambda = 0$ in the whole phase) and a massive (i.e., damped) mode. 
  We also show the absolute value of the angle between $\vec{v}_{A}$ and $\vec{v}_{B}$, which bifurcates into two branches in the chiral phase.
  The frequency of the steady-state in the chiral phase also undergoes a pitchfork bifurcation (not shown here, see Fig.~5 of Ref.~\cite{Hassan2015}).
  We have set $f_0 = \num{2}$ and $\gamma = \num{0.1}$, following Fig.~5 of Ref.~\cite{Hassan2015}.
  }
\end{figure}

Ref.~\cite{Hassan2015} considers two coupled ring-shaped laser cavities.
The evolution of the (nondimensionalized) complex amplitudes of light in each cavity $A_1$ and $A_2$ is described by
\begin{subequations}
\label{laser_equations}
\begin{align}
	\partial_t A_1 &= - \gamma A_1 + g_0 \, \frac{A_1}{1+|A_1|^2} + \ii A_2 \\
	\partial_t A_2 &= - \gamma A_2 - f_0 \, \frac{A_1}{1+|A_2|^2} + \ii A_1
\end{align}
\end{subequations}
(the quantities called $A_1$ and $A_2$ here are called $a_1$ and $a_2$ in Eq.~(3) of Ref.~\cite{Hassan2015}).
The linear system obtained by setting $|A_1|^2 = 0 = |A_2|^2$ in Eq.~\eqref{laser_equations} (i.e., linerizing around $A_1 = A_2 = 0$) has an EP when $|f_0 + g_0| = 2$.
The lasing transition occurs when the complex amplitudes $A_a$ acquire a finite value, spontaneously breaking the $U(1)$ symmetry $A_a \to \ee^{\ii \theta} A_a$ (see e.g. Ref.~\cite{DeGiorgio1970,Graham1970,Haken1975,Gartner2019}).
From our perspective, the square-root bifurcation of the steady-state frequency $\Omega_{\text{ss}}$ (called $\lambda$ in Fig.~5 of Ref.~\cite{Hassan2015}) corresponds to an exceptional transition.
This can be seen by computing the eigenvalues of the Jacobian of the dynamical system in Eq.~\eqref{laser_equations}, see Fig.~\ref{linear_operator_spectrum_laser}.
We point out that this transition is not associated with an EP of Eq.~\eqref{laser_equations} linerized around $A_1 = A_2 = 0$ (because it is a secondary bifurcation on top of the symmetry-broken state).
(In the current case, it does however coincide with an EP of the \emph{nonlinear} Hamiltonian, see the discussion in section~\ref{subsec_chiral}.)

\medskip

In Ref.~\cite{Hassan2015}, an experiment is carried out in which a transition is observed from one to two peaks in the emission spectrum of the laser (Fig.~8b and c of Ref.~\cite{Hassan2015}). The two resonances in the equivalent of the chiral phase is attributed to the noise, that excites the two clockwise and counterclockwise modes (in blue and red in Fig.~5  of Ref.~\cite{Hassan2015}). Because of the noise, the clockwise/counterclockwise symmetry is restored on average. We interpret this in terms of noise-assisted chirality flips in Sec.~\ref{arrhenius}, in which we show that the lifetime of the chiral phase (without flips) can be exponentially enhanced by many-body effects.
In principle, this could also be realized in non-Hermitian versions of photonic networks containing many coupled lasers~\cite{Nixon2013,Pal2017,Mahler2020,Parto2020,Ramos2020,HonariLatifpour2020,HonariLatifpour2020b,Acebron2005}.
A similar phenomenon can also occur in quantum polariton lasers~\cite{Hanai2019}.

\section{The case of multiple populations}
\label{sec_multiple_populations}

In this section, we discuss the case of systems composed of more than two species interacting non-reciprocally.
Most of the time, we expect their properties to reduce to the case of two species.
This is because the phase transitions (bifurcations) involve only the one or two most unstable modes. For instance, the chiral phase occurs when a damped transverse mode coalesce with a Goldstone mode (see Sec.~\ref{subsec_chiral}), while the swap phase emerges when the damping rate of a longitudinal mode vanishes.
These phase transitions should occur even when more than two populations are involved, provided that the most unstable modes correspond to the ones described above, while the other modes are more damped and effectively irrelevant.  

That being said, richer behaviors may possibly arise when collective modes unique to multi-species systems become unstable.
For example, we expect an additional phase transition within the chiral phase to arise when the \emph{third} transverse mode (which do not exist in the two-species system) becomes unstable. (We order the modes by decreasing non-positive growth rate.)
There, the three transverse modes are expected to affect each other non-reciprocally, which may potentially give rise to a more complex time-dependent phase in the order parameter dynamics, that might include chaotic behavior.  
Another possibility is the emergence of higher-order exceptional points (see e.g. Ref.~\cite{Ashida2020}), where more than two eigenmodes coalesce simultaneously (corresponding to a Jordan block of size $n > 2$), which may arise at higher codimension. 
From the point of view of bifurcation theory, this would be a special case of a Takens-Bogdanov type bifurcation with higher codimension as described in Ref.~\cite{Govaerts1993}.
In the vicinity of a $n$-th order exceptional point (where $n$ modes coalesce), the eigenvalues typically behave as 
$s\sim \ii\ee^{\ii 2\pi m/n} \Delta^{1/n}$ ($m=0,1,\dots,n-1$),
where $\Delta$ characterizes the distance from the exceptional point. 
As a result, we expect a higher-order exceptional point enforced pattern formation to occur in spatially extended many-body systems, even in systems \emph{without} flow (where there is no convective terms).
This is because the dispersion relation is expected to take the form $s(\bm k)\sim \ii\ee^{\ii 2\pi m/n} \bm k^{2/n}$ at the $n$-th order exceptional point (note that the finite $\bm k$ contribution $\Delta\sim \bm k^2$ characterizes the distance from the exceptional point).
Hence, a dynamical instability with the growth rate $\sigma(\bm k)=\text{Re}\, s(\bm k)>0$ should emerge at finite momentum for $n\ge 3$. 
This is in contrast to the ($n=2$) exceptional point enforced pattern formation discussed in Sec.~\ref{subsec exceptional point instability}, where the convective term of the flocking model plays an essential role. 
Finally, the higher-order exceptional point (with a higher codimension) could give rise to different types of phase transitions with more complicated hysteretic behavior.

\section{Pattern formation in the incompressible regime}
\label{incompressibility_constraint}

In this section, we discuss the influence of the incompressibility constraint on pattern formation. 
In the main text, we have entirely ignored the incompressibility constraint $\div(\vec{v}_{a}) = 0$ that would arise from Eq.~\eqref{density_eom_si} in a system where mass is conserved, while still assuming constant densities $\rho_a \approx \text{const.}$ in order to focus on the single hydrodynamic equation \eqref{eom_with_gradients_two_populations_si}.

We now show that the exceptional point enforced pattern formation (discussed in Sec. \ref{subsec exceptional point instability}) still occurs when the incompressibility constraint $\div(\vec{v}_{a}) = 0$ is enforced.

To do so, we project the hydrodynamic equation onto the subspace of divergence-free fields.
At the level of the linear stability analysis, this is done by applying the Leray projector~\cite{Leray1934,Ozanski2017} on the Jacobian $L(k)$.  

\begin{figure*}
  \centering
  \hspace*{-1cm}
  \includegraphics{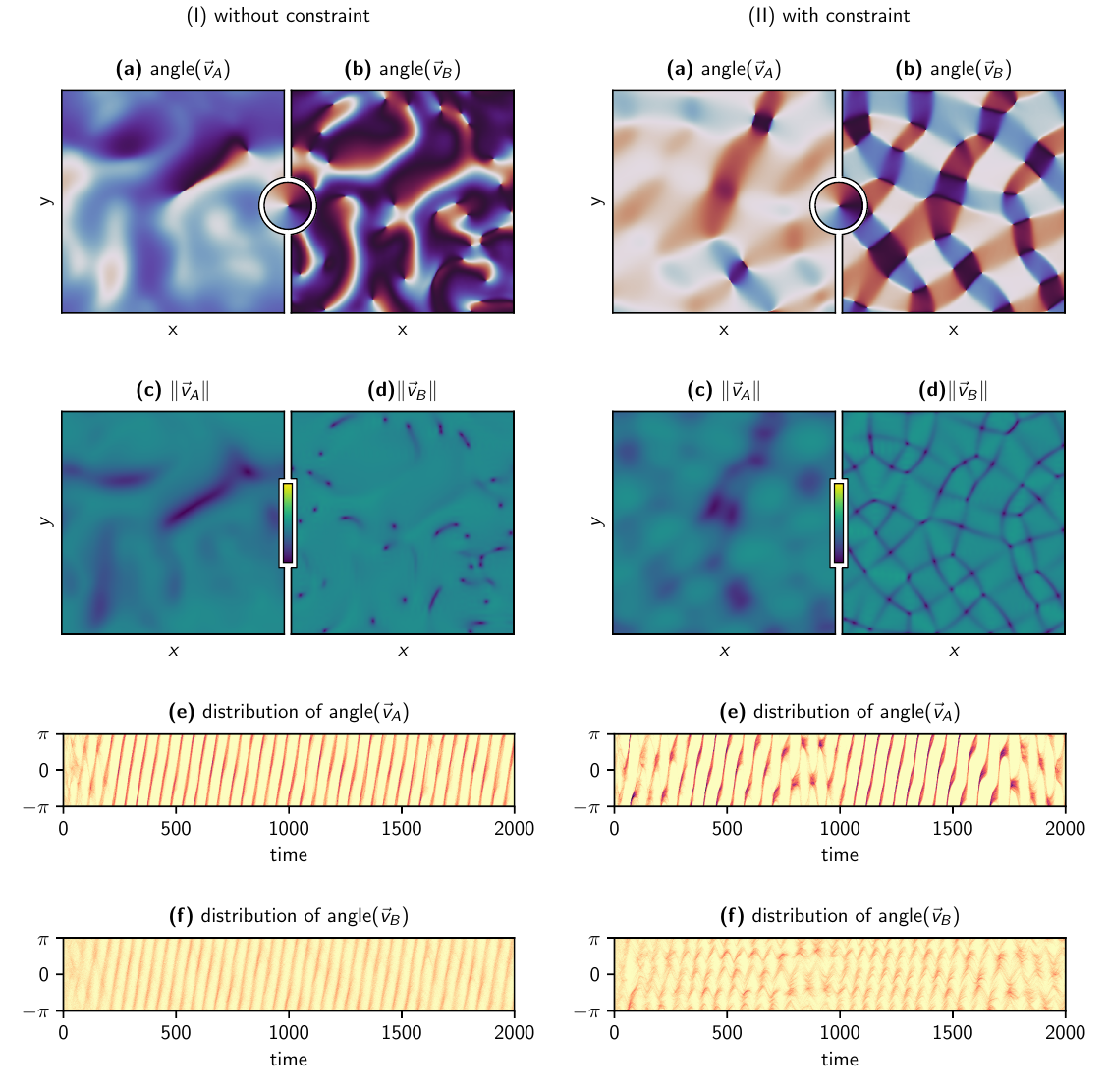}
  \caption{\label{si_figure_patterns_with_without_constraint}\strong{Pattern formation without and with incompressibility constraint}
  We used the same parameters as in Fig.~3d-f of the main text.
  }
\end{figure*}

\begin{figure}
  \centering
  \includegraphics[width=10cm]{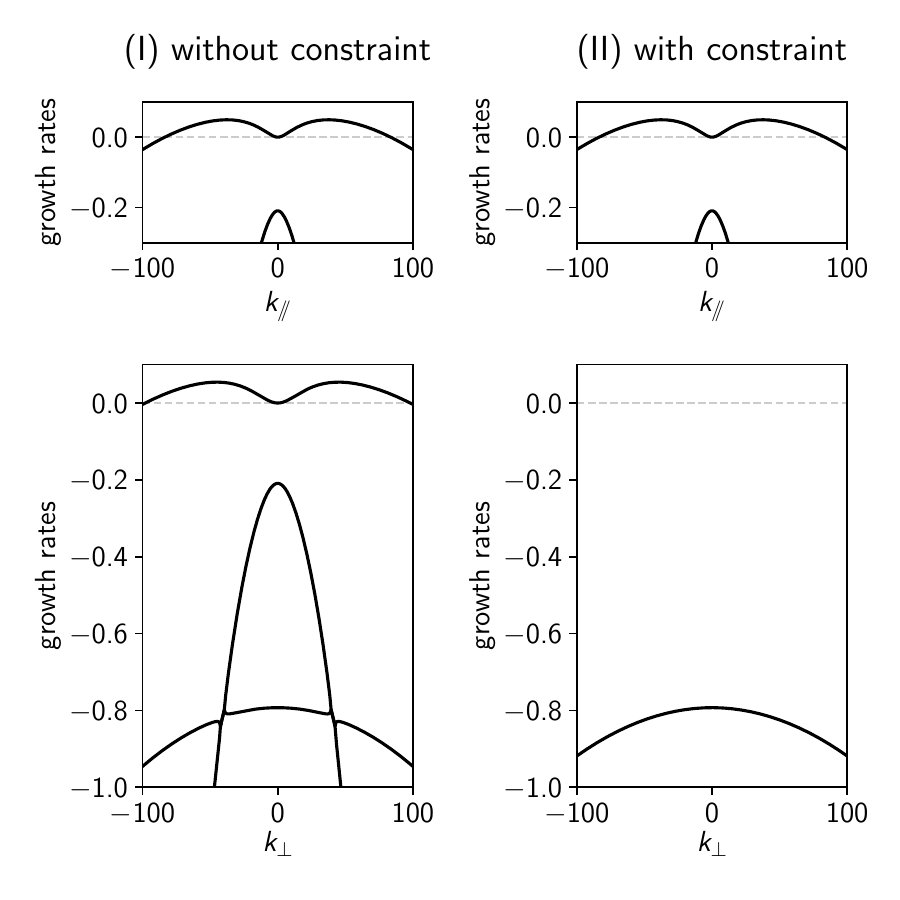}
  \caption{\label{si_figure_linear_stability_hydro_simus}\strong{Linear stability analysis without and with incompressibility constraint}
  We used the same parameters as in Fig.~3d-f of the main text.
  }
\end{figure}

The situation where the fluid is really incompressible can be analyzed by projecting the hydrodynamic equation onto the subspace of divergence-free fields. 
At the level of the linear stability analysis, we use the Leray projector
$\mathscr{P}(k) = \Id - \ket{\hat{k}}\!\bra{\hat{k}}$ (where $\hat{k} = \vec{k}/\lVert \vec{k} \rVert$), see Refs.~\cite{Leray1934,Ozanski2017}, and consider the projected linear operator
\begin{equation}
	L_{\text{incomp.}}(k) = \mathscr{P}(k) L(k) \mathscr{P}(k)
\end{equation}
restricted to the range of the orthogonal projection.
We find that the exceptional-point-driven instability predicted in the main text is still present in the linear stability analysis of the incompressible fluid, while some of the other instabilities can disappear (see Fig.~\ref{si_figure_linear_stability_hydro_simus} and SI Movie \mainref{movie_patterns_incompressible}{5}).
We have also performed simulations of the full hydrodynamic equation~\eqref{eom_with_gradients_two_populations_si} in the incompressible case (as in the main text, the simulations are performed under periodic boundary conditions using the open-source pseudospectral solver Dedalus~\cite{Burns2020}). 
To do so, two pressure fields $\Pi_a$ are added as Lagrange multipliers of the incompressibility constraints. 
Accordingly, we added a term $-\nabla \Pi_a$ to the RHS of the equation of motion for $\vec{v}_{a}$.
The simulations confirm the existence of pattern formation beyond the linear stability analysis, but the patterns are different, see Fig.~\ref{si_figure_patterns_with_without_constraint}).

We compare both situations in Figs.~\ref{si_figure_patterns_with_without_constraint} and \ref{si_figure_linear_stability_hydro_simus}, that respectively show the results of simulations of the non-linear hydrodynamic theory, and the corresponding linear stability analysis of the uniform flow.

In both cases, it is unclear whether the pattern will stabilize at long times. 
Interestingly, we find indications that stable or metastable lattices of vortices (somewhat similar to Abrikosov lattices) can occur in some situations. 
These questions will be analyzed elsewhere.

\clearpage
\twocolumngrid

\clearpage

\end{document}